\documentclass[article]{aa} 

\usepackage{graphicx}
\usepackage{longtable}
\usepackage{orcidlink}
\usepackage{multirow}
\usepackage{caption}
\usepackage{subcaption}
\usepackage[flushleft]{threeparttable}
\usepackage{subcaption}
\usepackage{float}
\usepackage{txfonts}
\usepackage{mwe}
\usepackage{siunitx}
\usepackage{indentfirst}
\usepackage{chemist}

\usepackage{ulem}
\usepackage{pdflscape}
\usepackage{hyperref}
\usepackage{threeparttablex}

\defcitealias{Rathour2024A&A...686A.268R}{RSR 24}

\begin{document}
\title{Non-evolutionary effects on period change in Magellanic Cepheids.}
\subtitle{II. Empirical constraints on non-linear period changes}

\titlerunning{Period change investigation of OGLE Magellanic Cepheids II}
\authorrunning{Rathour et al.}
\author{Rajeev Singh Rathour \inst{1}\orcidlink{0000-0002-7448-4285}, 
Radosław Smolec \inst{1}\orcidlink{0000-0001-7217-4884}, 
Gergely Hajdu \inst{1}\orcidlink{0000-0003-0594-9138},  
Oliwia Zi\'ołkowska \inst{1}\orcidlink{0000-0002-0696-2839}, 
Vincent Hocd\'e \inst{1}\orcidlink{0000-0002-3643-0366},\\
Igor Soszy\'nski \inst{2}\orcidlink{0000-0002-7777-0842}, 
Andrzej Udalski \inst{2}\orcidlink{0000-0001-5207-5619},
Paulina Karczmarek \inst{1,3}\orcidlink{0000-0002-0136-0046}}

\institute{Nicolaus Copernicus Astronomical Centre, Polish Academy of Sciences, Bartycka 18, 00-716 Warszawa, Poland\\
\email{rajeevsr@camk.edu.pl}
\and
Astronomical Observatory, University of Warsaw, Aleje Ujazdowskie 4, Warsaw, 00-478, Poland
\and
Departamento de Astronomía, Universidad de Concepción, Casilla 160-C, Concepción, Chile
} 

\date{Received ??; accepted ??}
\abstract
{Classical Cepheids are not only excellent standard candles, but also invaluable tools to test stellar evolution and pulsation theories. Rates of their pulsation period change, quantified usually through $O-C$ diagrams, can be confronted with predictions of stellar evolution theory. On the other hand, period changes on much shorter time scales ($\sim$10$^{2}$-10$^{4}$days), attributed to non-evolutionary effects are often detected and lack detailed explanation.
}
{We aim to provide a systematic and quantitative description of irregular or non-linear period changes in Cepheids. Such a study is crucial for a complete understanding of period changes in Cepheids and is key to decoupling the evolutionary aspects from the non-evolutionary ones. 
}
{We analysed part of the OGLE data for classical Cepheids in the Magellanic Clouds (MCs; from both Large Magellanic Cloud, LMC, and the Small Magellanic Cloud, SMC) using the modified Hertzsprung $O-C$ technique. A sample of 3658 stars, with the best quality data and void of additional low-amplitude periodicities (e.g. due to non-radial pulsations), that could impact the results, was selected for analysis. Based on $O-C$ shapes, stars were classified into three categories: no period change (class 1), linear period change (class 2), and irregular change (class 3). The Eddington-Plakidis test, wavelet analysis, Stetson index, and instantaneous period method were used to characterise class 3 candidates. We also investigated the correlation between the irregular period change in Cepheids and their metallicity environment
}
{In our investigation, $33.5\pm0.7$\%  of analysed stars show irregular period changes. Considering the pulsation mode,  irregular period changes were detected in $16.5\pm0.7$\% of the analysed fundamental mode stars and in $68.1\pm1.2$\% of the first overtone stars. The amplitude of variability in the $O-C$ diagrams increases with the pulsation period, and at a given pulsation period, it is larger for first overtone stars. While the increase is linear for first overtone stars, for fundamental mode stars it becomes steeper as the pulsation period increases. Time scales of the observed variability range from a few hundred to a few thousand days. 

}
{Irregular period changes are a ubiquitous property of classical Cepheids and may impact the derivation of secular, evolutionary period change rates; hence their quantitative characterisation is essential. The nature of these changes is still unknown. Our research provides observational constraints on their modelling. The markedly higher frequency of irregular period variations in first overtone Cepheids is a key observation that must be accounted for by the models. 
}

\maketitle


\section{Introduction}
\label{sec: Introduction}
Classical Cepheids, also known as Type-I Cepheids (hereafter referred to as Cepheids), are pulsating stars with a range of masses, $\sim3-13$\, M$_{\odot}$, and periods $\sim1-100$\,d. The evolutionary phases of these stars cover a short post-main sequence phase (shell hydrogen burning) and the significantly longer core helium-burning stage. They are very well known to be regular pulsators and follow tight Period-Luminosity ($P-L$) relationships (\citealt{leavitt1908}; also known as Leavitt law), which makes them the backbone of the extragalactic distance scale \citep[e.g.][and references therein]{Freedman2001ApJ...553...47F,Riess2021ApJ...908L...6R}, to eventually measure the Hubble constant. Apart from cosmological applications, Cepheids are excellent laboratories for testing both pulsation theory \citep[e.g.][]{Buchler2009AIPC.1170...51B, Marconi2013ApJ...768L...6M} and evolution theory \citep[e.g.][]{CassisiSalaris2011ApJ...728L..43C}. These bright young stars with stable light curves also exist in eclipsing binaries, giving precise stellar parameters \citep[e.g.][]{Pietrzynski2010Natur.468..542P,Pilecki2013MNRAS.436..953P,Pilecki2015ApJ...806...29P}.

The $\kappa$-$\gamma$ mechanism acting in the partially ionised helium and hydrogen zones \citep[e.g.][]{Cox1980tsp..book.....C} excites pulsations in Cepheids. The mechanism is operational in a specific region in the Hertzprung-Russell diagram called the instability strip (IS). Cepheids are known to have 
up to three crossings through the IS, depending primarily on their mass. The first crossing occurs during the post-main sequence phase while the star is burning hydrogen in the shell. This crossing is significantly faster ($\sim10-100$ times) than the subsequent ones. The second and third crossings comprise the `blue loop' (core helium burning phase). During these two crossings, Cepheid evolution occurs on the nuclear time scale and as a consequence, stars move slowly through the IS. In the first and the third crossings, Cepheids evolve redward, whereas during the second crossing, they move blueward. The extent of the blue loops is sensitive to micro-physical input such as nuclear reaction rates \citep[e.g.][]{Weiss2005A&A...441.1129W,Morel2010A&A...520A..41M,Ziolkowska2024ApJS..274...30Z} as well as the treatment of macro-physical details such as mass loss \citep[e.g.][]{Bono2006MmSAI..77..207B,Neilson2011A&A...529L...9N}, convective overshooting \citep[e.g.][]{Neilson2011A&A...529L...9N}, and rotation of the star \citep[e.g.][]{Anderson_Richard2014A&A...564A.100A,Smiljanic2018A&A...616A.112S}. Early evolutionary models \citep{Iben1965ApJ...142.1447I,Becker1977ApJ...218..633B,Becker1985cto..conf..104B, Xu2004A&A...418..213X} suggested two more possible crossings (fourth and fifth) during the shell helium-burning phase; however, with advancement in the opacity tables, newer models \citep[e.g.][]{Meynet2000A&A...361..101M,Meynet2002A&A...390..561M,Bono2000ApJ...543..955B,Salasnich2000A&A...361.1023S} suggest that shell helium burning occurs after the ignition of core oxygen-burning, hence, only allowing for first three IS crossings.

The measure of the decade-long changes in the observed period of classical pulsators gives a window to probe into the stellar evolutionary effects causing them. A long-known traditional technique to carry out such a measurement is the $O-C$ method, where `$O$' stands for observed and `$C$' stands for calculated. A linear change in period is characterised by a parabolic shape of the $O-C$ curve; however, there is no physical reason for evolutionary period change behavior to be strictly linear \citep{Fernie1990PASP..102..905F}. The very first convincing reason is that since the Cepheid evolution along a track in the HR diagram is in itself nonlinear in time, the period evolution is not linear either, giving rise to a departure from parabolic shape in the $O-C$ diagram.

More than a century ago, the very first study of pulsation period variation in Cepheids (for $\delta$~Cephei) was reported by \cite{Hertzsprung1919AN....210...17H}, using observations from 1785 to 1911. Since then, period change has played a key role in investigating the crossing number and testing stellar evolutionary models \citep[e.g.][]{Turner1998JAVSO..26..101T,Turner2003A&A...407..325T,Turner2004A&A...423..335T,Neilson2012ApJ...760L..18N, Anderson_Richard2014A&A...564A.100A,Anderson_Richard2016A&A...591A...8A}. Numerous developments have been made on both the observation and theoretical fronts to investigate the evolutionary period changes in Cepheids. 

In the last couple of decades, observed period change rates for sizable samples were reported in the LMC \citep{PietrukowiczLMC2001AcA....51..247P,Poleski2008AcA....58..313P,Karczmarek2011AcA....61..303K}, SMC \citep{PietrukowiczSMC2002AcA....52..177P}, and Galactic fields \citep[e.g.][]{Turner2006PASP..118..410T,Csornyei2022MNRAS.511.2125C}. The study by \cite{Turner2006PASP..118..410T} gave empirical evidence that Cepheids with a positive period change rate constitute nearly two-thirds of the sample. The most recent comprehensive study of Cepheid period change rates for the LMC \citep{Rodriguez-Segovia2022MNRAS.509.2885R} uses data covering nearly a century-long baseline combining Digital Access to a Sky Century @ Harvard \citep[DASCH;][]{Grindlay2012IAUS..285...29G}, Optical Gravitational Lensing Experiment \citep[OGLE;][]{udalski2015ogle}, All-Sky Automated Survey \citep[ASAS;][]{Pojmanski1997AcA....47..467P} and the MAssive Compact Halo Object \citep[MACHO;][]{Alcock1996ApJ...461...84A,Alcock2000ApJ...542..281A} data.

On the theoretical front, a long-standing problem is the Cepheid mass discrepancy. Masses of classical Cepheids, as predicted by stellar evolutionary calculations, are significantly larger (10--20\% at present; e.g. \citealt{Keller2008ApJ...677..483K}) than predicted by stellar pulsation calculations \citep[e.g.][]{Cox1980tsp..book.....C,Moskalik1992ApJ...385..685M,Bono2000ApJ...543..955B,Keller2008ApJ...677..483K}. Recent Cepheid dynamical mass measurements based on eclipsing binary systems \citep{Pietrzynski2010Natur.468..542P, Pilecki2018ApJ...862...43P} agree with pulsation theory predictions. The discrepancy with the evolution theory predictions may be lifted when a mild amount of overshooting or pulsation-driven mass loss are included \citep[see e.g.][]{Neilson2011A&A...529L...9N}. Observed evolutionary period change rates also give an important constraint for this long-standing enigma. Stellar evolution and pulsation models for the LMC \citep{Fadeyev2013AstL...39..746F} and Galactic metallicity \citep{Fadeyev2014AstL...40..301F} considering convective core overshooting have resulted in good agreement with a large collection of period change rate observations \citep{Turner1998JAVSO..26..101T,Turner2006PASP..118..410T,PietrukowiczLMC2001AcA....51..247P,Poleski2008AcA....58..313P}. Evolutionary models and population synthesis were used in a study by \cite{Neilson2012ApJ...760L..18N} and they concluded that enhanced mass loss is necessary for a better agreement with observed period change rates from \cite{Turner2006PASP..118..410T}. Still, reproducing the ratio of positive to negative period change rates remains an issue. The impact of rotation on Cepheid evolution and its consequences on the predicted period change rates was investigated by \cite{Anderson_Richard2014A&A...564A.100A,Anderson_Richard2016A&A...591A...8A}. Their models, including mild convective core overshoot, standard (non-enhanced) mass loss, and rotation satisfactorily reproduced observed period change rates. On the other hand, \cite{Miller2020ApJ...896..128M}, incorporating convective core overshooting and stellar rotation, they concluded that rotation and/or overshooting alone, cannot account for the observed period change rates. Hence, they hint toward a need for enhanced pulsation-driven mass loss as missing physics. \cite{Espinoza2022MNRAS.tmp.2545E} used the MESA evolution and the MESA-RSP pulsation codes \citep{Paxton2019ApJS..243...10P} to generate theoretical models including metallicity, convective overshooting, and rotation effects to compute theoretical period change rates. Their predictions are limited to $4-7$\,M$_{\odot}$, and barring the short-period regime, they agree well with LMC period change rate observations \citep{Rodriguez-Segovia2022MNRAS.509.2885R}.

In the first paper of this series \citep[][hereinafter RSR24]{Rathour2024A&A...686A.268R}, we presented a sample of Cepheids exhibiting non-evolutionary period changes which were due to the light travel time effect \citep[LTTE,][]{Irwin1952ApJ...116..211I,Irwin1959AJ.....64..149I} indicating the likely presence of a binary companion. In this work we move on to the second kind of non-evolutionary period changes, which is much less understood. For a long time, the literature has reported certain effects in the $O-C$ diagrams that are too fast (order of thousand days) to be attributed to stellar evolution. Such effects were mentioned as irregularities in the expected parabolic shape of the $O-C$ diagrams, and sometimes referred to as {\it period change noise} \citep{Szabados1983Ap&SS..96..185S,Zhou1999PBeiO..33...17Z}. S~Vul is a classic example of a well-studied Cepheid with a positive period change rate with some wave-like features in the $O-C$ diagram \citep[see e.g.  fig.~25 from][]{Csornyei2022MNRAS.511.2125C}. Such quasi-periodic period fluctuations become more prominent for longer pulsation periods \citep{Percy1997PASP..109..264P,Percy1999PASP..111...94P,Molnar2019ApJ...879...62M,Csornyei2022MNRAS.511.2125C}. One of the earliest studies of period changes in the Magellanic Cloud Cepheids which revealed some irregular period change candidates was done by \cite{Deasy1985MNRAS.212..395D}. The authors attributed the irregularities to probable phase discontinuities due to small atmospheric changes. In a series of works \cite{Szabados1983Ap&SS..96..185S,Szabados1984IAUS..105..445S,Szabados1989CoKon..94....1S,Szabados1991CoKon..96..123S,Szabados1992ASPC...32..255S} collected a sizable sample of Cepheids showing extreme cases of irregular period changes, which are characterised by broken linear fits to the $O-C$ diagrams. These are mainly `\textit{phase jump}' (stepwise $O-C$) and `\textit{phase slip}' (sawtooth-like $O-C$) phenomena with earliest mention in \cite{Szabados1989CoKon..94....1S}. In the former case, the pulsation phase experiences abrupt change keeping the period the same, whereas the latter case is shown to have a sudden jump in the pulsation period followed by re-jump to the previous period  \citep[for examples see][]{Csornyei2022MNRAS.511.2125C}. However, circumstantial evidence of these phase jumps observed in only binary Cepheids yet points towards it being an extrinsically induced phenomenon \cite{Szabados1989CoKon..94....1S,Szabados1992ASPC...32..255S}; a hypothesis that still needs to be validated.

There are other processes that may explain the non-evolutionary period change. For example, in the theoretical work by \cite{Sweigart1979A&A....71...66S}, it was proposed that the deviations from the evolutionary period change could be a consequence of several discrete mixing events leading to the redistribution of composition in the core. Such mixing may be due to semiconvective zones and can be random, leading to irregularities superimposed on evolutionary period changes. Conversely, these abrupt period changes are a direct window to probe the semiconvective process in classical pulsators. \cite{Stothers1980PASP...92..475S} advocated hydromagnetic effects causing abrupt period changes, as a consequence of sudden generation or destruction of the magnetic field inside pulsators (in the context of RR~Lyrae stars). \cite{Cox1998ApJ...496..246C} suggests that the small variations in helium abundance gradients below the hydrogen and helium convection zones could lead to sudden period changes in Cepheids. The work further proposes that convective overshooting induces intermittent helium dredge-up on much short time scales ($\sim$days). mass loss episodes comprise yet another non-evolutionary mechanism proposed for classical pulsators to cause pulsation period changes on shorter time scales \citep{Laskarides1974Ap&SS..27..485L,Koopmann1994ApJ...423..380K}. The recent efforts in modelling stochastic oscillations for pulsating stars, ranging from solar-like to Mira stars \citep{Avelino2020MNRAS.492.4477A,Cunha2020MNRAS.499.4687C}, also seem to be an interesting avenue.

Testing these models for classical Cepheids, as a first step, requires a substantial observational sample of stars showing irregular period changes. Hence, compiling a homogeneous sample of such stars and a quantitative description of the effect is the motivation for this work. For this purpose, we analyze OGLE photometry for MC Cepheids. The structure of the paper is the following. In Sect.~\ref{sec: Analysis} we provide a discussion on the pre-processing of the data, sample cuts and $O-C$ methodology. Sect.~\ref{sec: Classification Methodology} entails the $O-C$ classification based on statistical techniques to filter the Cepheids under investigation. In Sect.~\ref{sec: characterisation} we explain in detail the various methods used to characterise the irregular period changes. Sect.~\ref{sec: Results} presents our results, in particular incidence rates and characterisation of irregular period change stars. Lastly, in Sect.~\ref{sec: Discussion} we expand upon our findings and present a comparative analysis of irregular period change in both galaxies and end with conclusions in Sect.~\ref{sec: Conclusions}.

\section{Data analysis}
\label{sec: Analysis}
\subsection{Choice of survey}
\label{subsec: Choice of survey}

Before discussing the technical context of the data, we justify the choice of the survey in our study. Traditionally, century-long data, and in some cases a few decades, are required to measure evolutionary period change rates. The usage of long-term collection of pre-CCD era archival data, for example the DASCH survey \citep{Grindlay2012IAUS..285...29G, Tang2013PASP..125..857T} is a reasonable approach to measure such period changes. However, these data were collected with different telescopes on photographic plates, lack precision ($\sim$0.1 mag) and time resolution, which limits their suitability for our purpose. On the other hand, the near-continuous and high-precision observations of space telescopes such as CoRoT \citep{Baglin2002ESASP.485...17B}, \textit{MOST} \citep{Walker2003PASP..115.1023W}, \textit{TESS} \citep{Ricker2015JATIS...1a4003R} and \textit{Kepler} \citep{Borucki2010Sci...327..977B} provide a window to probe into short-timescale instabilities and low amplitude modulations manifested as a cycle-to-cycle variation of Cepheid light curves  \citep[e.g.][]{Poretti2015MNRAS.454..849P,Evans2015MNRAS.446.4008E} or period jitter \citep[e.g.][]{Derekas2012MNRAS.425.1312D,Derekas2017MNRAS.464.1553D}. The time span of observations and the sample of Cepheids observed with space telescopes is however very limited; too small to investigate changes on time scales longer than few tens of cycles on a statistically significant sample. For our purpose, data of the OGLE project \citep{Udalski1999AcA....49..223U,Udalski1999bAcA....49..437U,soszynski2008optical,soszynski2010optical,Soszynski2015AcA....65..297S,soszynski2017AcA....67..297S,Soszynski2019AcA....69...87S} that continues to monitor virtually all Cepheids in the Magellanic Clouds (4656 in the LMC, 4931 in the SMC) with high temporal resolution and for a time span of $\sim 20$\,yrs is the best choice, which in addition assures homogeneity of the data. 

OGLE data also give us chance to capture Cepheids during the first crossing, when their period change rate is predicted to be high \citep[e.g.][]{Anderson_Richard2014A&A...564A.100A}. Therefore, these first-crossing Cepheids are quite a rare occurrence, and only a large sample of stars gives a good chance of detecting them.

\subsection{OGLE data}
\label{subsec: Survey description}

We use publicly available photometric data\footnote{\url{http://www.astrouw.edu.pl/ogle/}} from  OGLE-III \citep{Udalski2008AcA....58...69U,soszynski2008optical,soszynski2010optical} and OGLE-IV \citep{udalski2015ogle,Soszynski2015AcA....65..297S,soszynski2017AcA....67..297S,Soszynski2019AcA....69...87S} survey. The span of observations for OGLE-III is 2001-2009 and for OGLE-IV is 2010-present, hereby providing nearly 20+ years of data. The survey observed the MC during OGLE-II \citep[1997–2000;][]{Udalski1997AcA....47..319U,Udalski1999AcA....49..223U,Udalski1999bAcA....49..437U} phase as well, and these observations were eventually merged with OGLE-III observations increasing the data span. We also use non-public, OGLE-IV extended survey data (starting from August 12, 2022) for the final list of non-evolutionary period change candidates. Therefore, the temporal span of our main sample was increased by $\sim$1.3 year with the new OGLE observations. 

The data for the ongoing OGLE survey is collected using the 1.3-m Warsaw telescope at Las Campanas Observatory in Chile. There were technical upgrades to the instrument in terms of filters and the CCD camera from 8 chip 2048$\times$4096-pixel mosaic in OGLE-III to 32 chip with 2048$\times$4102-pixel mosaic in OGLE-IV survey. These upgrades result in differences in the zero-point magnitude between phases III and IV of the survey. 

In terms of filters, the telescope uses Johnson-Kron-Cousins for \textit{I}-band and \textit{V}-band. We utilize only the \textit{I}-band data since it is much more densely sampled than the \textit{V}-band data. OGLE data is highly suitable for period change studies as has been shown in many works  \citep[e.g.][]{PietrukowiczLMC2001AcA....51..247P,PietrukowiczSMC2002AcA....52..177P,Kubiak2006AcA....56..253K,Poleski2008AcA....58..313P,Hajdu2015MNRAS.449L.113H,Prudil2019MNRAS.487L...1P,Hajdu2021ApJ...915...50H,Rodriguez-Segovia2022MNRAS.509.2885R}.

A schematic workflow for our analysis described in the following sections is presented in Figs.~\ref{fig:pipeline1} and \ref{fig:pipeline2}.

\subsection{Combining data}
\label{subsec: Combining data}

As a first step in our analysis we need to carefully combine OGLE-III and OGLE-IV data, which differ in zero point as mentioned in Sect.~\ref{subsec: Survey description}. For this, the \textit{I}-band magnitude data is first converted to an arbitrary intensity scale. Then, a Fourier series is fitted individually to both data sets and we correct for shift in the mean intensity. The order of the Fourier series is decided by requiring $A_{k}/\sigma(A_{k})>4$, for the first $k$ terms of the fit with ${A_{k}}$ and $\sigma(A_{k})$ being the amplitude and uncertainty of the Fourier series terms. We tested that such a criterion prevents over-fitting of the light curves and typically results in higher order ($\sim10$) fit for the fundamental mode stars and lower order ($\sim 6$) fit for the first overtone stars. Then, as a final step, we transform the combined OGLE-III and IV data set back to the magnitude scale.

\subsection{Sample cuts}
\label{subsec: Sample selection}

For several reasons, not all Cepheids in the OGLE collection are suitable for $O-C$ analysis, hence we start with several sample cuts. These are based not only on data quality (e.g. we reject stars that lack data in one of the OGLE phases) but are also motivated by the urge to have a clean sample of single-periodic pulsators. Hence stars with additional pulsation modes and modulations, that could affect the $O-C$ and its interpretation, are rejected.

Firstly (Step 1) we remove targets for which data are missing for either OGLE-III or OGLE-IV, since we need a longer baseline. In this process, 1651 Cepheids were omitted from the analysis. Then (Step 2), we reject targets for which 'remarks' are present in the OGLE catalogue. Most of these remarks are about blending, the presence of modulations or the presence of a secondary period. Since we aimed to analyze purely single radial mode stars in this work, it was necessary to make this cut. We also remove targets that show periodic modulation of the pulsation \citep[][Step 3]{Smolec2017MNRAS.468.4299S,Smolec2023MNRAS.519.4010S}, additional radial mode(s) \citep[][Step 4]{Smolec2016MNRAS.458.3561S,Smolec2023MNRAS.519.4010S} and other low amplitude variability \citep[][Step 5]{Smolec2016MNRAS.458.3561S,Smolec2023MNRAS.519.4010S}. During steps 3--5, we rejected 374 targets from our analysis. At last we also rejected the targets that were reported as Cepheid binary candidates in our previous study \citepalias[Step 6,][]{Rathour2024A&A...686A.268R}. The summary of sample cuts is presented in Tab~\ref{tab:data sample} and the flow of the first part of the pipeline is presented in Fig.~\ref{fig:pipeline1}.

\subsection{$O-C$ procedure}
\label{subsec: O-C procedure}
The technique of $O-C$ diagram has been long known in the literature to quantify period changes \cite[see reviews by][]{Zhou1999PBeiO..33...17Z,Sterken2005ASPC..335....3S}. It is estimated by computing differences between observed and expected occurrence of a specific variability phase, the latter computed assuming a constant period. For pulsating stars and eclipsing binaries, these variability phases are usually the maxima or minima of the light curve. To correctly determine the $O-C$ diagram, one would require a high cadence around these phases. This is not the case for OGLE data, which have long time span, but are sparse. For such data a modified version of the \cite{Hertzsprung1919AN....210...17H} method is well suited. This method uses a template light curve to measure phase shifts at arbitrary epochs across the data span \citep[see examples in][and \citetalias{Rathour2024A&A...686A.268R}]{Hajdu2021ApJ...915...50H, Rodriguez-Segovia2022MNRAS.509.2885R}.

\begin{figure*}
\begin{center}
{\includegraphics[height=5.5cm,width=\linewidth]{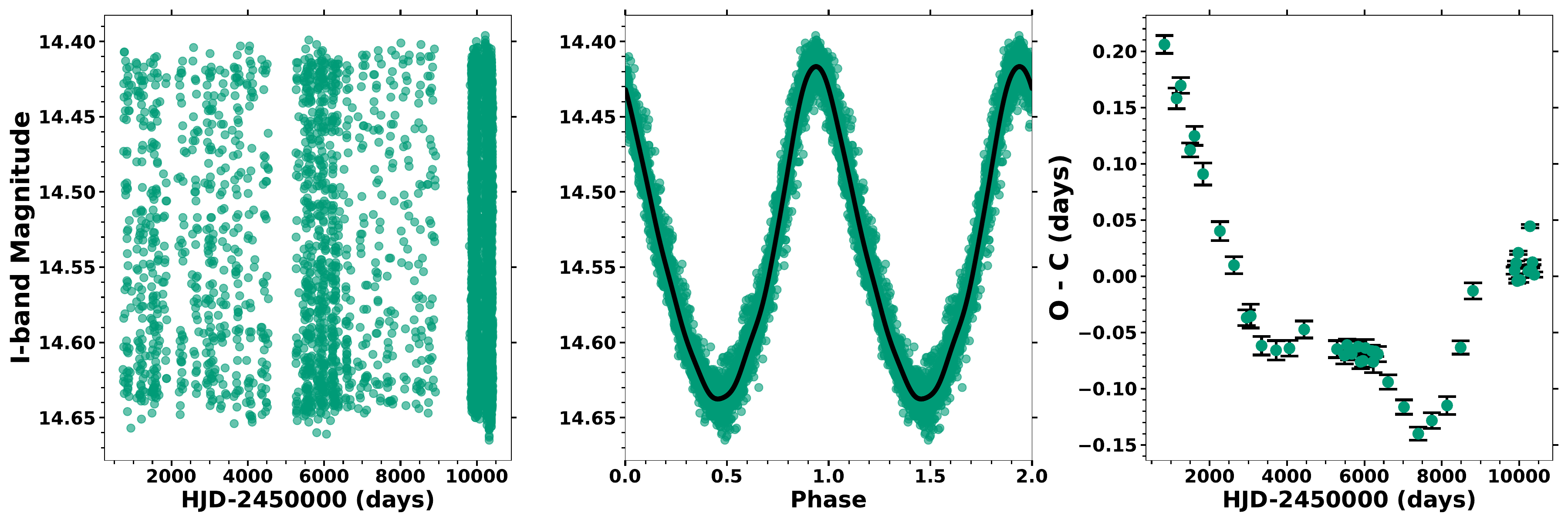}}
\caption{$O-C$ analysis for OGLE-LMC-CEP-0614 ($P=3.60057495$\,d) showing combined OGLE photometry (left panel), phased light curve with suitable order (five in this case) Fourier fit (the template in black; middle panel), and computed $O-C$ diagram (right panel).}
\label{fig:oc_example_figure}
\end{center}
\end{figure*}

The detailed methodology to compute the $O-C$ diagram, division of data for finer time resolution, and estimation of bootstrapped errors on individual $O-C$ points are described in \citetalias{Rathour2024A&A...686A.268R}. The reader is directed to the paper for specific details. An example of $O-C$ extraction is presented in Fig.~\ref{fig:oc_example_figure}. 

Based on visual inspection of the $O-C$ diagrams, additional targets were rejected (Step 7, the final cut). This cut was based on several grounds, such as a few missing seasons in the data, problems with the Fourier template, strong amplitude changes visible in the photometry data, or the pulsation period being very close to the integer value causing problems with the analysis.

\begin{table*}
\caption{Stepwise rejected MC Cepheids with distribution relative to pulsation mode.}
\label{tab:data sample}
\centering\footnotesize
\begin{tabular}{lc|ccccccc|ccc}
\hline
\hline
\textbf{Galaxy/Mode} & \textbf{OGLE} & \multicolumn{7}{c|}{\textbf{Sample Cleaning Steps}} & \multicolumn{3}{c}{\textbf{Final Sample}} \\
\hline
& & \textbf{Step 1} & \textbf{Step 2} & \textbf{Step 3} & \textbf{Step 4} & \textbf{Step 5} & \textbf{Step 6} & \textbf{Step 7} & \textbf{Class 1} & \textbf{Class 2} & \textbf{Class 3} \\
\hline
\hline
\textbf{SMC F}  & 2783  & 190 & 40   & 17  & -   & -    & 85  & 599   & 1483 & 138 & 231\\
\textbf{SMC 1O} & 1819  & 199 & 220  &  3   & 9   & 120  & 50  &  284   & 142 & 133 & 659\\
\hline
\textbf{LMC F}  & 2478  & 678 & 22   &  18  & -   & -    & 30  & 417    & 910 & 113 & 290\\ 
\textbf{LMC 1O} & 1797  & 584 & 120  &  21  & 36  & 150  & 17  & 239    & 125& 100 & 405\\ 
\hline
\hline
\textbf{Total}  & 8877  &1651 & 402  &  59 & 45  & 270  &182  & 1539   & 2660& 484 & 1585\\ 
\hline
\end{tabular}
\tablefoot{Step 1: Crossmatch between OGLE-III and IV; Step 2: Candidates listed with remarks in OGLE database; Step 3: Candidates listed with periodic modulation of pulsation; Step 4: Candidates containing additional radial mode; Step 5: Candidates with additional low amplitude variability; Step 6: Candidates for binary Cepheids; Step 7: Filtered or rejected during the analysis.}
\end{table*}

\section{Classification methodology}
\label{sec: Classification Methodology}

Our goal is to investigate Cepheids that exhibit irregular period changes. To select those stars, we divide the sample into three classes: class 1, with apparently no period change at all (flat $O-C$ diagrams), class 2, with linear period change (parabolic $O-C$) and class 3, that shows irregular period change (anything more complex than flat/parabolic $O-C$). To make this classification we investigate $O-C$ diagrams for a sample of 3658 Cepheids (sample after cuts described in Sect.~\ref{subsec: Sample selection} and Sect.~\ref{subsec: O-C procedure}).

For the initial classification we use several statistical tests as outlined below. Then, results of this initial classification serve as an input to final classification based on Bayesian model fitting. In practice, we do not test for irregular variability. Rather we test whether linear and parabolic models are sufficient to describe the data.

To this end we used a python library \texttt{lmfit} \citep{Newville2016ascl.soft06014N} for non-linear least-squares minimization and curve fitting, to fit the computed $O-C$ diagrams with linear and parabolic models. Then, we used various statistical tests as diagnostics, such as for each model we calculated reduced chi-square, Akaike information criterion \citep[AIC;][]{Akaike1974ITAC...19..716A} and Bayesian information criterion \citep[BIC;][]{Schwarz1978_10.1214/aos/1176344136} for estimating the goodness of fit. We also apply the \cite{Jurcsik2001AJ....121..951J} criterion to the $O-C$ fits \citep[see also][]{Prudil2020A&A...635A..66P}. This criterion aims to filter out the constant period candidates from the significant period change rate ones and is defined by:
\begin{eqnarray}
\label{eq:jurcsik_eq}
\frac{|a_{2}|}{\sigma (a_{2})} > 2, 
\end{eqnarray}
where $a_{2}$ is the coefficient of the quadratic component and $\sigma(a_{2})$ is its uncertainty. In addition to that, we also simultaneously conduct statistical diagnostic tests on the residuals to confirm if there are any significant features left. The tests are \cite{Anderson&Darling10.1214/aoms/1177729437} (A-D) and \cite{Ljung&Box10.1093/biomet/65.2.297} (L-B). The former tests whether the residuals are distributed normally and the latter checks whether the residuals are uncorrelated. Both these tests reflect the goodness of the model in predicting the behavior of the underlying data. Each of the above tests quantifies whether the two models to describe $O-C$, linear and parabolic, are adequate and sufficient to describe the data. If all the tests are in favor of the linear model then our initial assignment is class 1. If all tests are in favor of quadratic model, then our initial assignment is class 2. Otherwise we assign class 3.

The above analysis provided not just the test set with prior classification, but also the range of coefficients for linear and parabolic models can take, which forms the basis for a much more robust and sophisticated final classification. In the next step, we utilized the \texttt{UltraNest} package \citep{Buchner2021JOSS....6.3001B} for Bayesian inference, leveraging its nested sampling approach to efficiently explore parameter spaces for our linear and parabolic models. This method is well-suited for selecting the best model when dealing with complex parameter relationships in the $O-C$ curves. We employed uniform priors for both models and assumed a Gaussian likelihood to evaluate the fit to the observational data. 
For the nested sampling, we employed default value of 400 live points, which effectively act as `walkers' in the parameter space, ensuring thorough exploration of the posterior distribution. For our purpose, 400 walkers ensured computational efficiency with the accuracy needed for precise model fitting. The likelihood function, $\mathcal{L}$, and the evidence, $\mathcal{Z}$, are defined as:
\begin{eqnarray}
\label{eq:ultranest_eq3}
   \mathcal{L} = -\frac{1}{2} \sum_{j} \left(\frac{y_j - \eta_j}{y_{e,j}}\right)^{2}\,,
\end{eqnarray}

\begin{eqnarray}
\label{eq:ultranest_eq4}
   \mathcal{Z} = \int_{\Omega_{\Theta}} \mathcal{L} (\Theta) \hspace{0.1cm} \Pi (\Theta) \hspace{0.1cm} d\Theta\,.
\end{eqnarray}
In the above equation \eqref{eq:ultranest_eq3}, $y_j$ is the observed $O-C$ value, $\eta_j$ is the model prediction, and $y_{e,j}$ is the error on the observed $O-C$ value obtained from the bootstrapping method. Eq.~\eqref{eq:ultranest_eq4} is the integral over parameter space, $\Omega_{\Theta}$, of the product of the likelihood and its priors, $\Pi (\Theta)$. Based on preliminary checks provided by the earlier classification, we keep a single set of uniform flat priors for the full data set for linear and parabolic models respectively. The number of likelihood evaluations and efficiency rates for both models indicated that \texttt{UltraNest} efficiently explored the parameter space, with the effective sample size  well above the required threshold, confirming well-sampled posterior distributions. Convergence was verified using the Kullback-Leibler divergence \citep{Kullback1951}, demonstrating that the posterior uncertainty strategy was satisfied. The details are provided in the Appendix~\ref{appendix: Bayesian formalism}. 

Model selection is based on the value of evidence ($\mathcal{Z}$) and the Bayes factor, defined as: 
\begin{eqnarray}
\label{eq:Bayes factor}
K = \frac{\mathcal{Z}{1}}{\mathcal{Z}{2}}, 
\end{eqnarray}
where $\mathcal{Z}{1}$ and $\mathcal{Z}{2}$ are the evidence values for the linear and parabolic models, respectively. A value of $K > 1$ indicates a preference for the linear model, while $K < 1$ favors the parabolic model. However, the Bayes factor alone does not guarantee that either model captures the intrinsic data characteristics or that the fit is robust. The absolute evidence values, $\mathcal{Z}{1}$ and $\mathcal{Z}{2}$, can provide insight into how well the models represent the data, but this comparison is only one aspect of assessing model quality. For final model assessment we investigated the residuals with A-D and L-B tests. Failure to satisfy these tests suggests that the residuals exhibit irregular behavior, and these stars were flagged as candidate irregular period change ones. Thus, our approach combines Bayesian evidence with residual analysis to classify the Cepheid $O-C$ curves into class 1 and class 2, and those that did not satisfactorily fit either linear or parabolic models into class 3.

After the complete analysis, the distribution of class 3 targets comprised of 231 SMC F-mode, 659 SMC 1O-mode, 290 LMC F-mode and 405 LMC 1O-mode Cepheids (see Tab.~\ref{tab:data sample}). For this sample, we update the data with new, extended non-public photometry and reanalyzed the $O-C$ diagrams. The new data spans $\sim$1.3 yrs and is typically much more densely sampled (see left panel of Fig.~\ref{fig:oc_example_figure}).


\begin{figure*}
\begin{center}
{\includegraphics[height=3.5cm,width=0.24\linewidth]{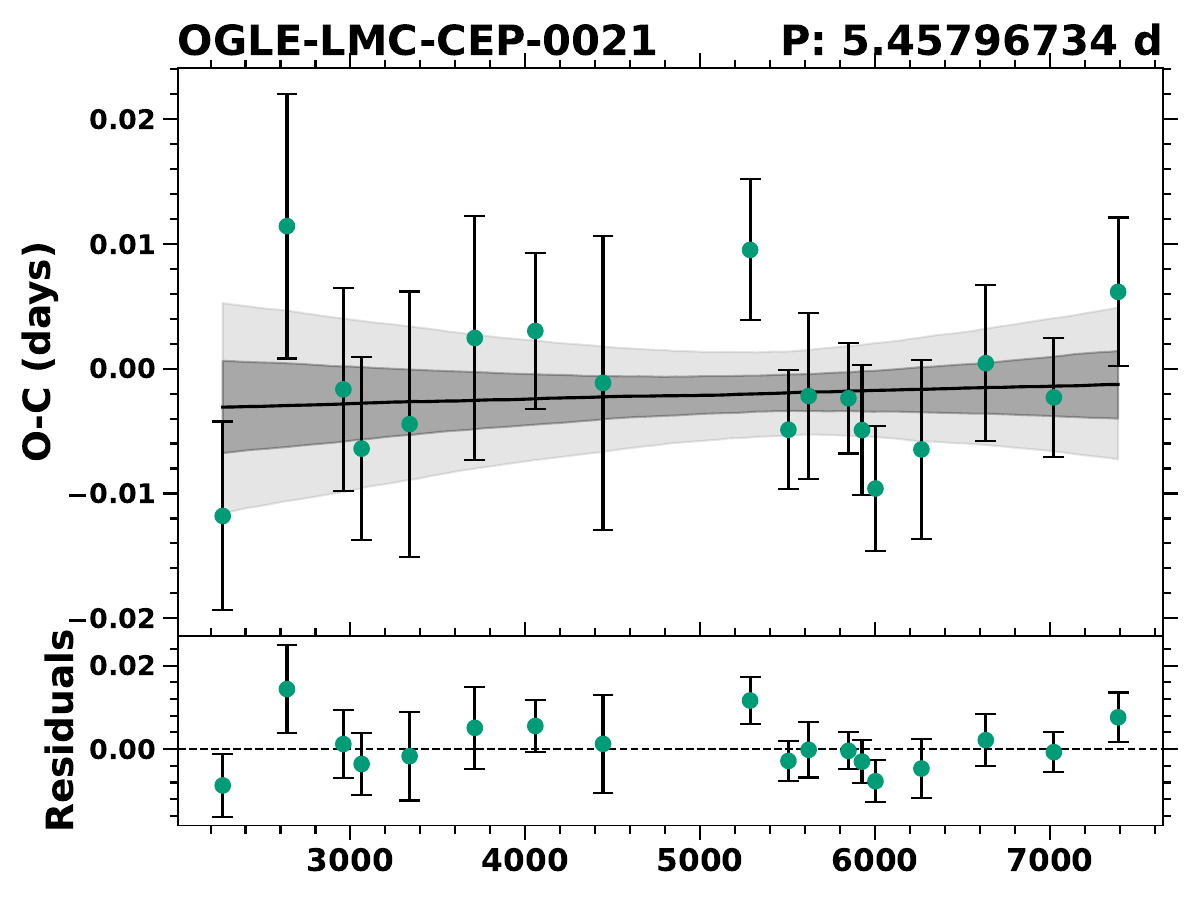}}
{\includegraphics[height=3.5cm,width=0.24\linewidth]{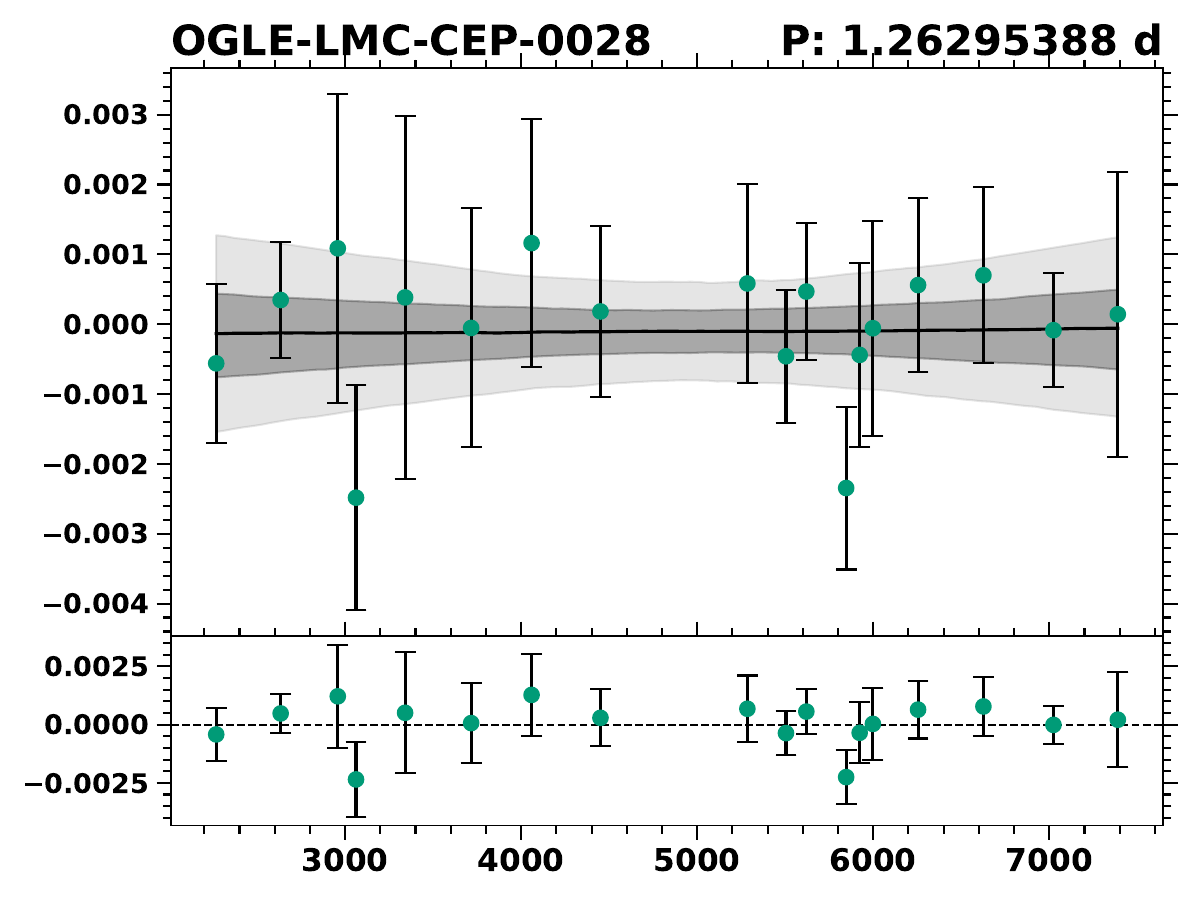}}
{\includegraphics[height=3.5cm,width=0.24\linewidth]{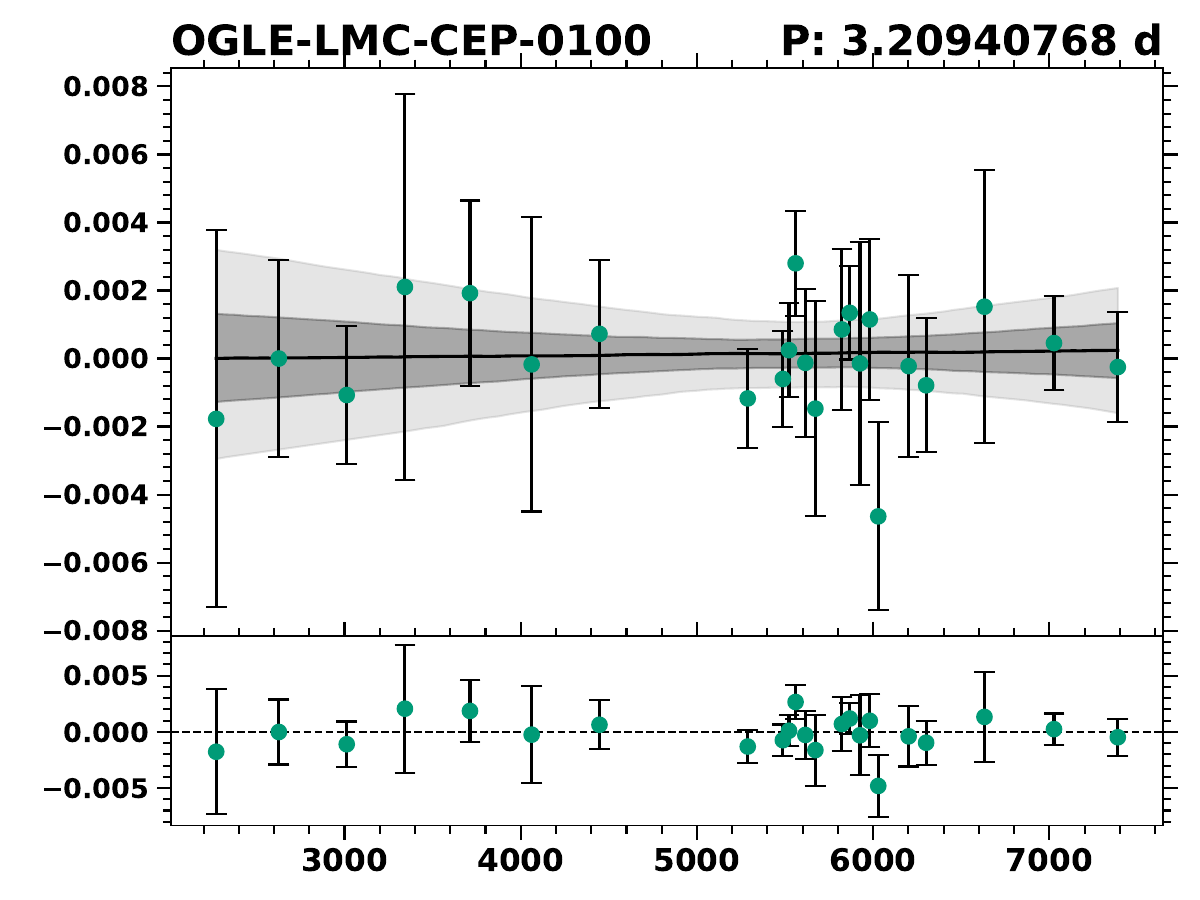}}
{\includegraphics[height=3.5cm,width=0.24\linewidth]{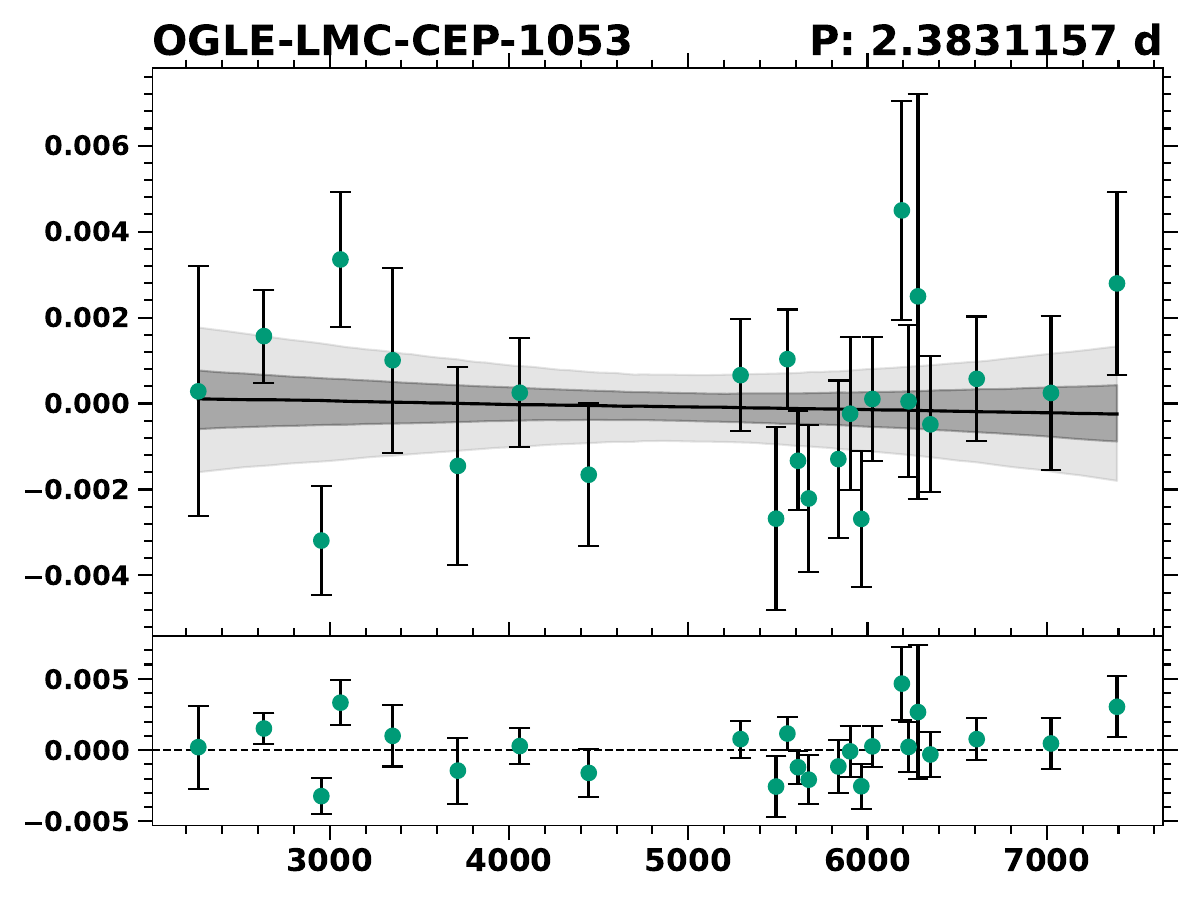}}

{\includegraphics[height=3.5cm,width=0.24\linewidth]{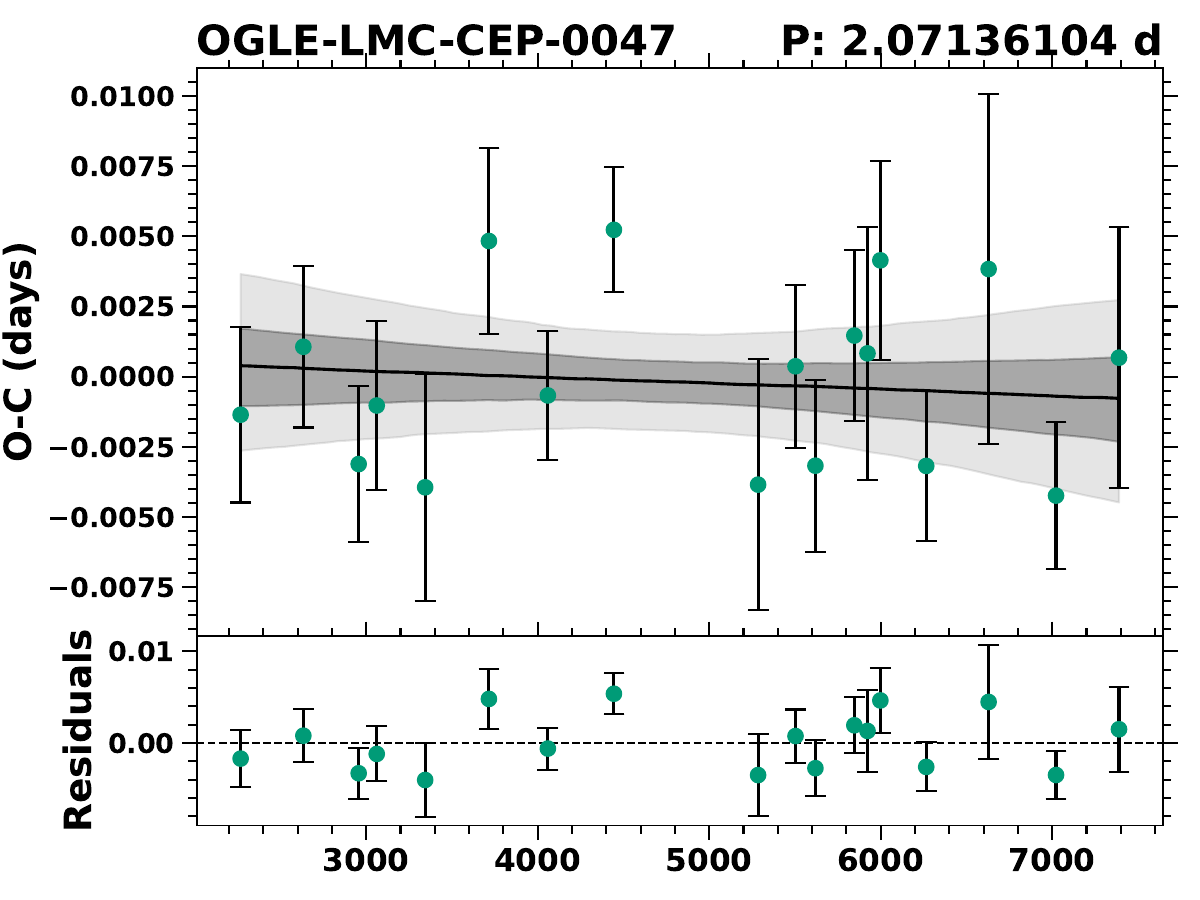}}
{\includegraphics[height=3.5cm,width=0.24\linewidth]{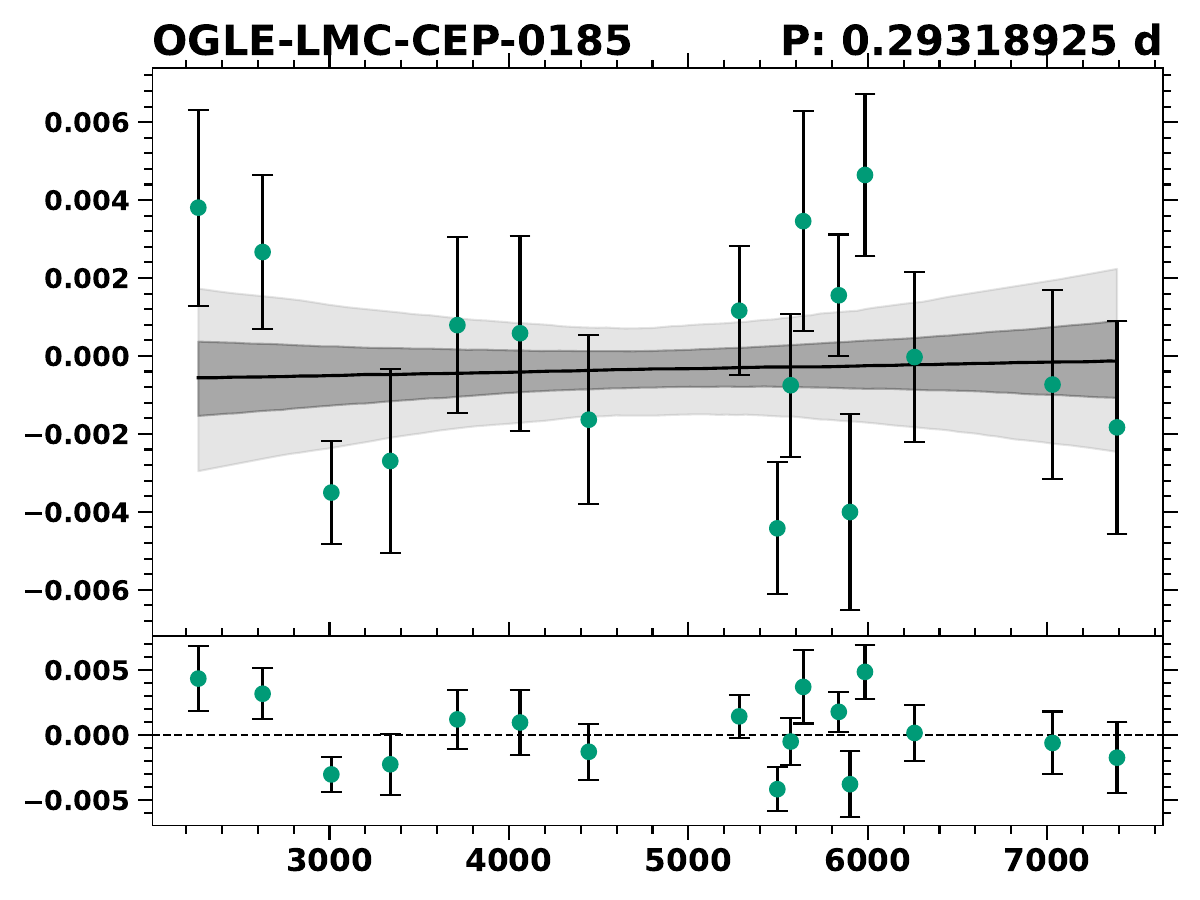}}
{\includegraphics[height=3.5cm,width=0.24\linewidth]{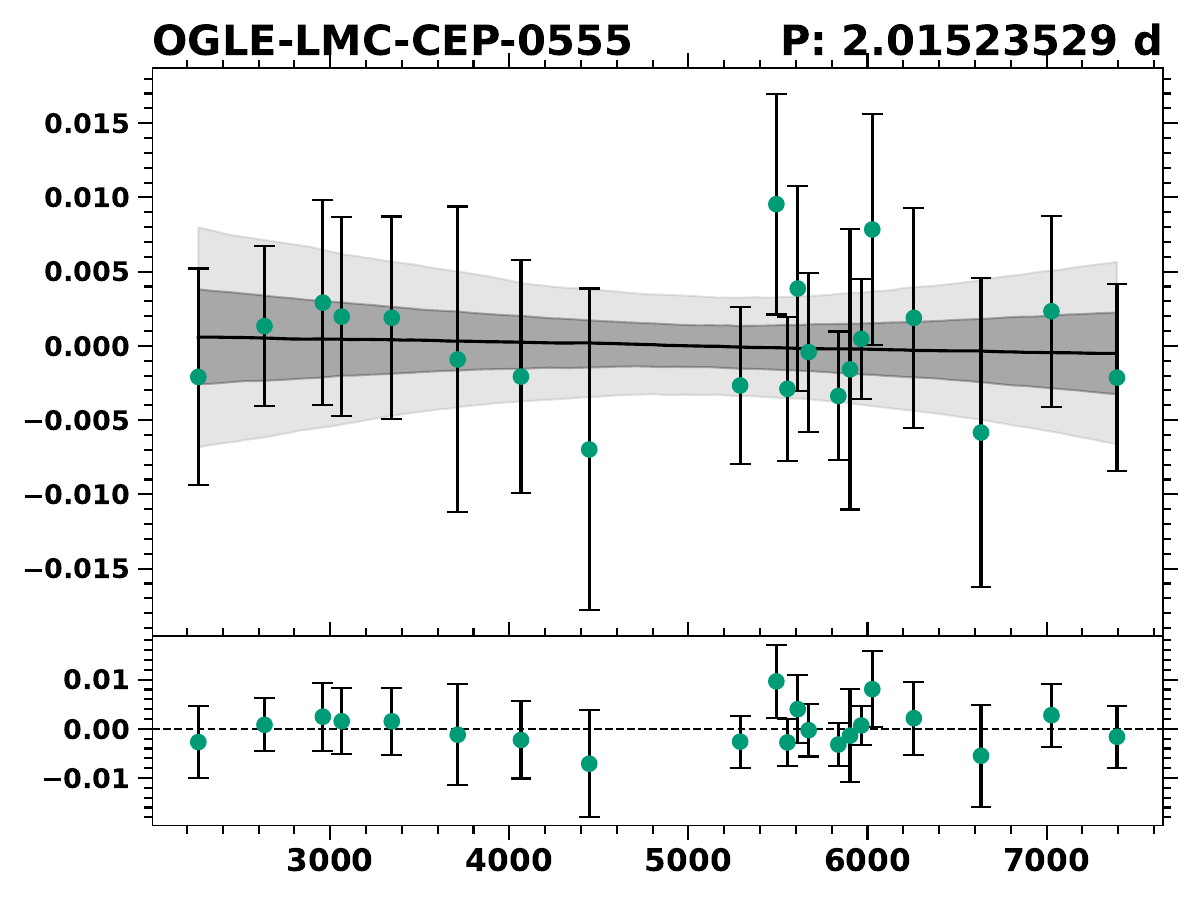}}
{\includegraphics[height=3.5cm,width=0.24\linewidth]{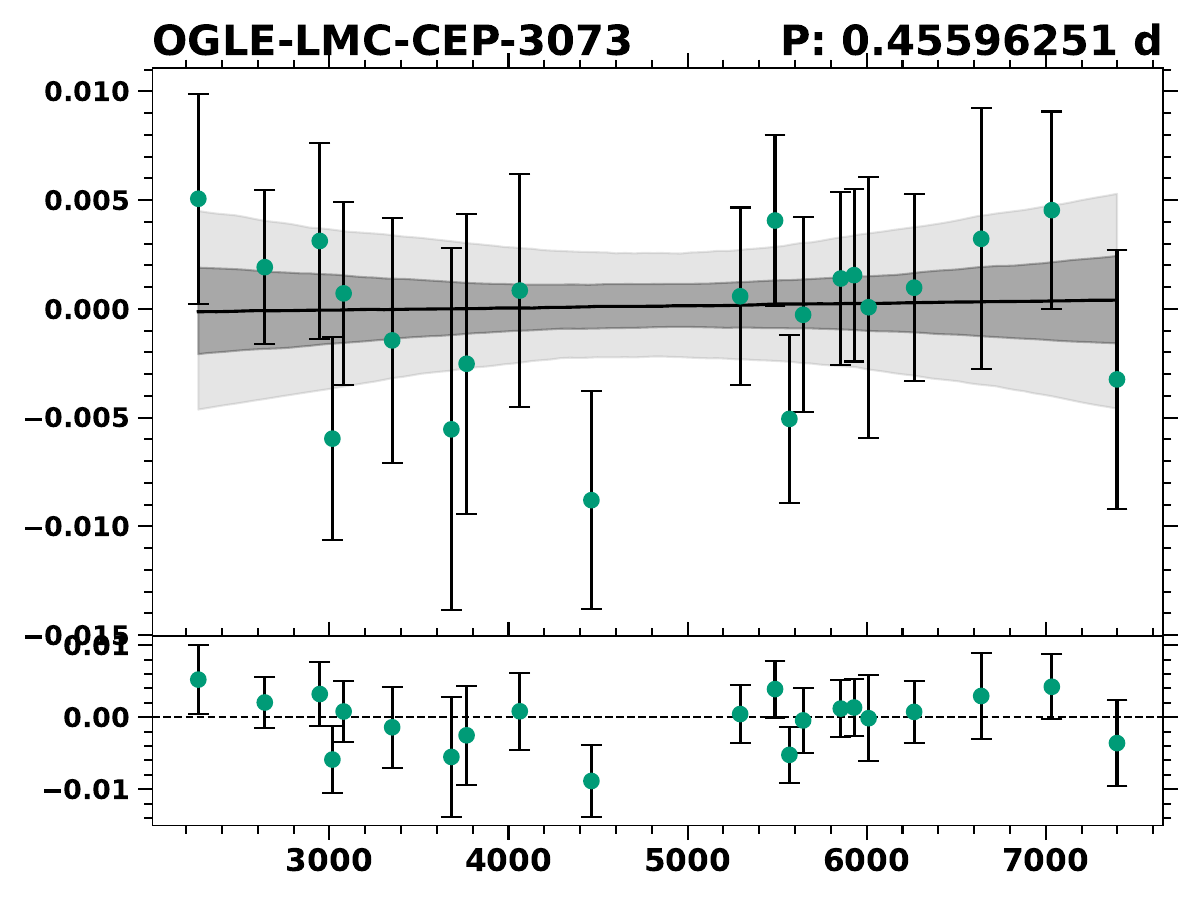}}

\caption{Examples of flat shape $O-C$ diagrams (class 1) over-plotted with their MCMC linear fit solution (in gray) showing LMC F-mode (row 1) and LMC 1O-mode (row 2) candidates. Above each panel the OGLE-ID and pulsation period are shown.}
\label{fig:ocplot_linear_examples}
\end{center}
\end{figure*}


\begin{figure*}
\begin{center}
{\includegraphics[height=3.5cm,width=0.24\linewidth]{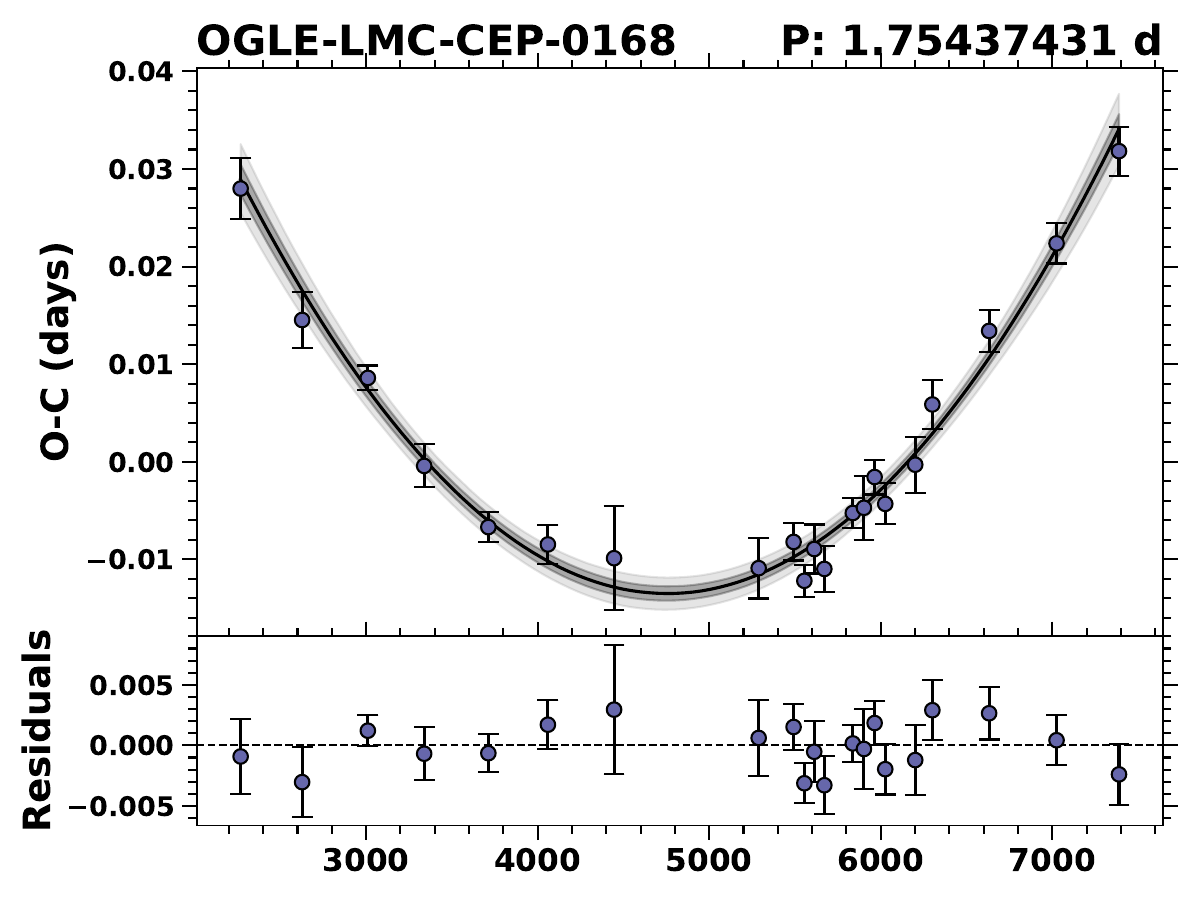}}
{\includegraphics[height=3.5cm,width=0.24\linewidth]{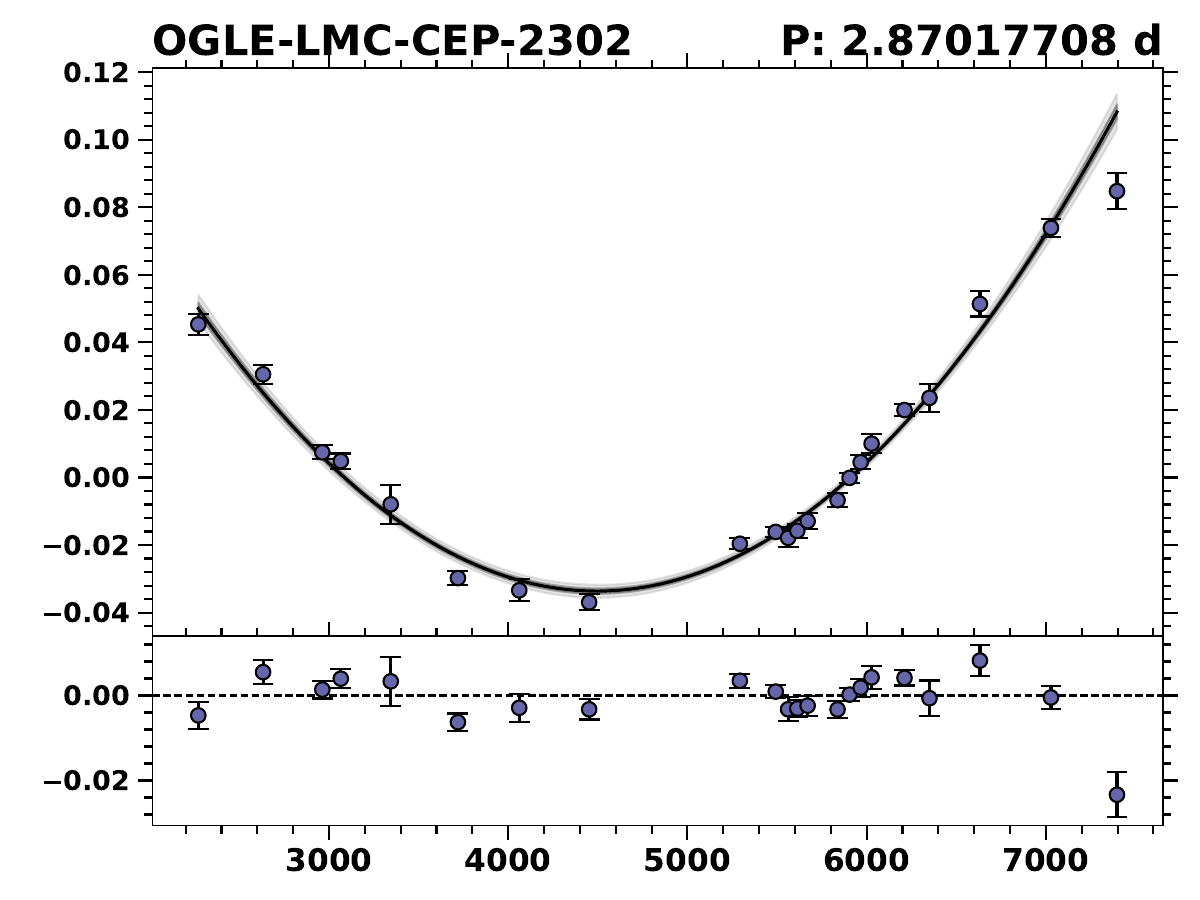}}
{\includegraphics[height=3.5cm,width=0.24\linewidth]{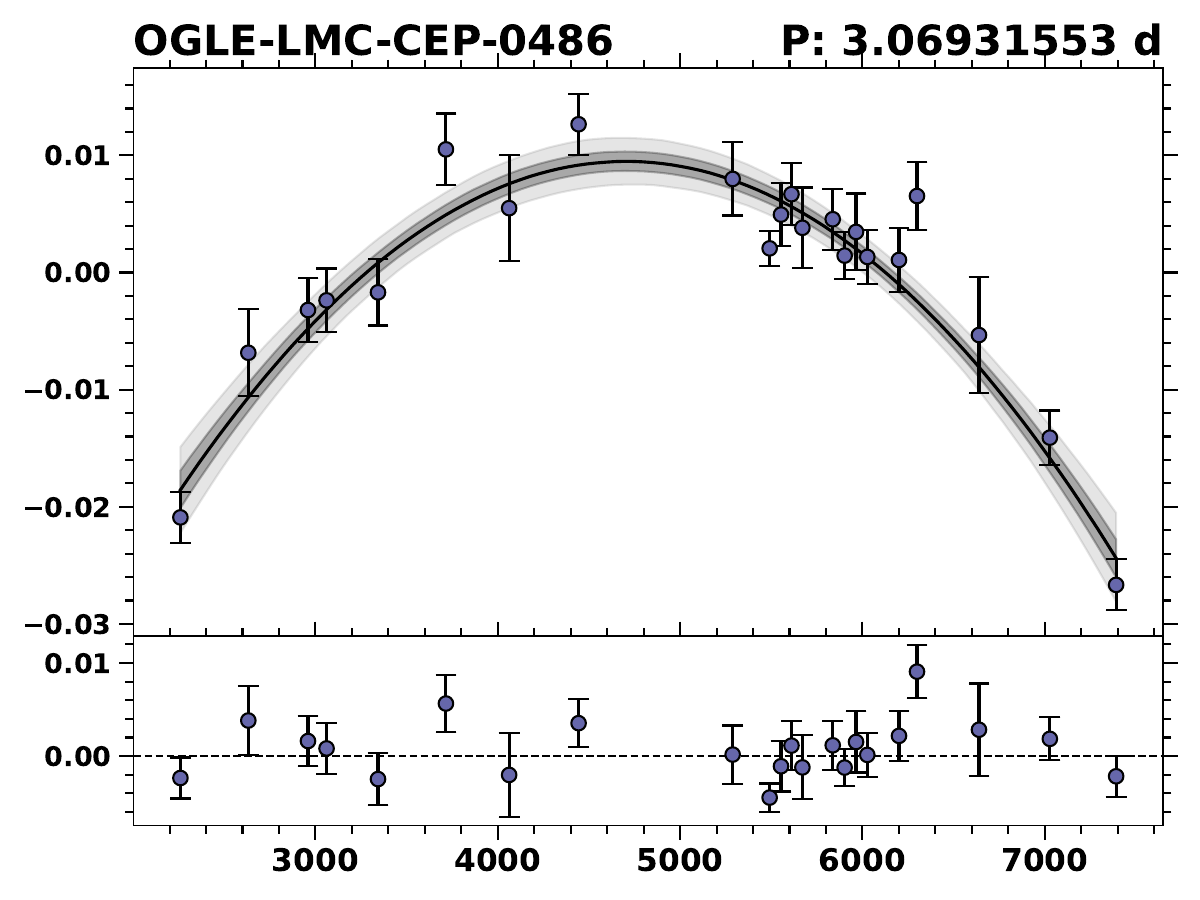}}
{\includegraphics[height=3.5cm,width=0.24\linewidth]{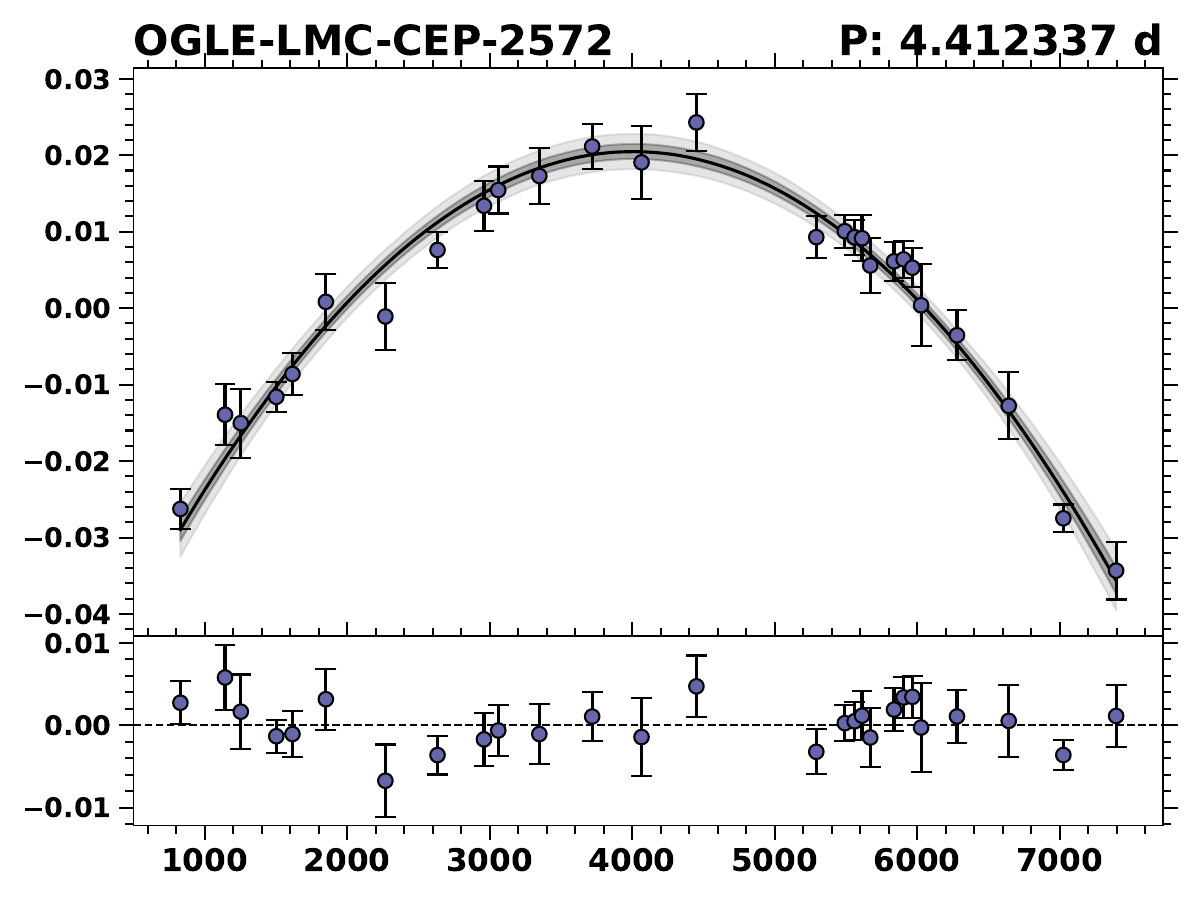}}

{\includegraphics[height=3.5cm,width=0.24\linewidth]{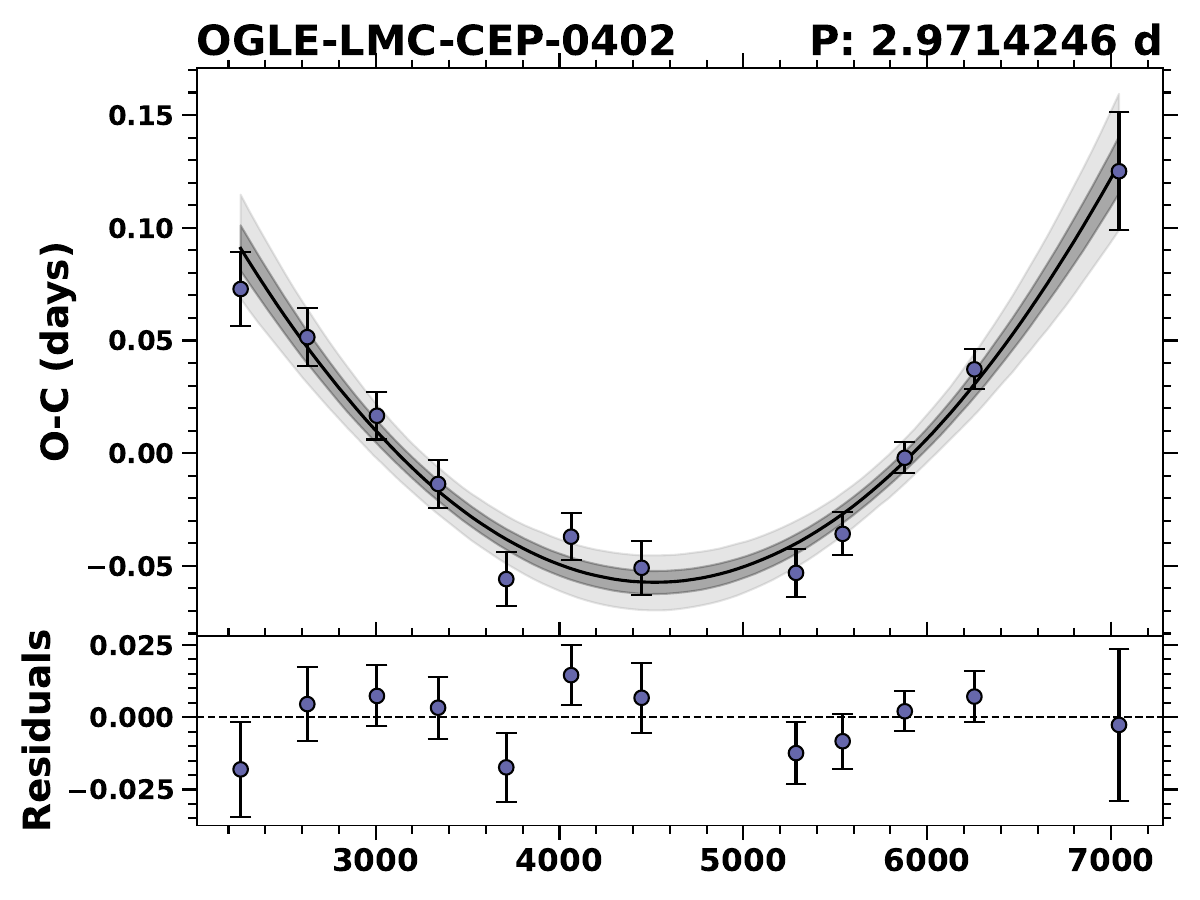}}
{\includegraphics[height=3.5cm,width=0.24\linewidth]{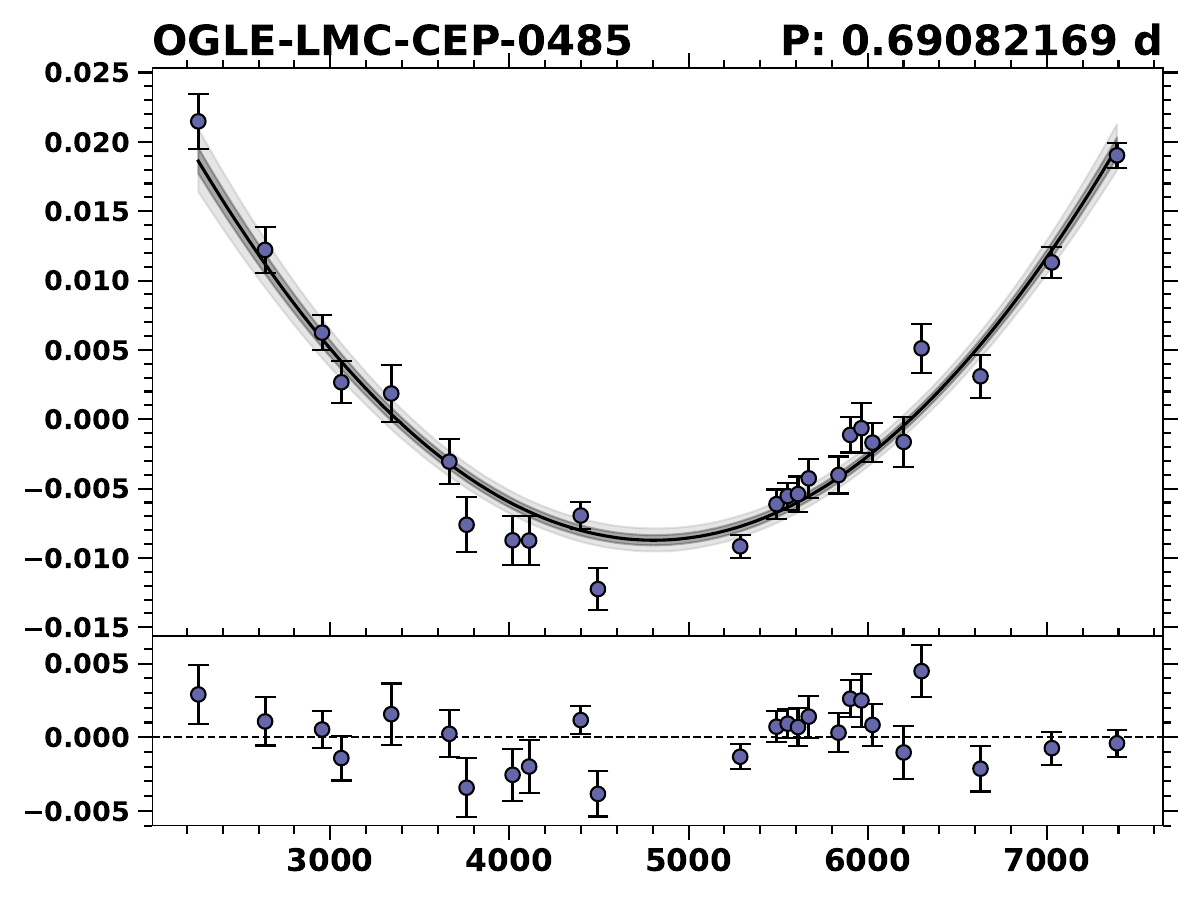}}
{\includegraphics[height=3.5cm,width=0.24\linewidth]{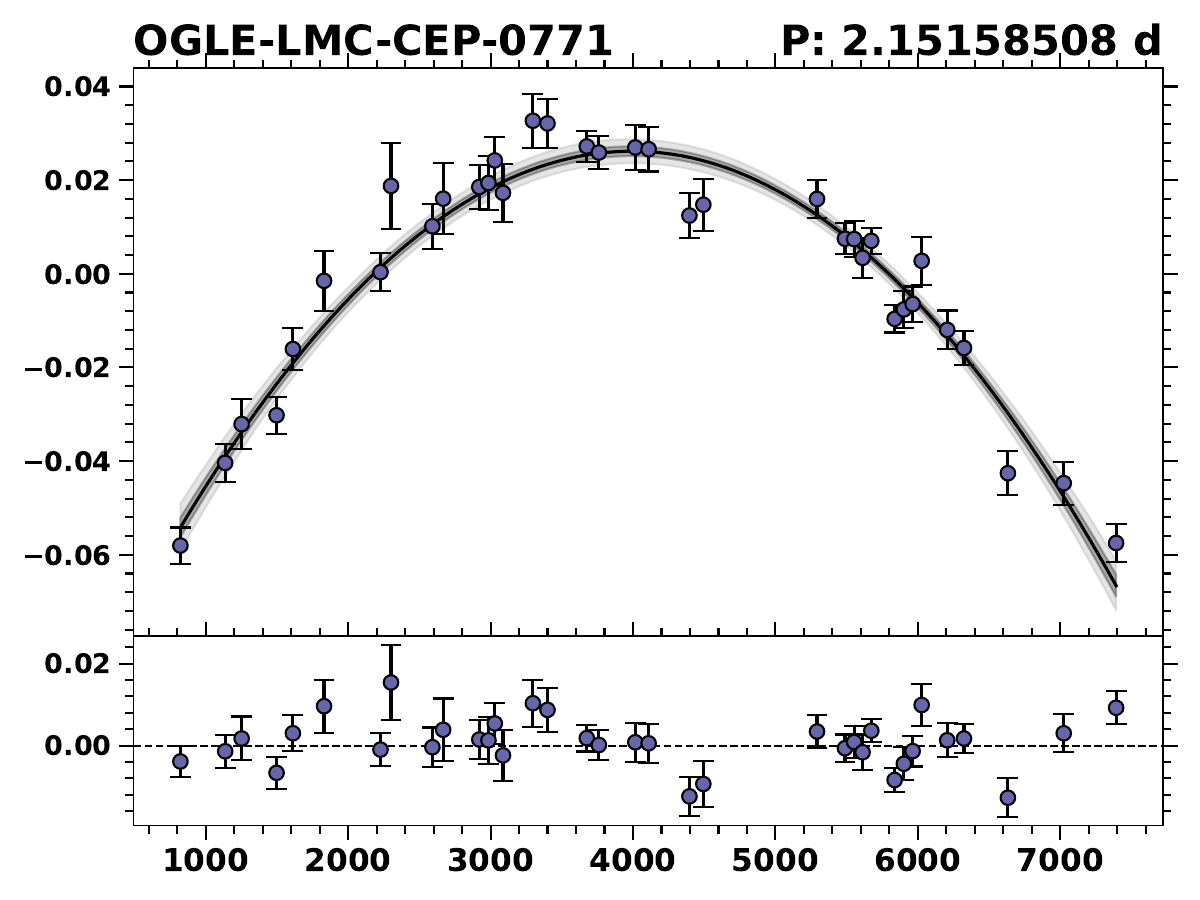}}
{\includegraphics[height=3.5cm,width=0.24\linewidth]{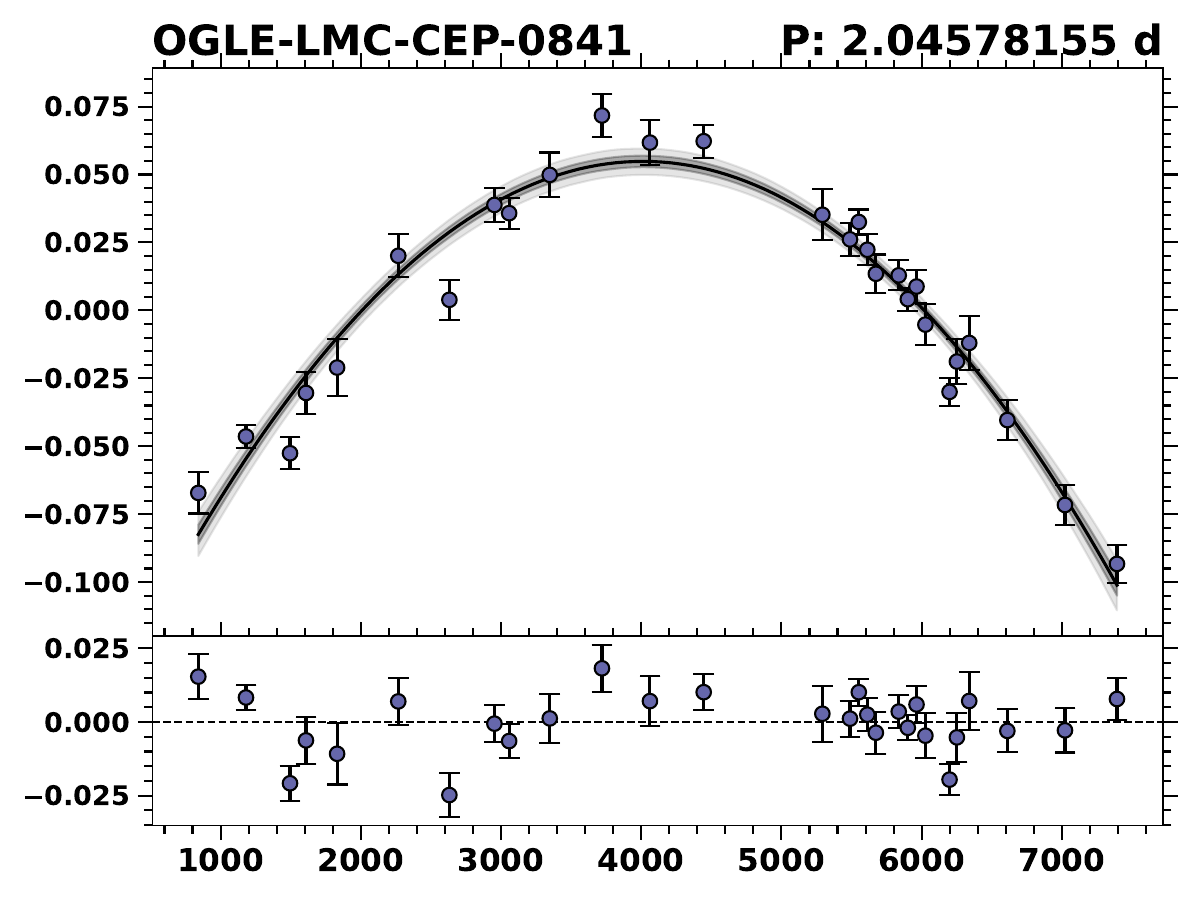}}

\caption{Examples of parabolic shape $O-C$ diagrams (class 2) over-plotted with their MCMC linear fit solution (in gray) showing LMC F-mode (row 1) and LMC 1O-mode (row 2) candidates. Above each panel, the OGLE-ID and pulsation period are shown.
}
\label{fig:ocplot_parabola_examples}
\end{center}
\end{figure*}


\begin{figure*}
\begin{center}
{\includegraphics[height=3.5cm,width=0.24\linewidth]{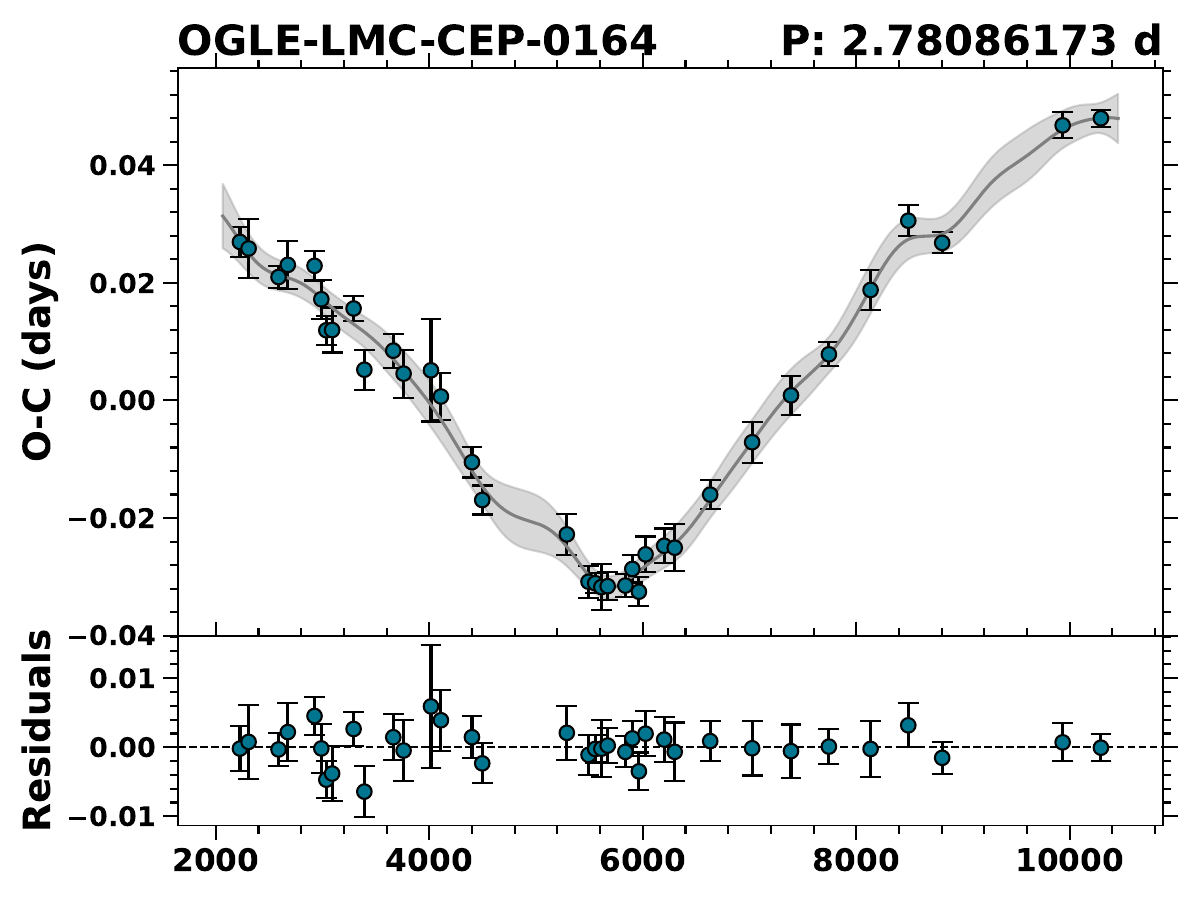}}
{\includegraphics[height=3.5cm,width=0.24\linewidth]{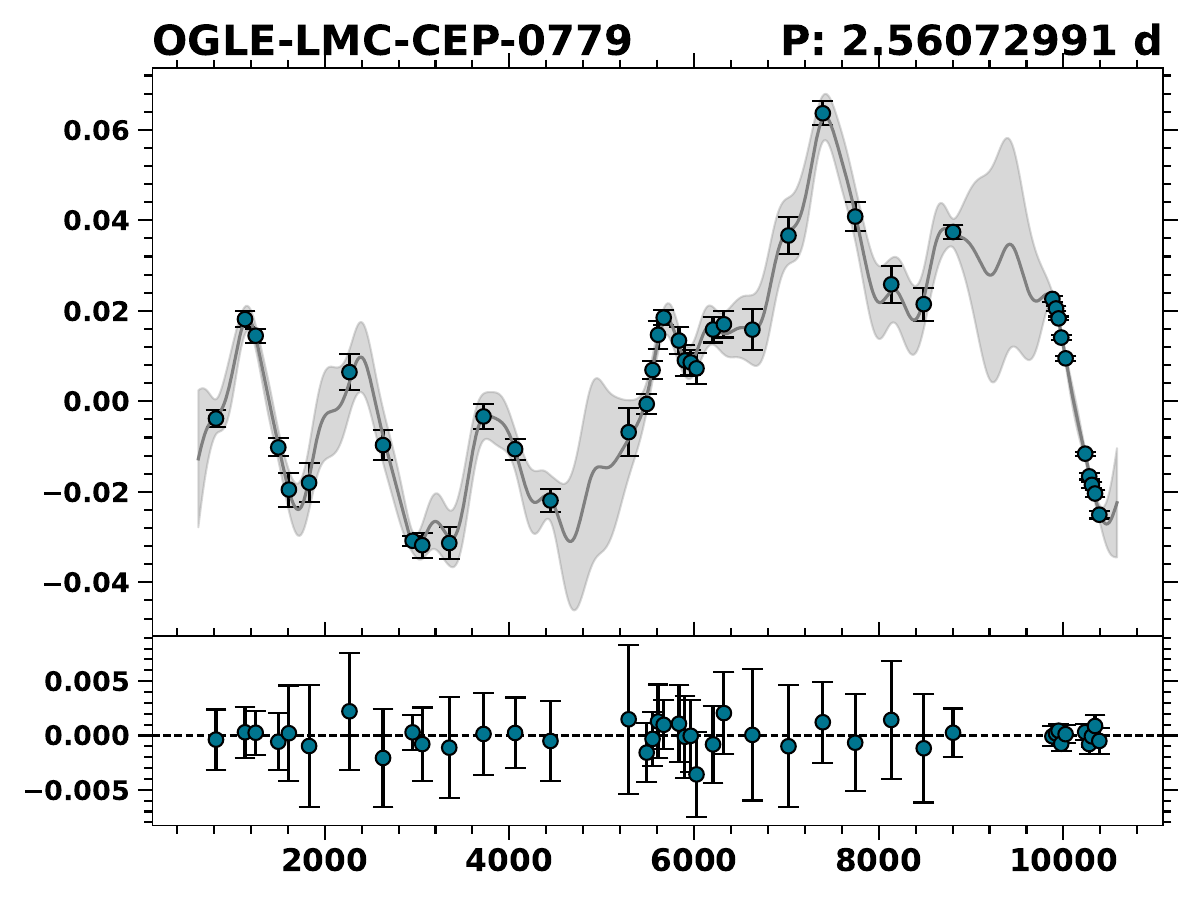}}
{\includegraphics[height=3.5cm,width=0.24\linewidth]{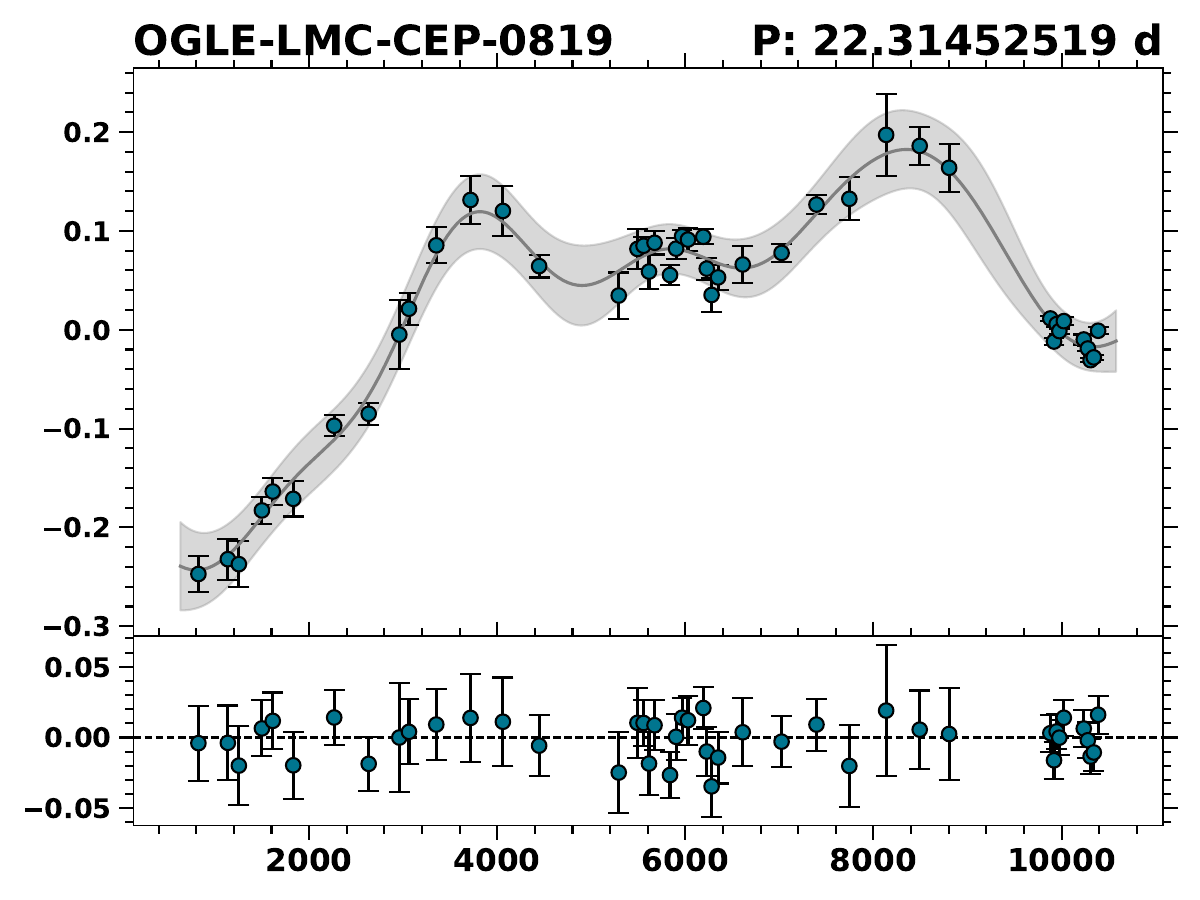}}
{\includegraphics[height=3.5cm,width=0.24\linewidth]{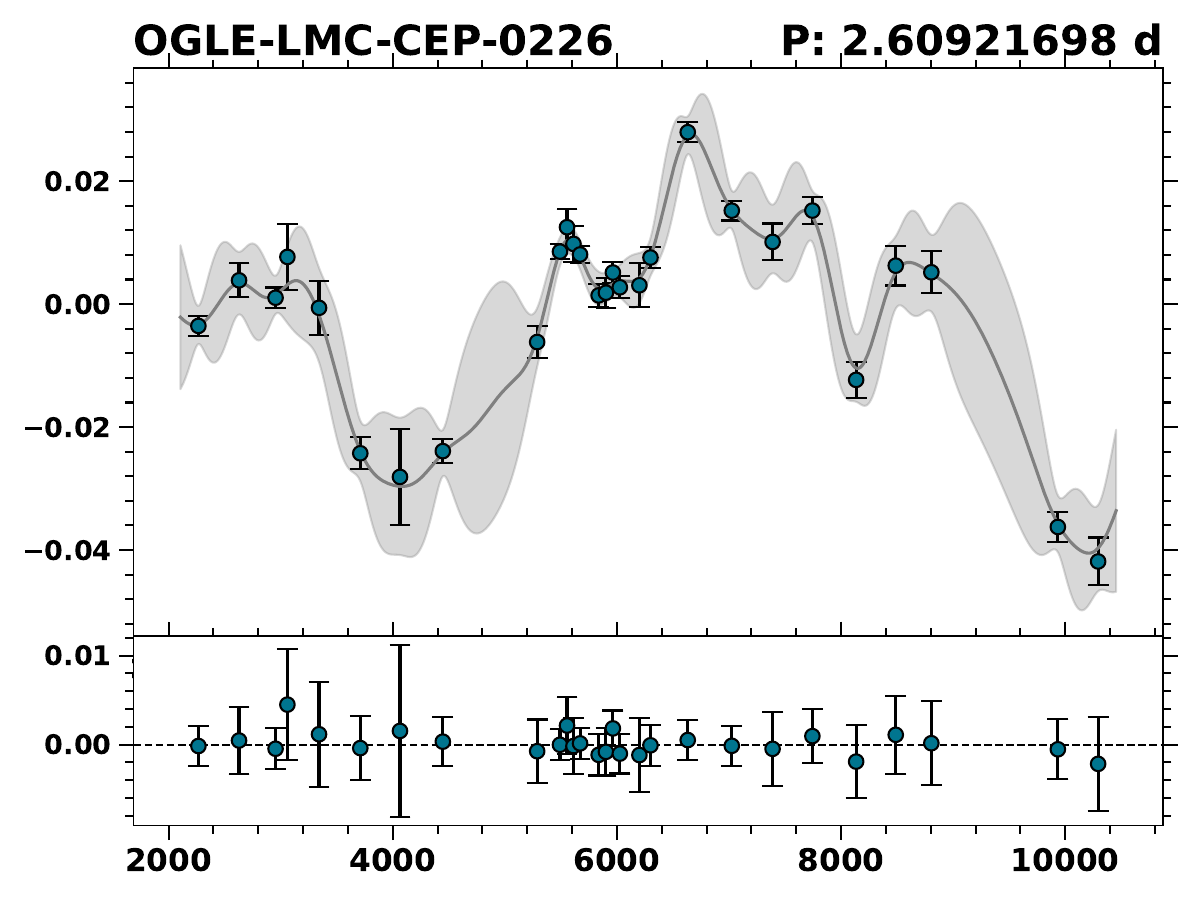}}
{\includegraphics[height=3.5cm,width=0.24\linewidth]{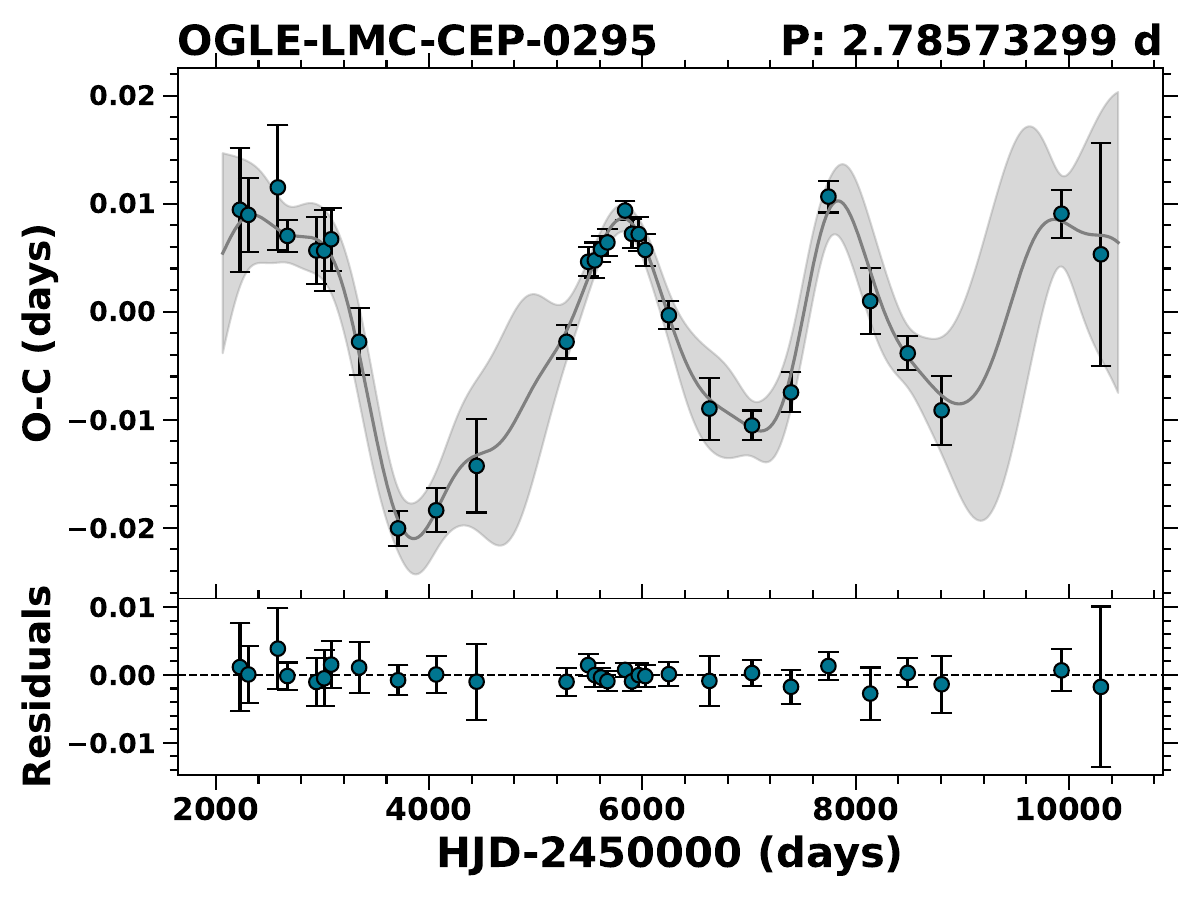}}
{\includegraphics[height=3.5cm,width=0.24\linewidth]{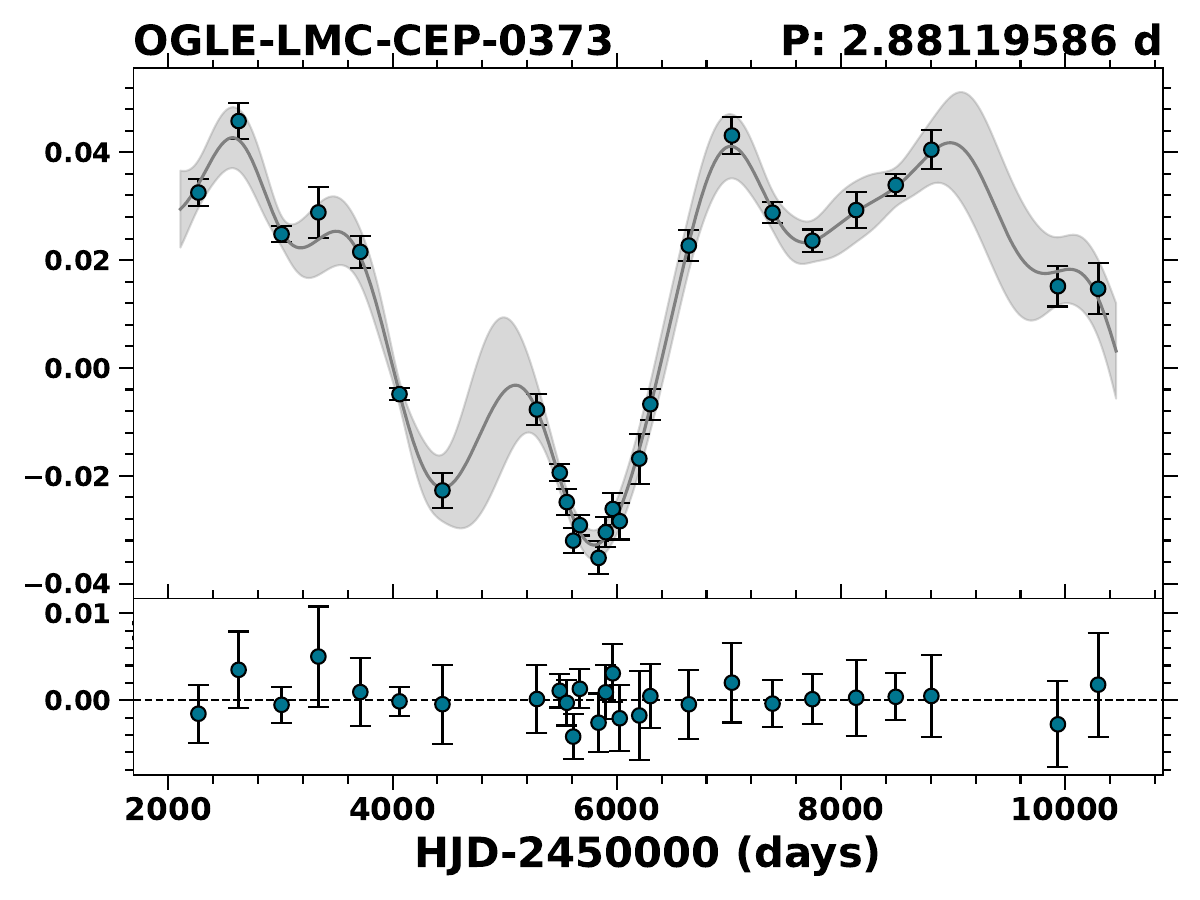}}
{\includegraphics[height=3.5cm,width=0.24\linewidth]{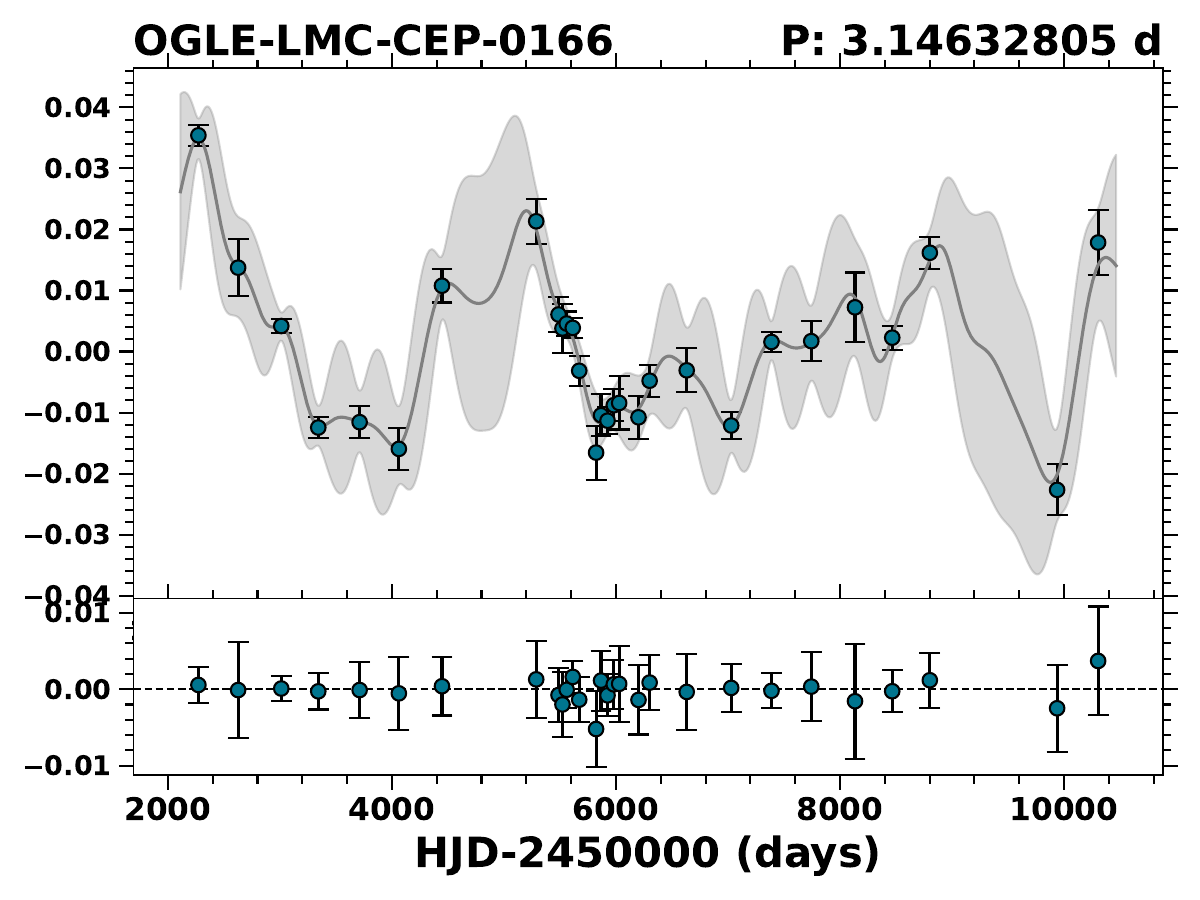}}
{\includegraphics[height=3.5cm,width=0.24\linewidth]{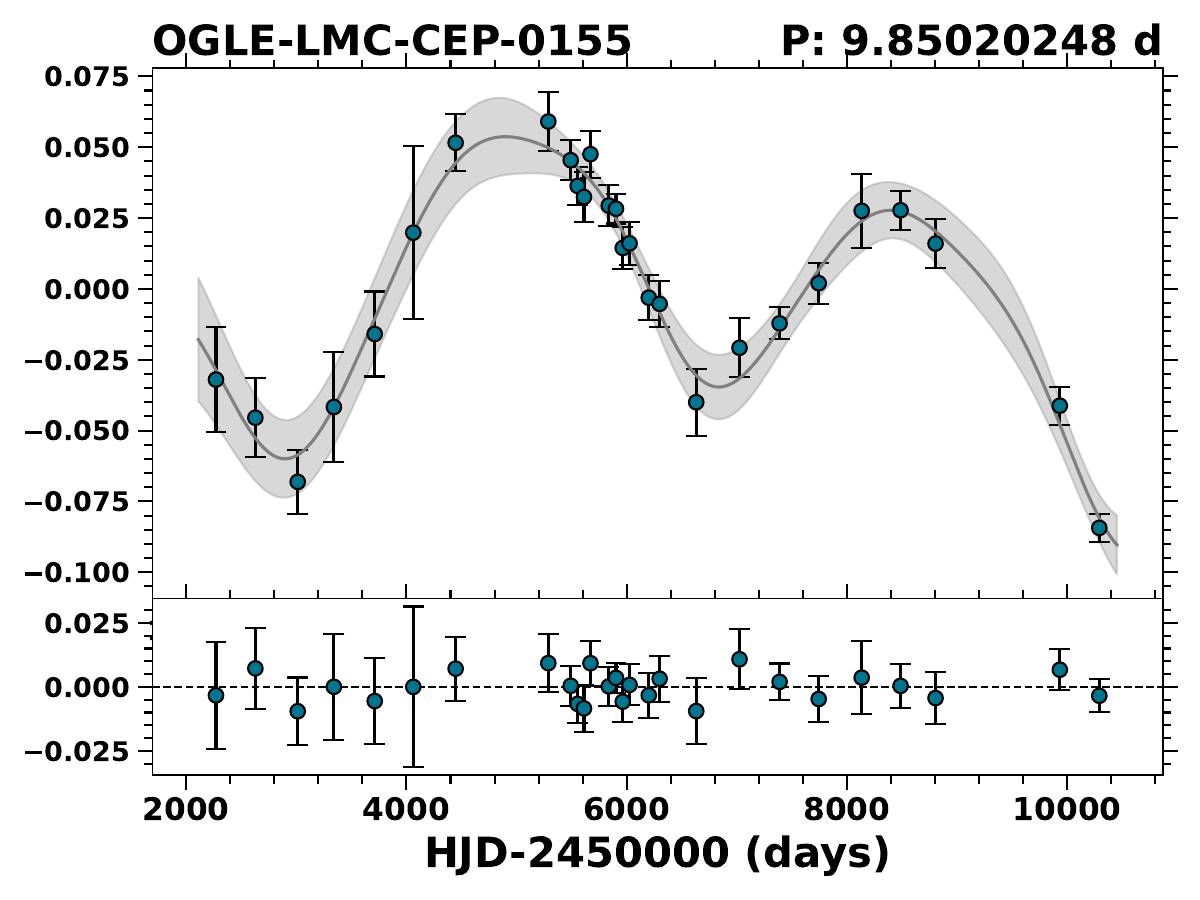}}
\caption{Examples of irregular shape $O-C$ diagrams (class 3) over-plotted with their GP fit solution (in gray) showing LMC F-mode candidates. Above each panel, the OGLE-ID and pulsation period are shown.}
\label{fig:ocplot_irregular_examples_LMCF}
\end{center}
\end{figure*}


\begin{figure*}
\begin{center}
{\includegraphics[height=3.5cm,width=0.24\linewidth]{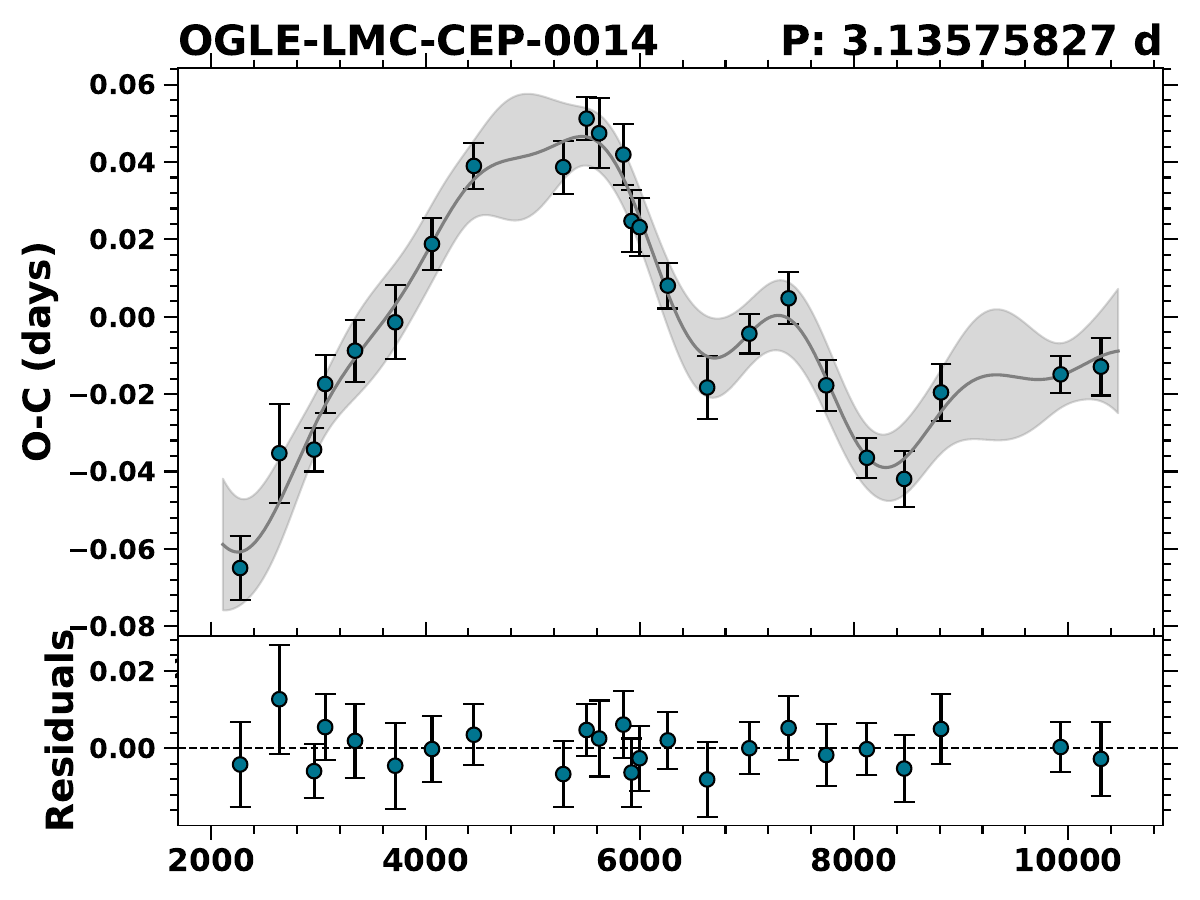}}
{\includegraphics[height=3.5cm,width=0.24\linewidth]{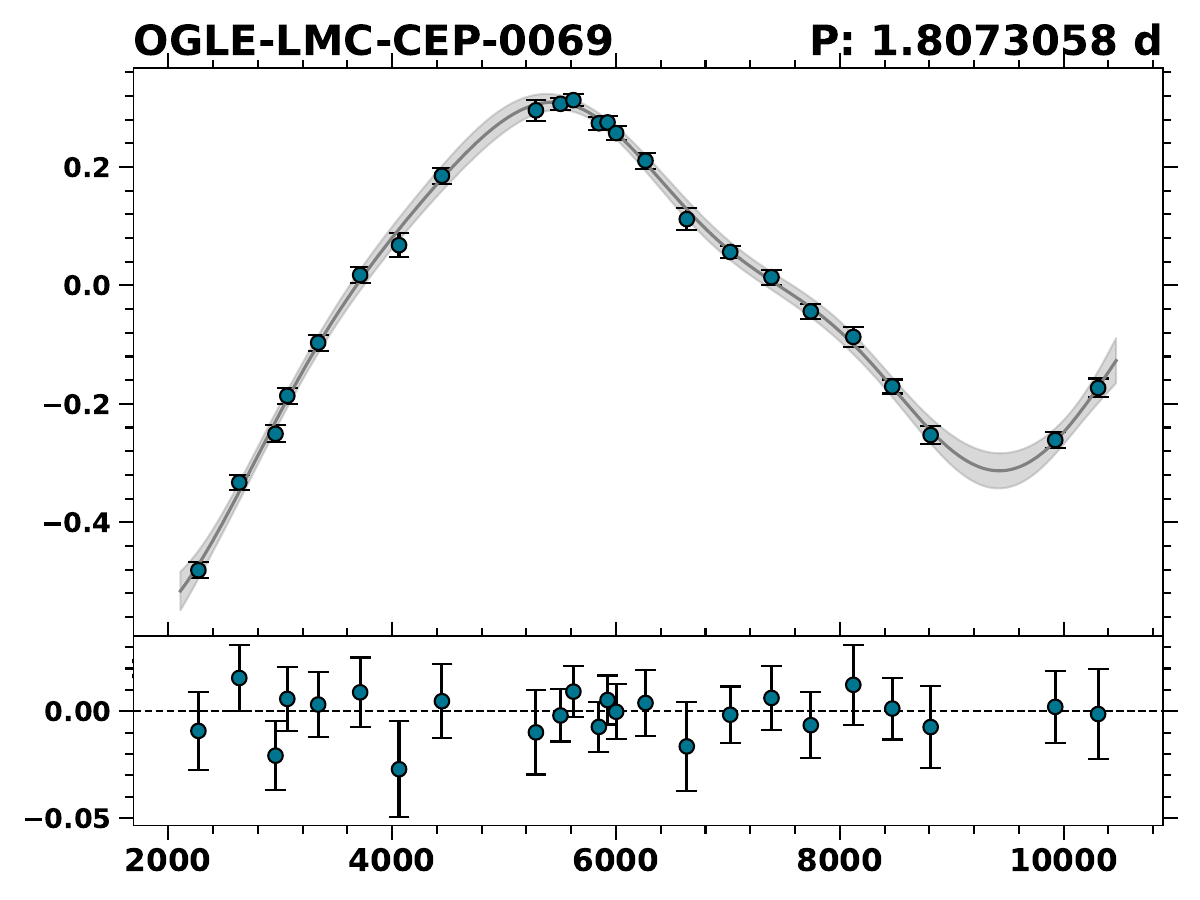}}
{\includegraphics[height=3.5cm,width=0.24\linewidth]{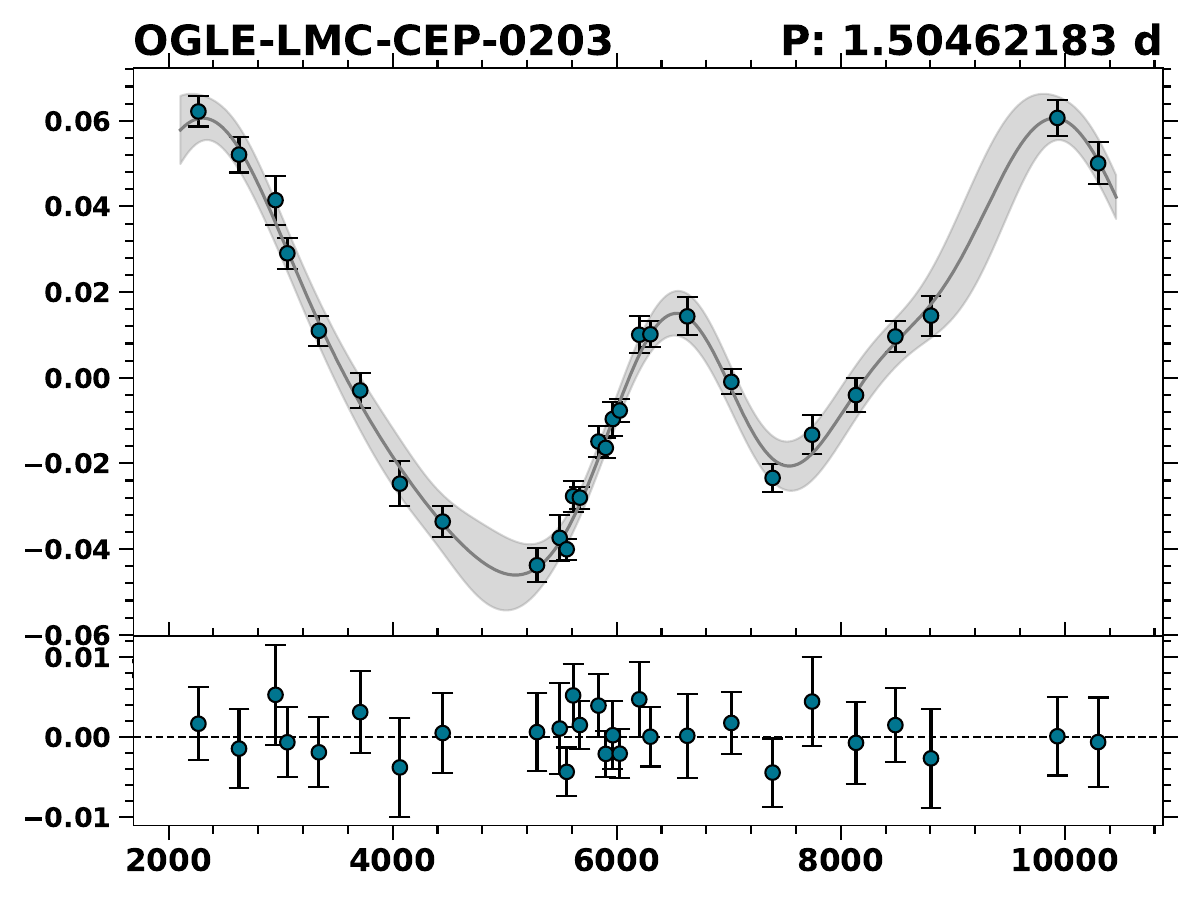}}
{\includegraphics[height=3.5cm,width=0.24\linewidth]{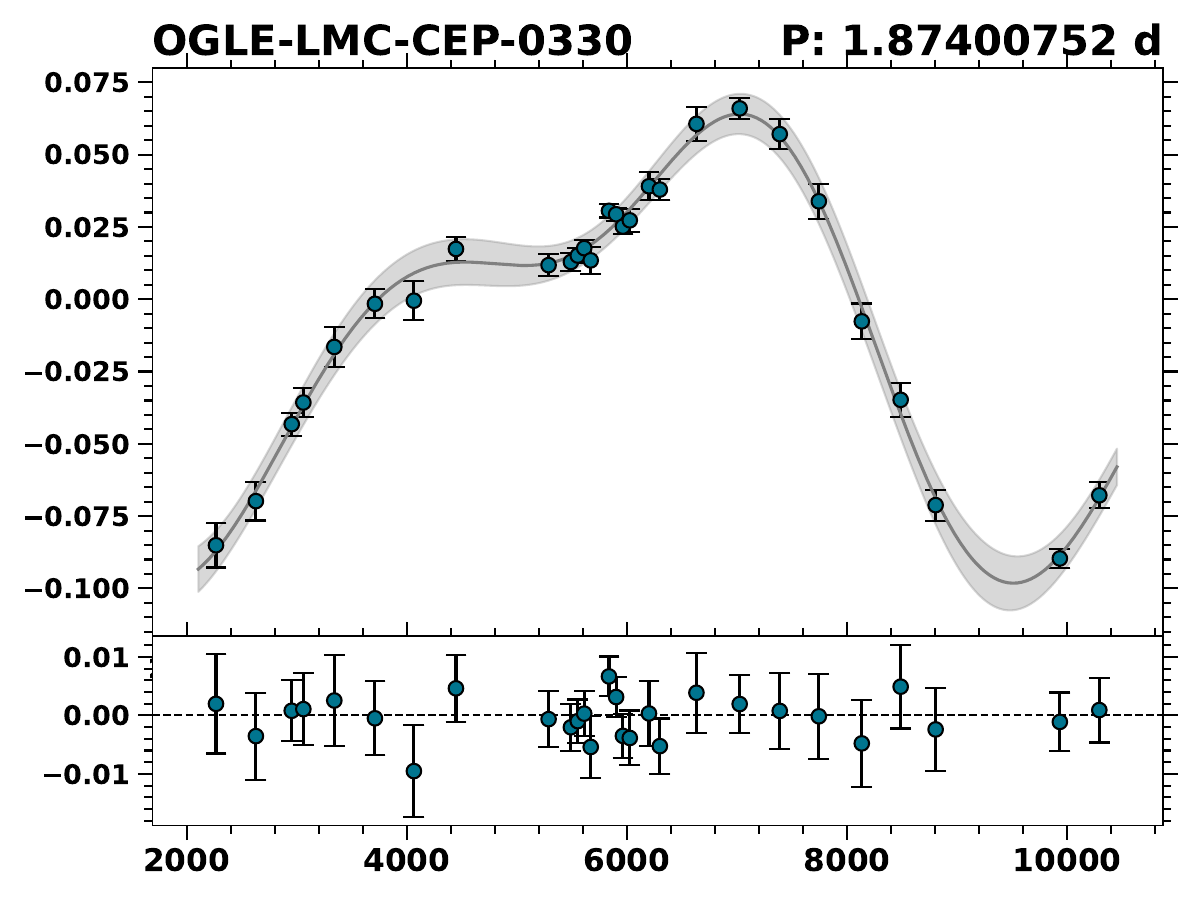}}
{\includegraphics[height=3.5cm,width=0.24\linewidth]{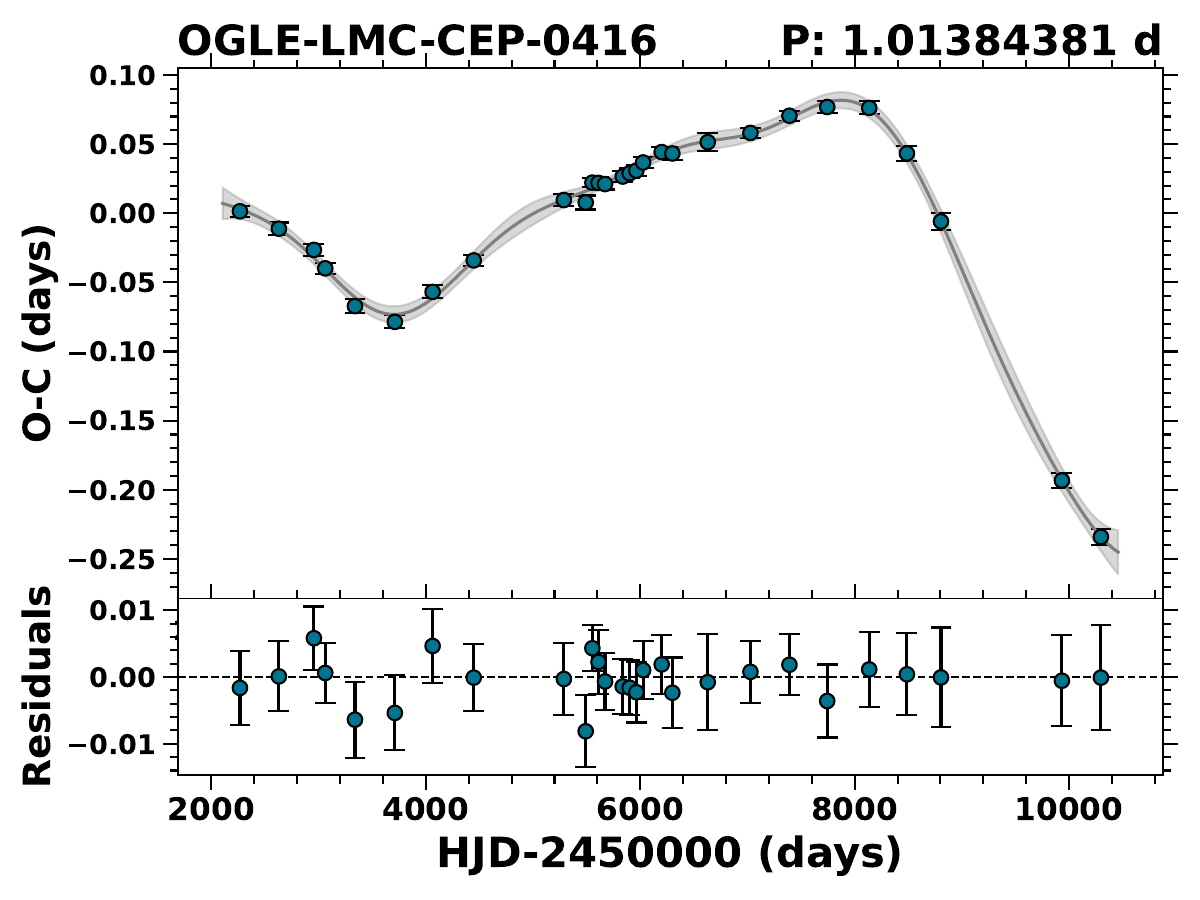}}
{\includegraphics[height=3.5cm,width=0.24\linewidth]{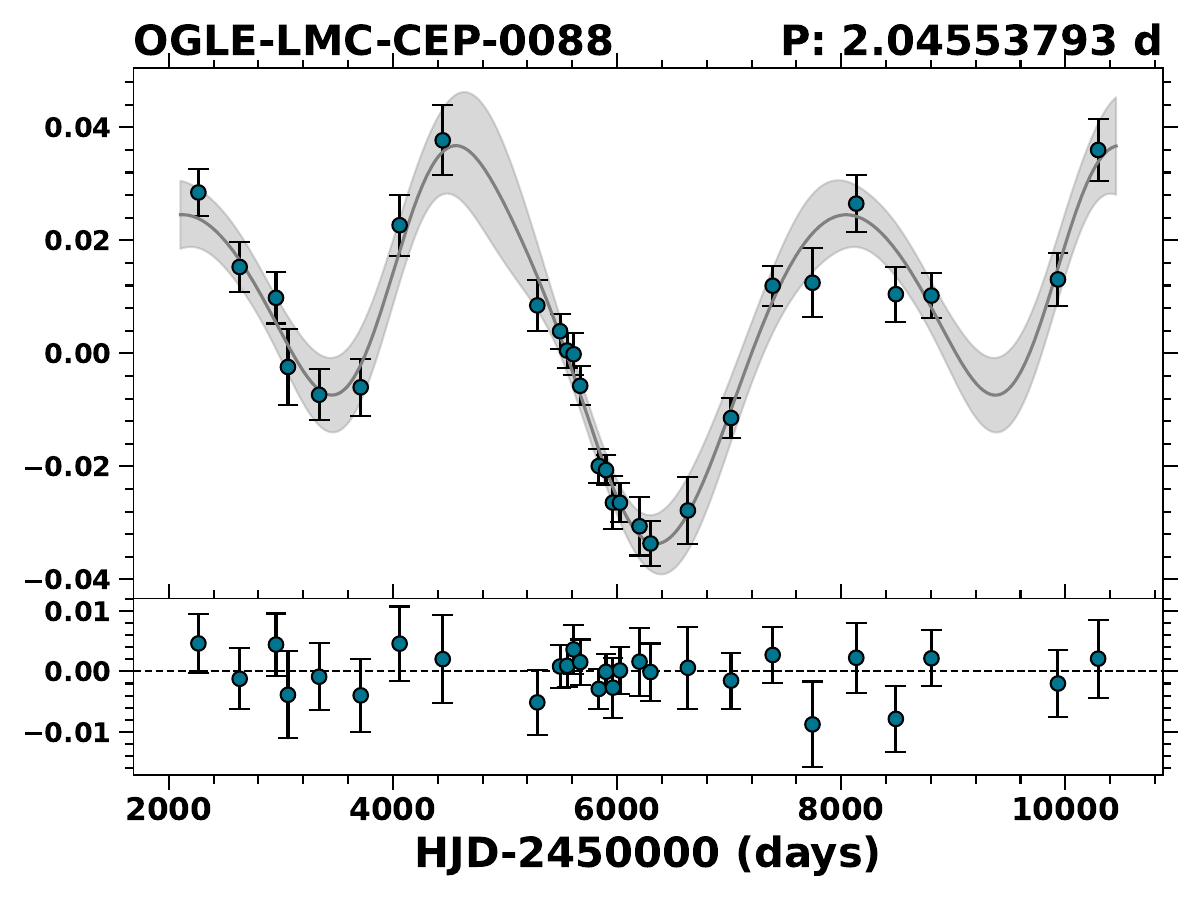}}
{\includegraphics[height=3.5cm,width=0.24\linewidth]{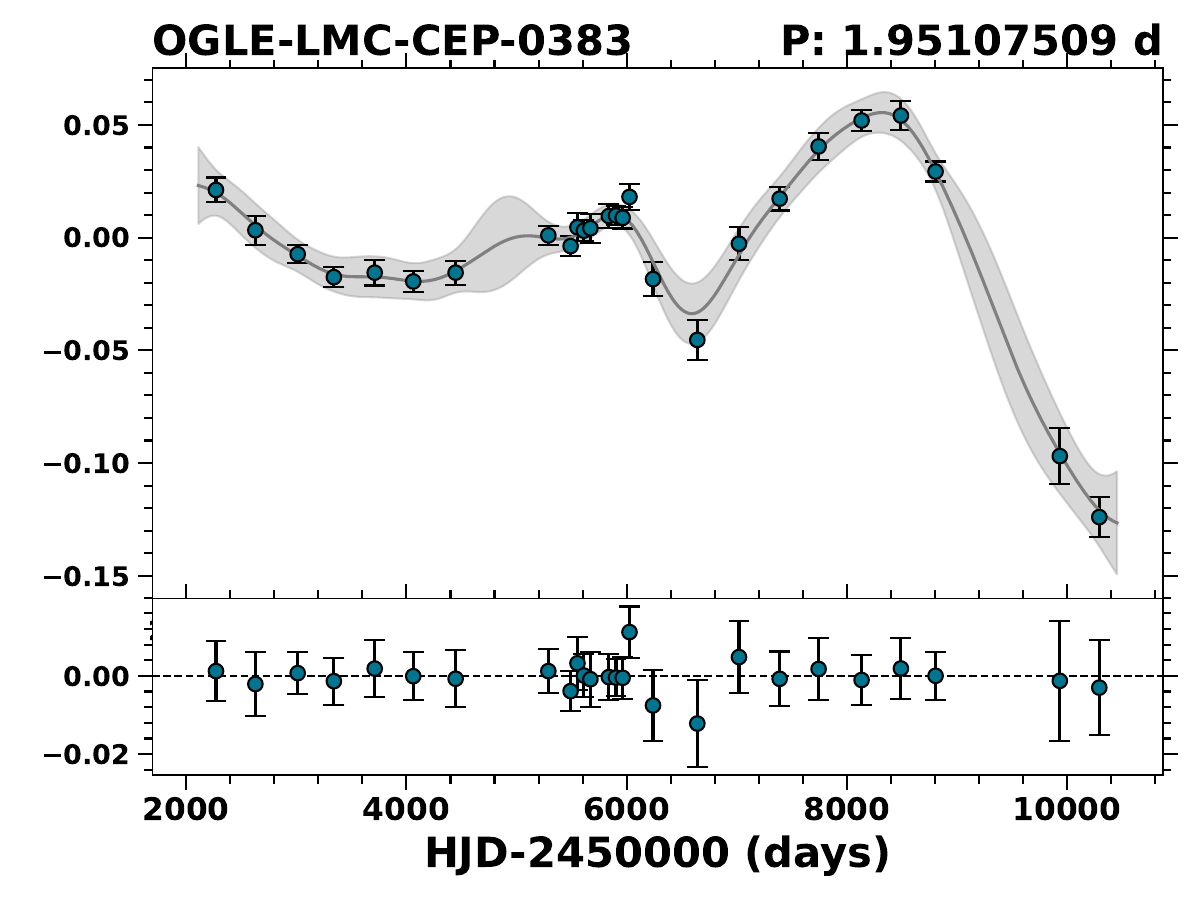}}
{\includegraphics[height=3.5cm,width=0.24\linewidth]{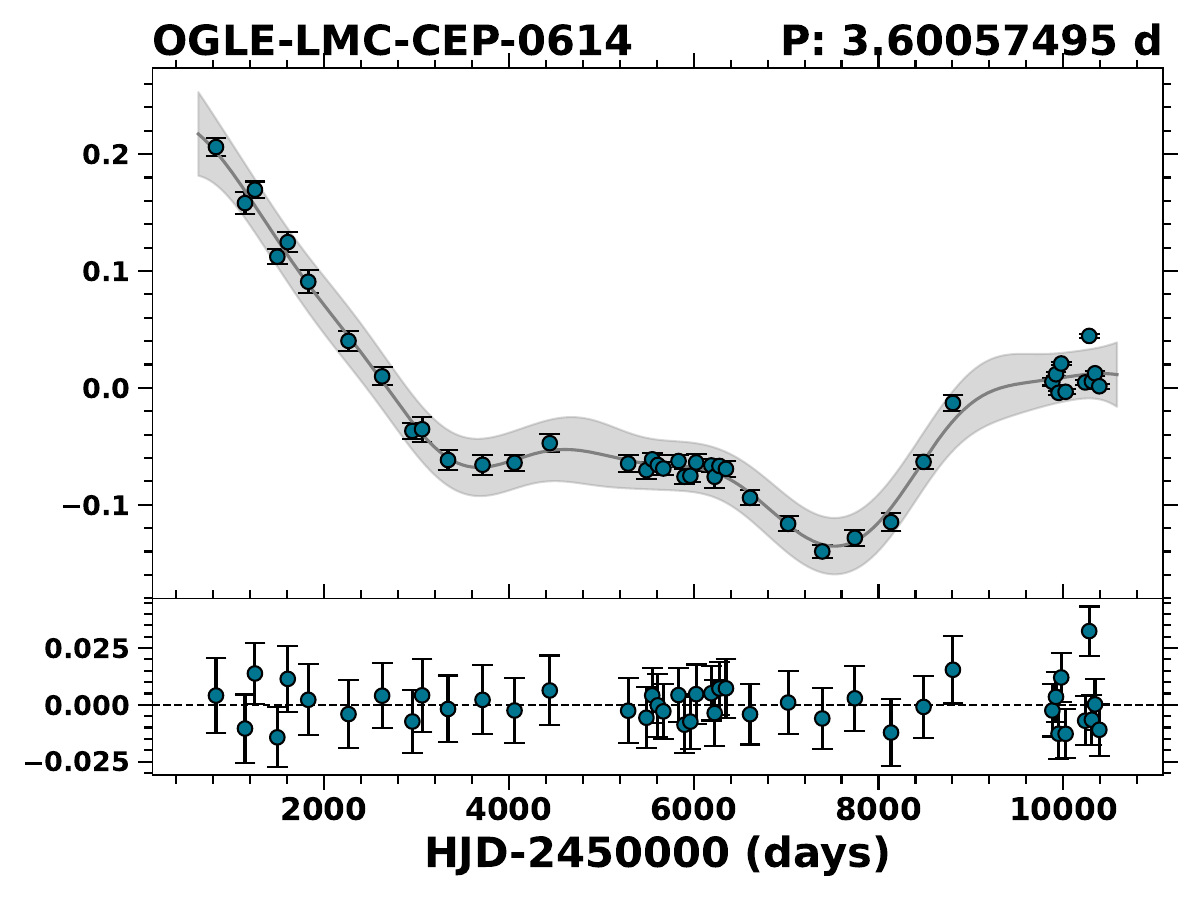}}
\caption{Examples of irregular shape $O-C$ diagrams (class 3) over-plotted with their GP fit solution (in gray) showing LMC 1O-mode candidates. Above each panel, the OGLE-ID and pulsation period are shown.}
\label{fig:ocplot_irregular_examples_LMC1O}
\end{center}
\end{figure*}


\begin{figure*}
\begin{center}
{\includegraphics[height=3.5cm,width=0.24\linewidth]{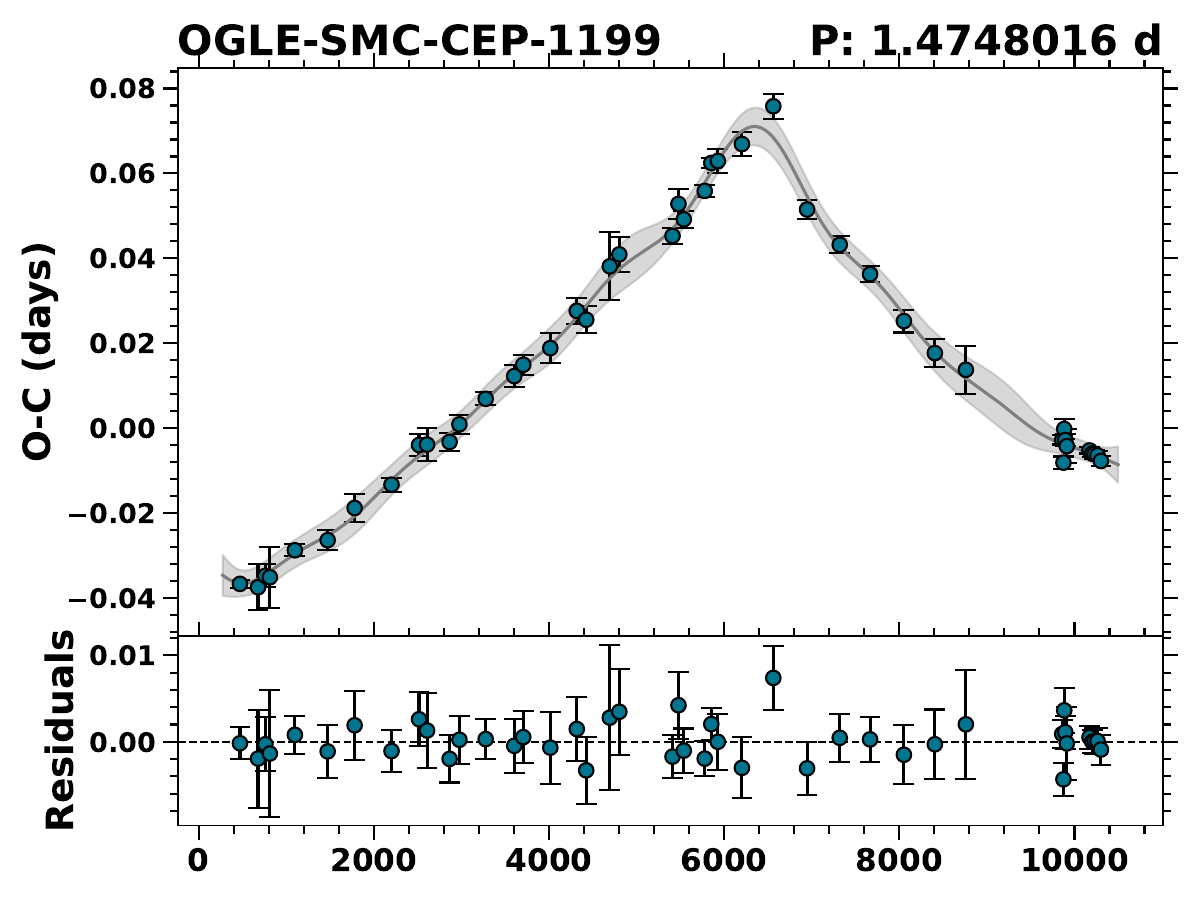}}
{\includegraphics[height=3.5cm,width=0.24\linewidth]{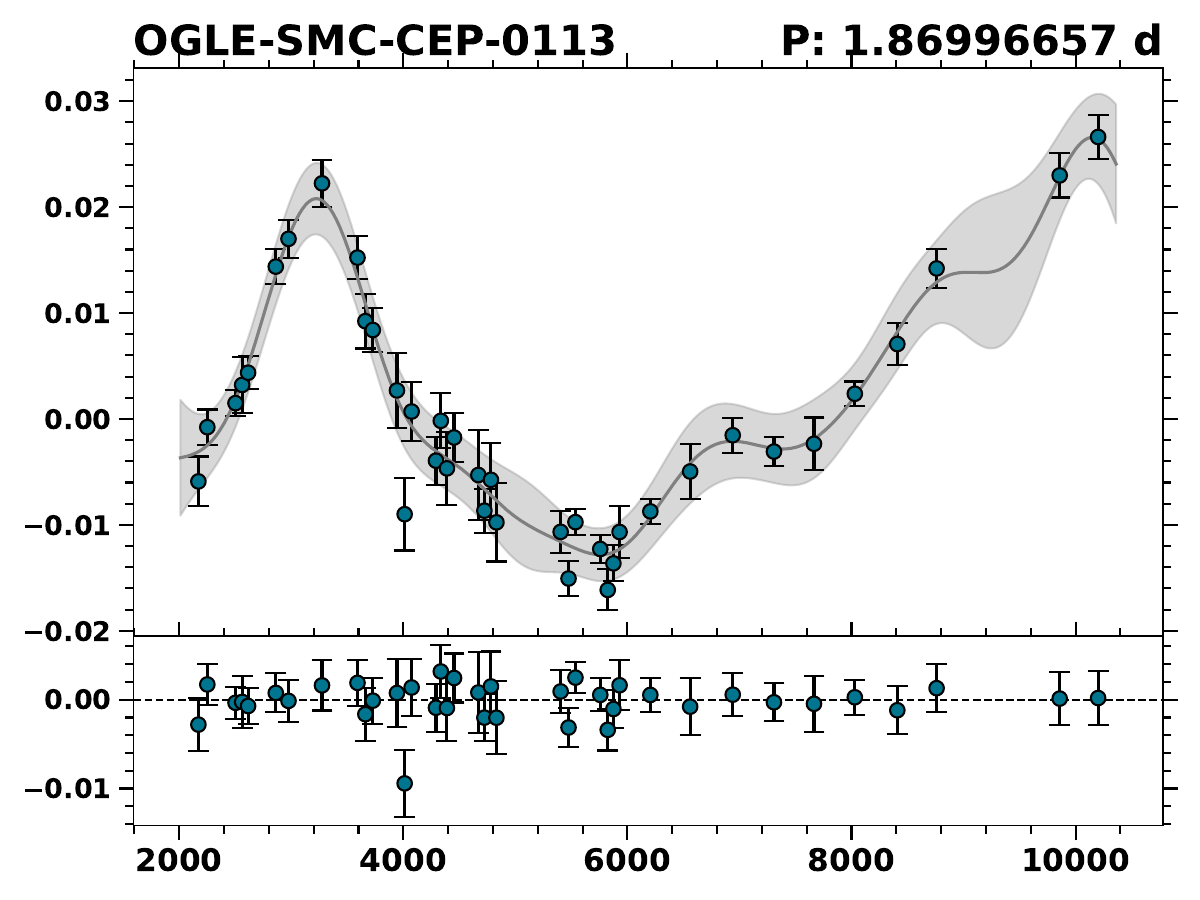}}
{\includegraphics[height=3.5cm,width=0.24\linewidth]{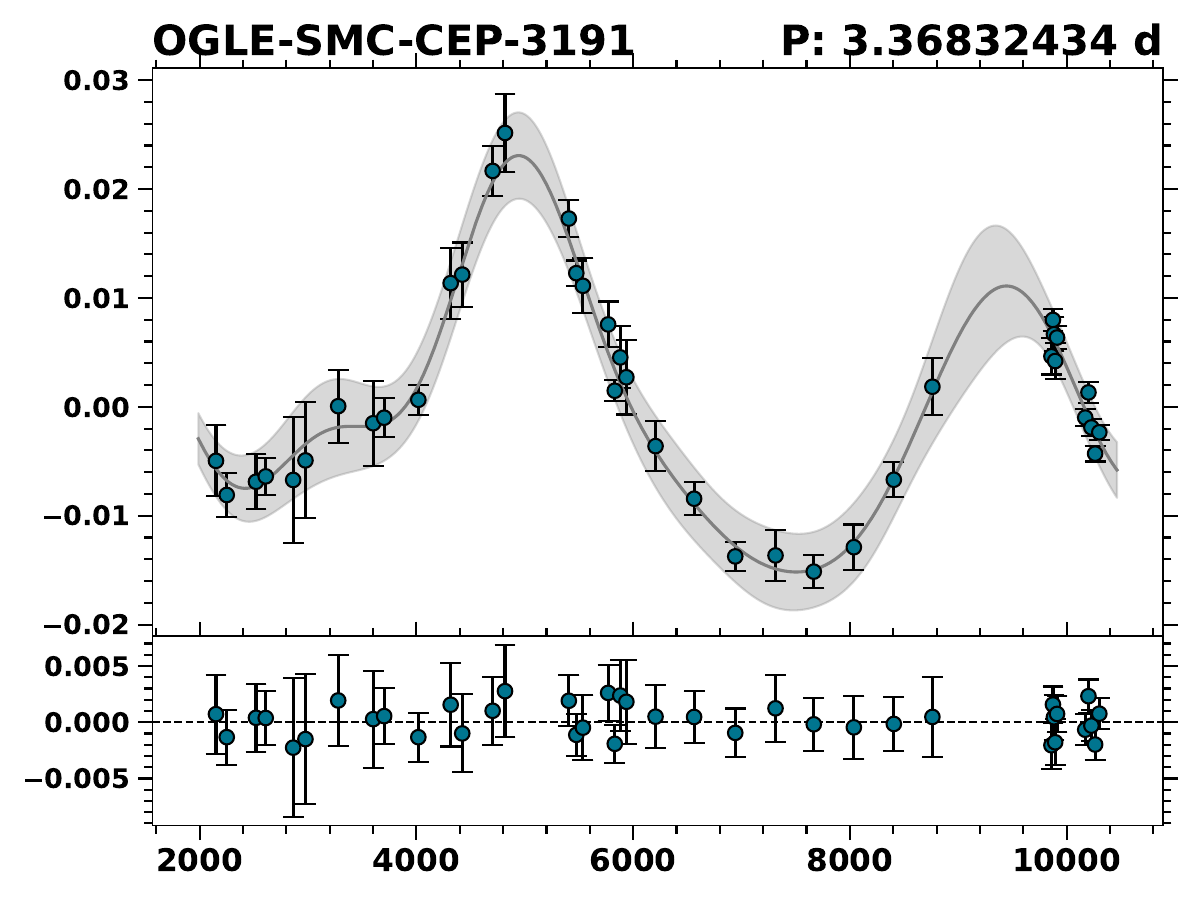}}
{\includegraphics[height=3.5cm,width=0.24\linewidth]{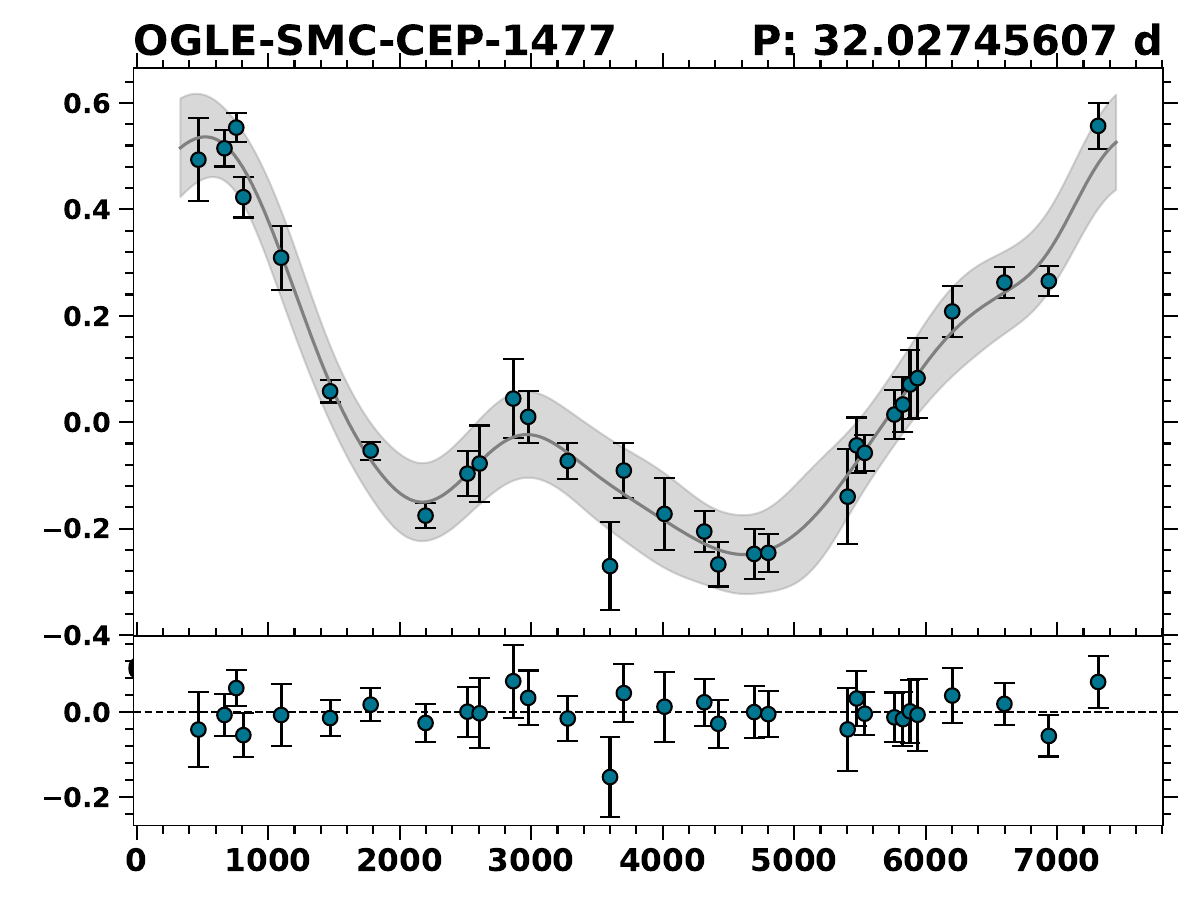}}
{\includegraphics[height=3.5cm,width=0.24\linewidth]{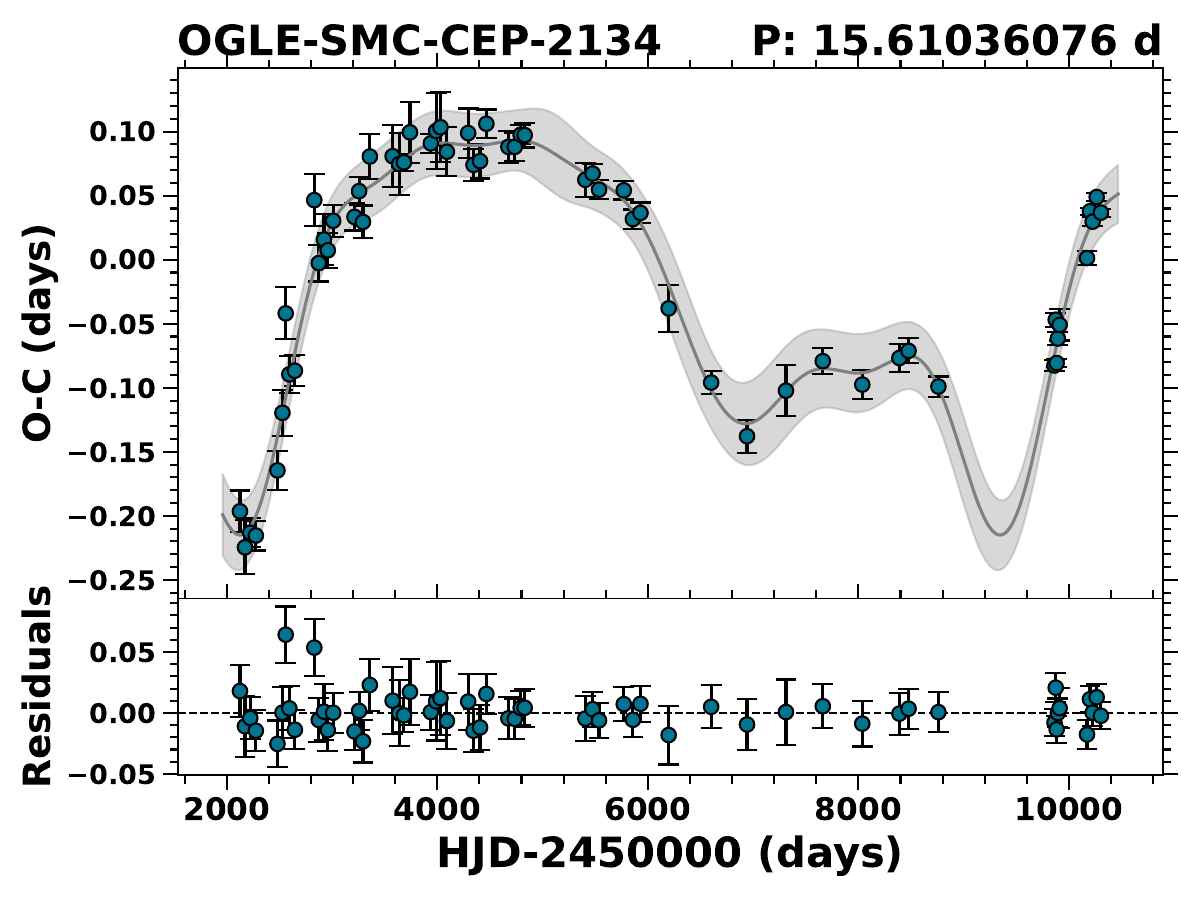}}
{\includegraphics[height=3.5cm,width=0.24\linewidth]{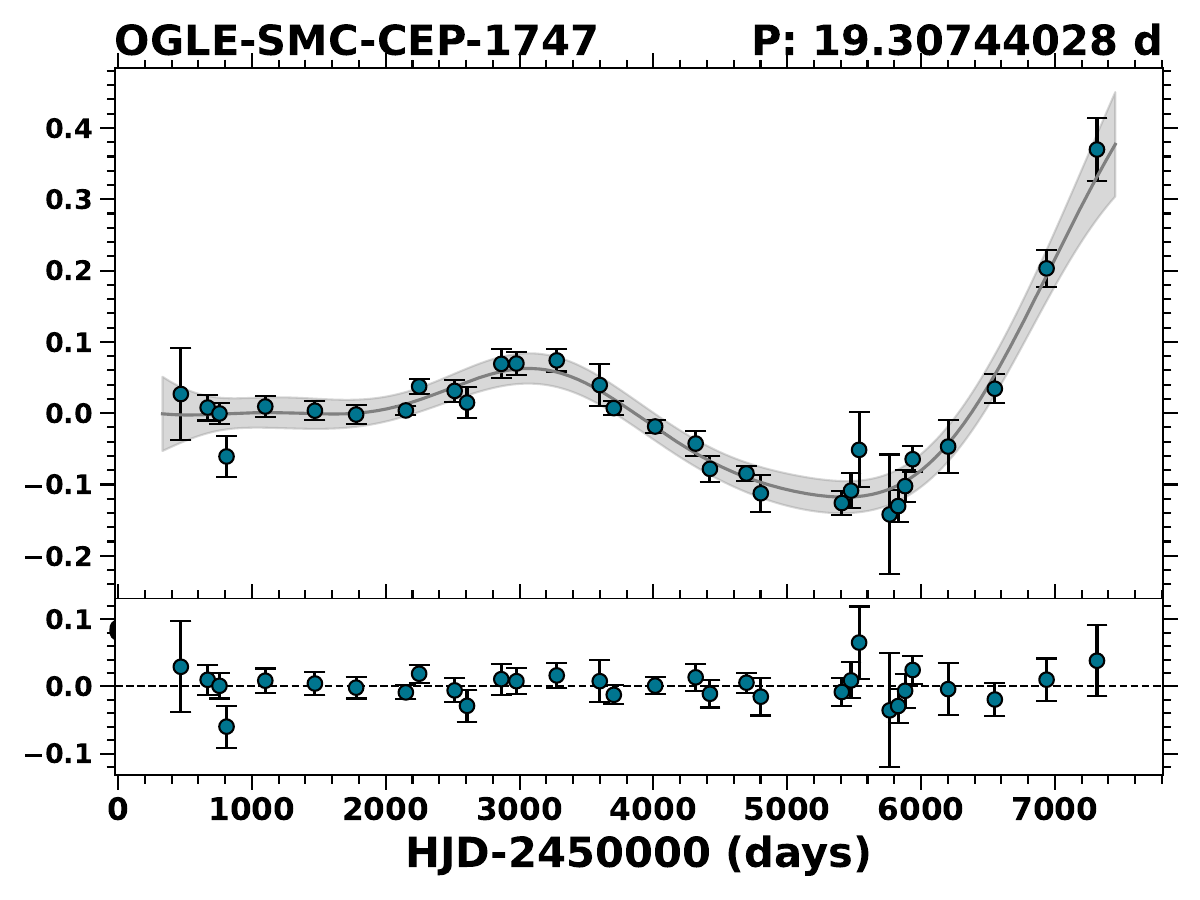}}
{\includegraphics[height=3.5cm,width=0.24\linewidth]{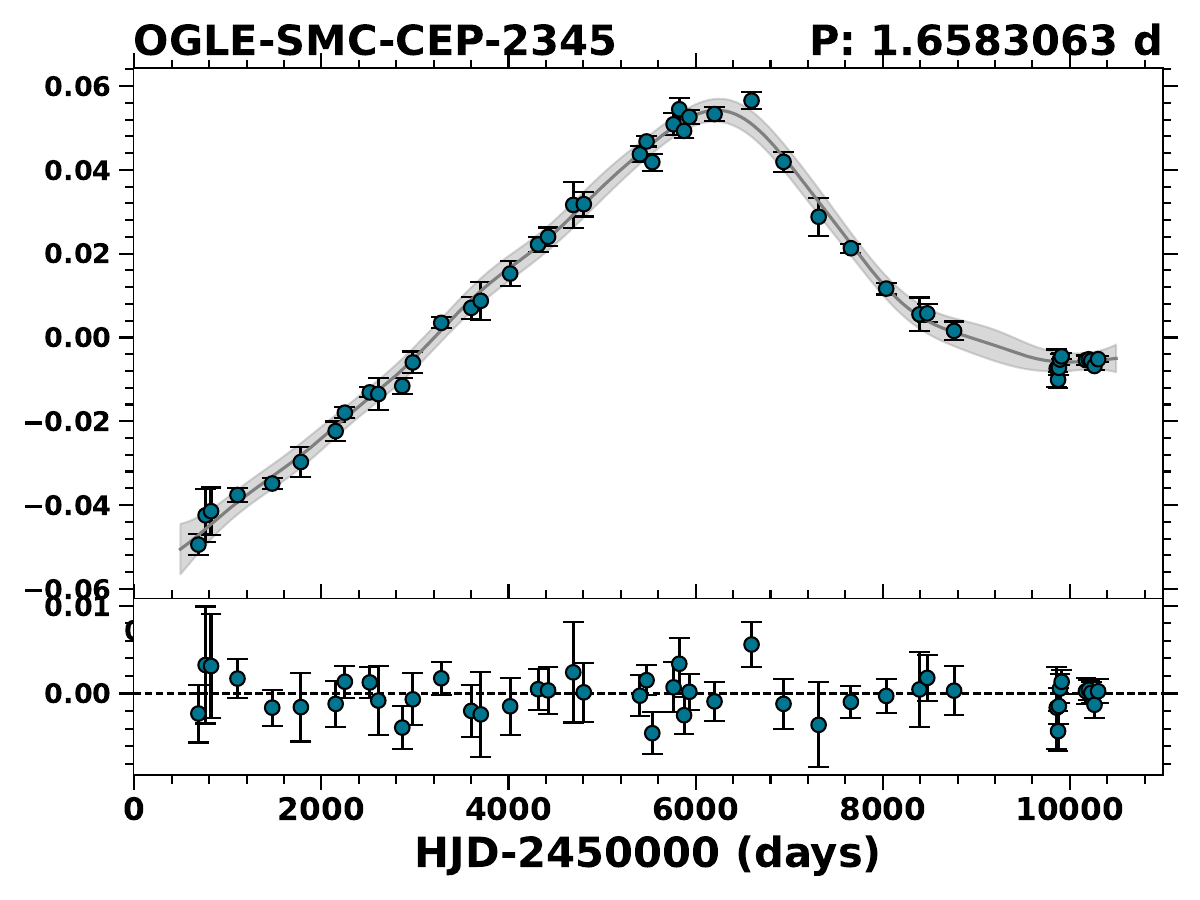}}
{\includegraphics[height=3.5cm,width=0.24\linewidth]{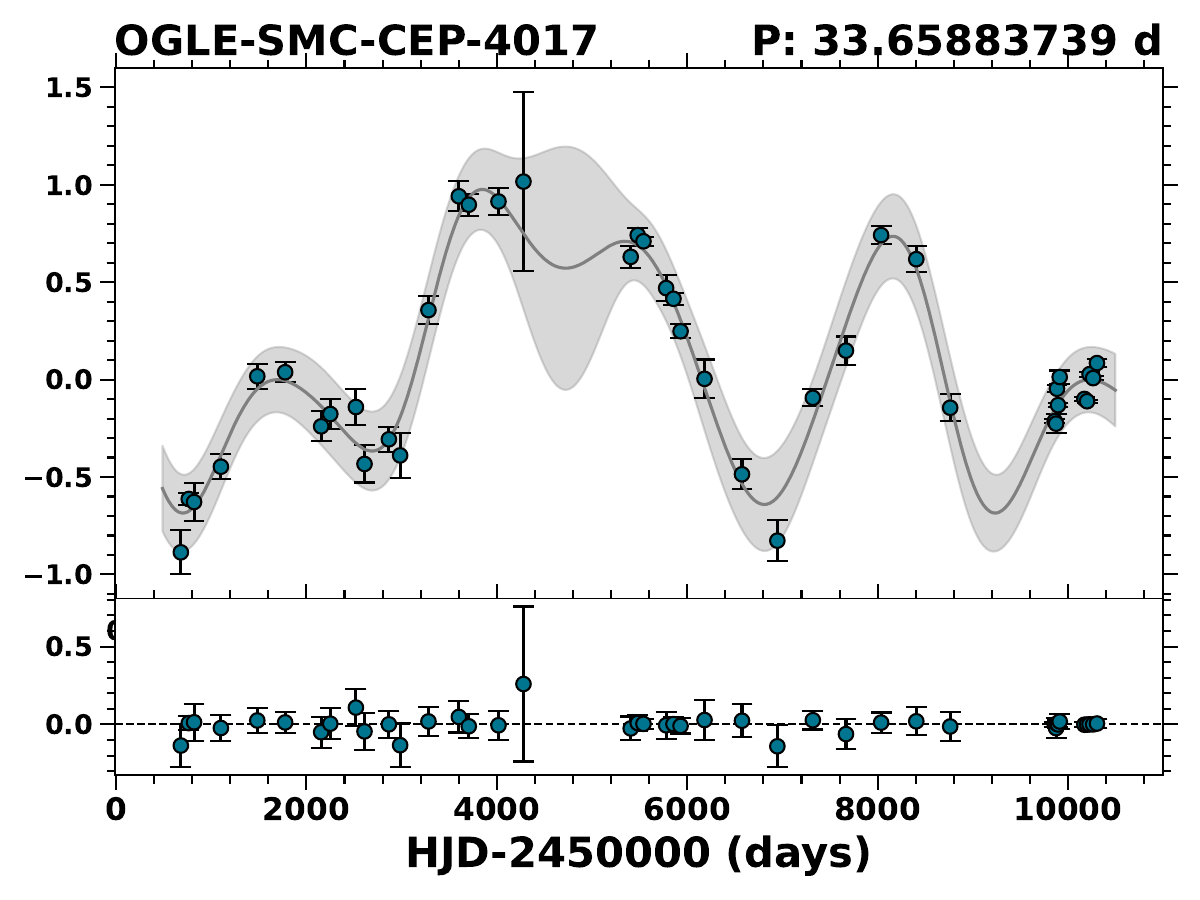}}
\caption{Examples of irregular shape $O-C$ diagrams (class 3) over-plotted with their GP fit solution (in gray) showing SMC F-mode candidates. Above each panel, the OGLE-ID and pulsation period are shown.}
\label{fig:ocplot_irregular_examples_SMCF}
\end{center}
\end{figure*}


\begin{figure*}
\begin{center}
{\includegraphics[height=3.5cm,width=0.24\linewidth]{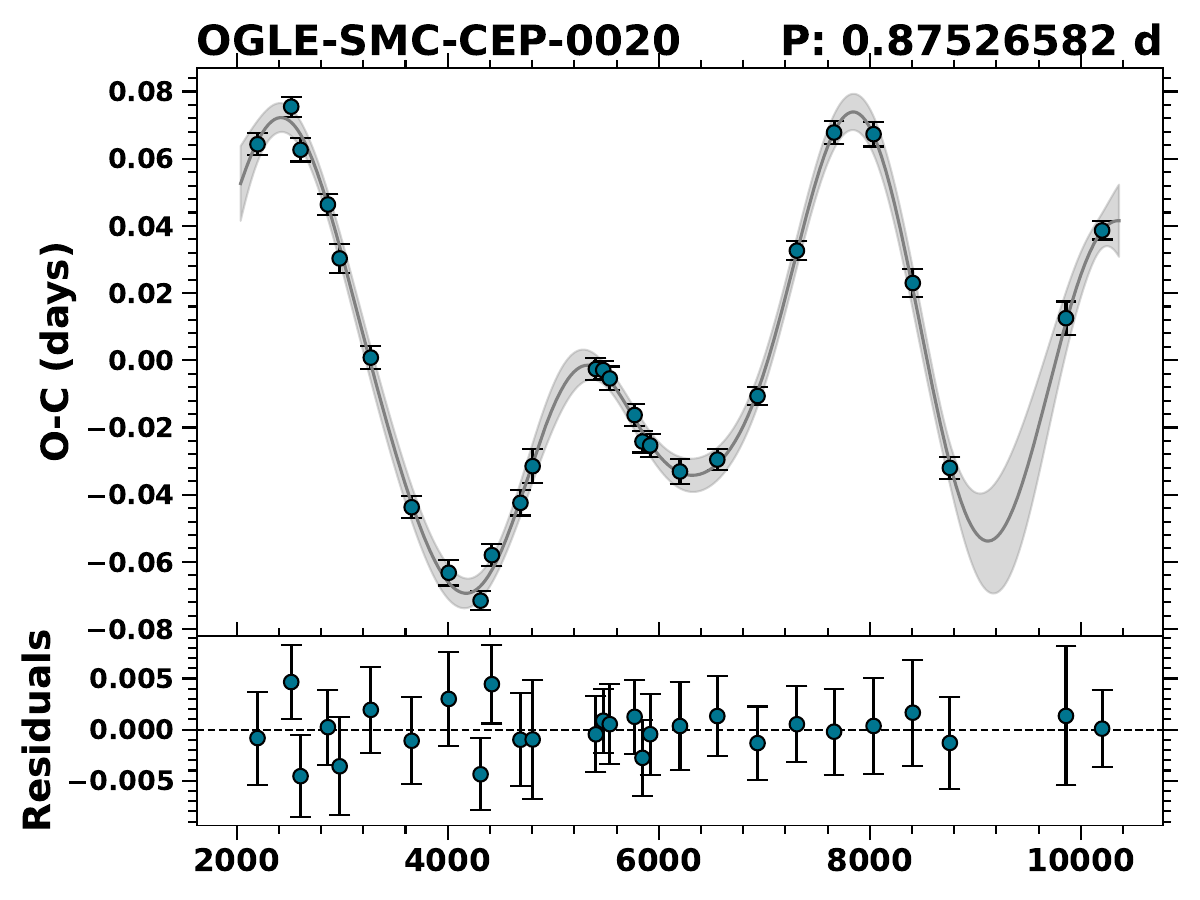}}
{\includegraphics[height=3.5cm,width=0.24\linewidth]{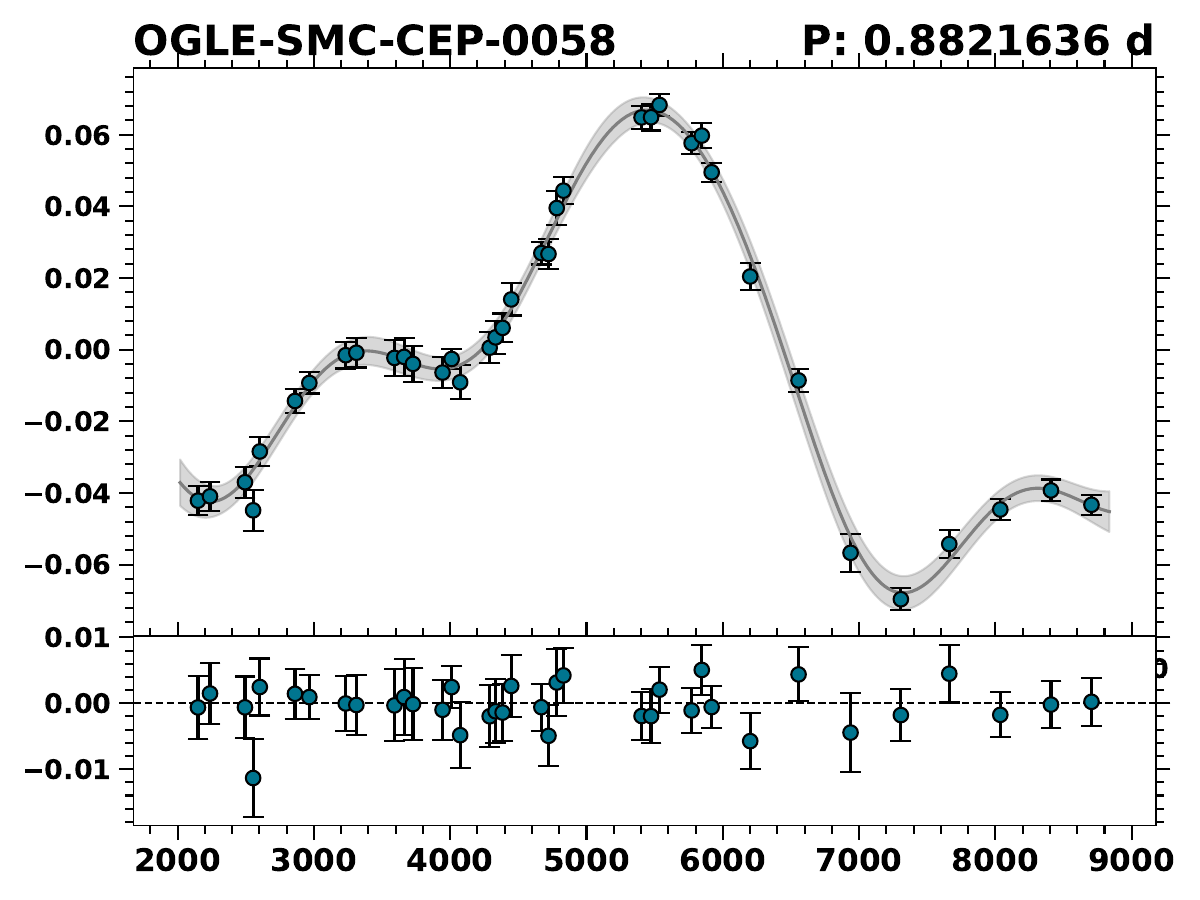}}
{\includegraphics[height=3.5cm,width=0.24\linewidth]{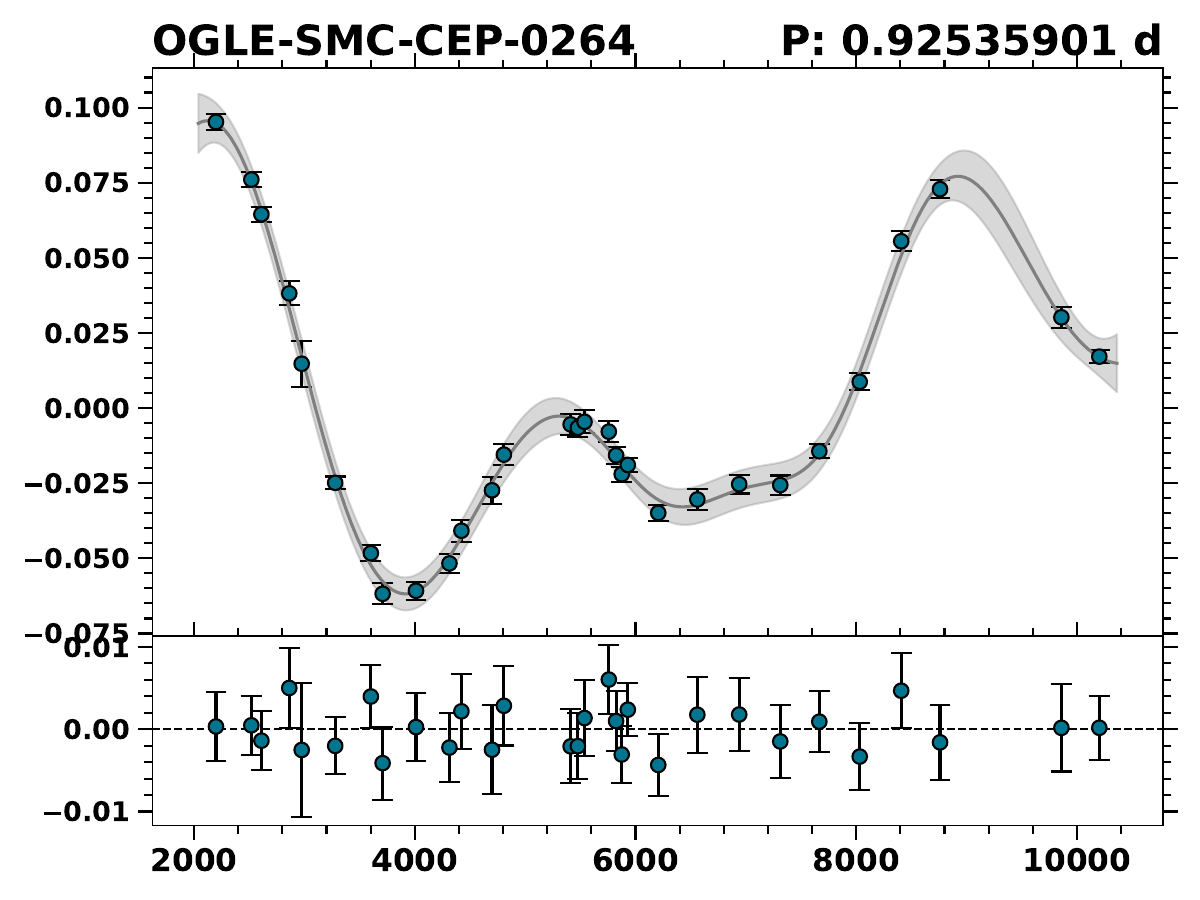}}
{\includegraphics[height=3.5cm,width=0.24\linewidth]{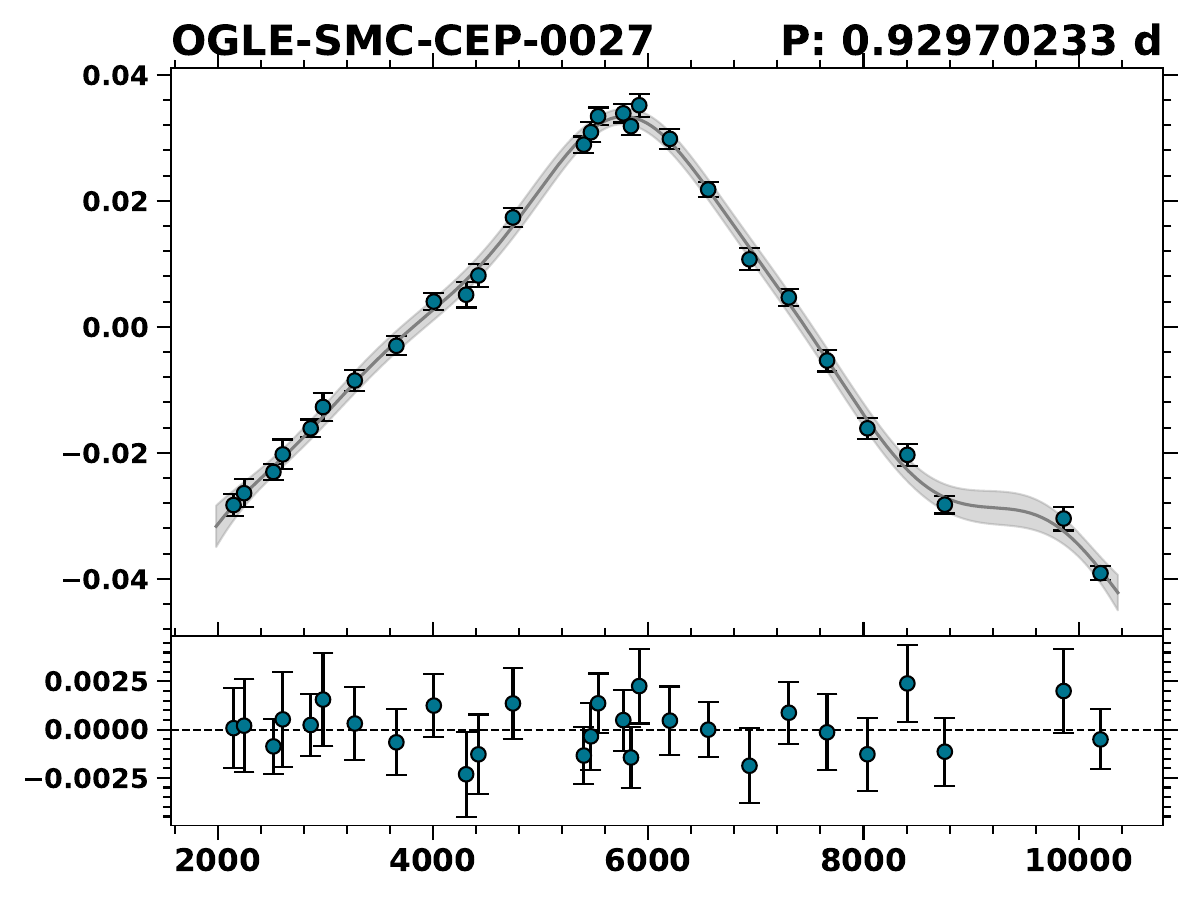}}
{\includegraphics[height=3.5cm,width=0.24\linewidth]{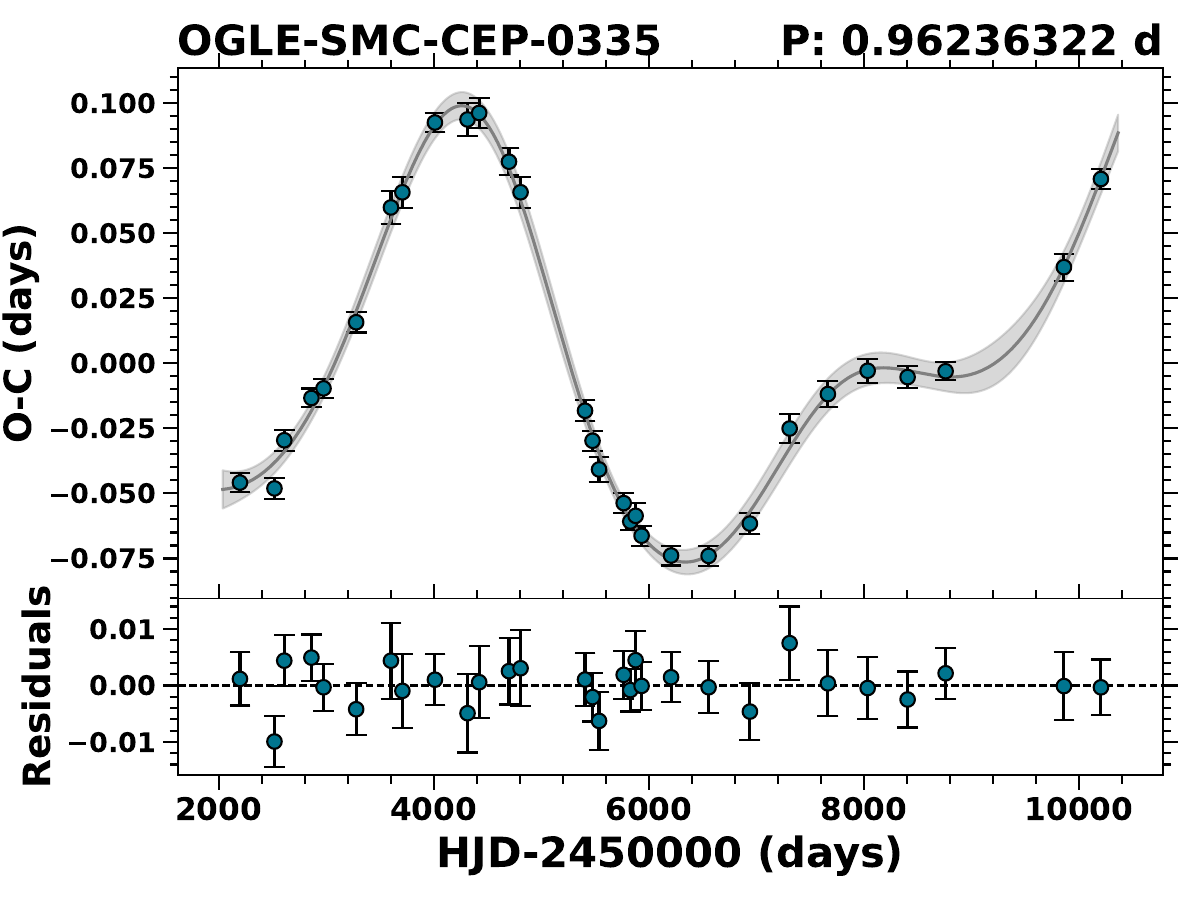}}
{\includegraphics[height=3.5cm,width=0.24\linewidth]{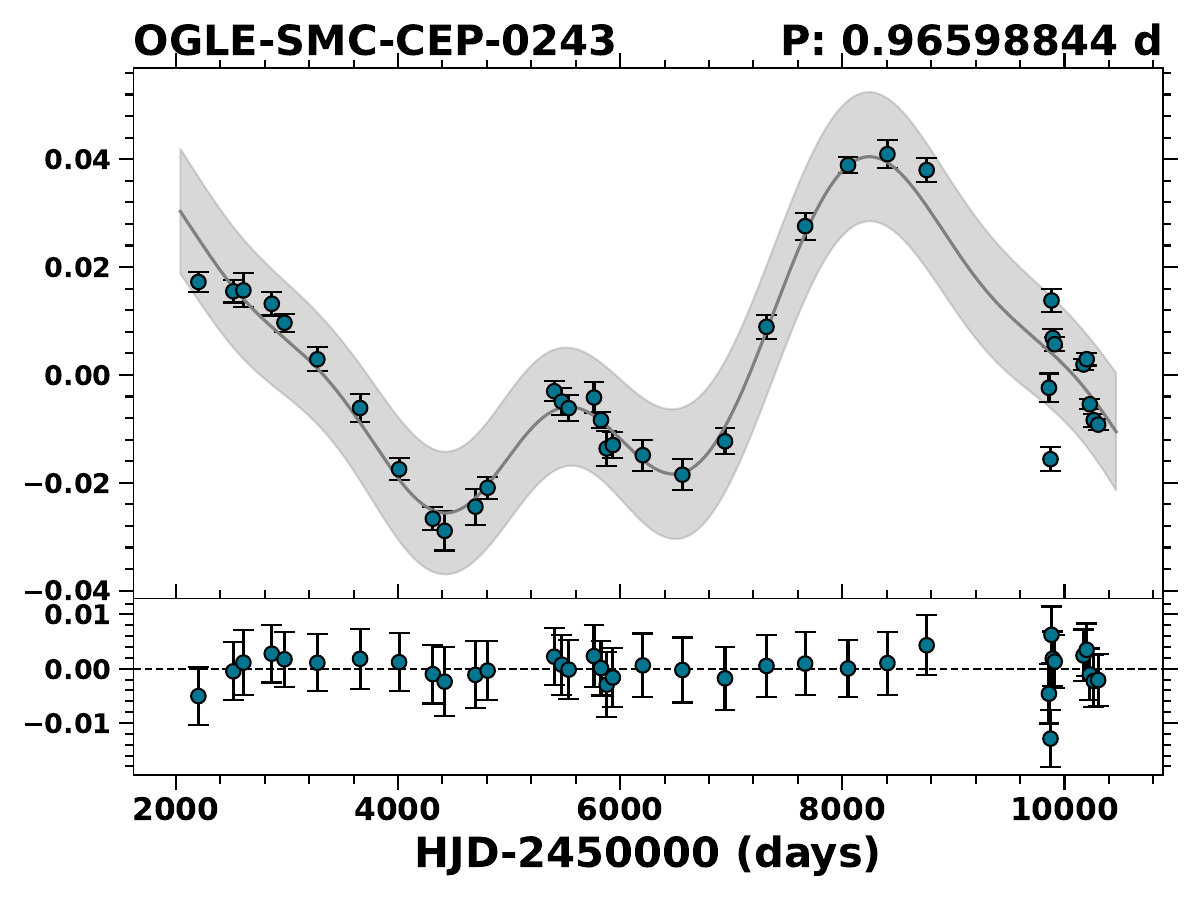}}
{\includegraphics[height=3.5cm,width=0.24\linewidth]{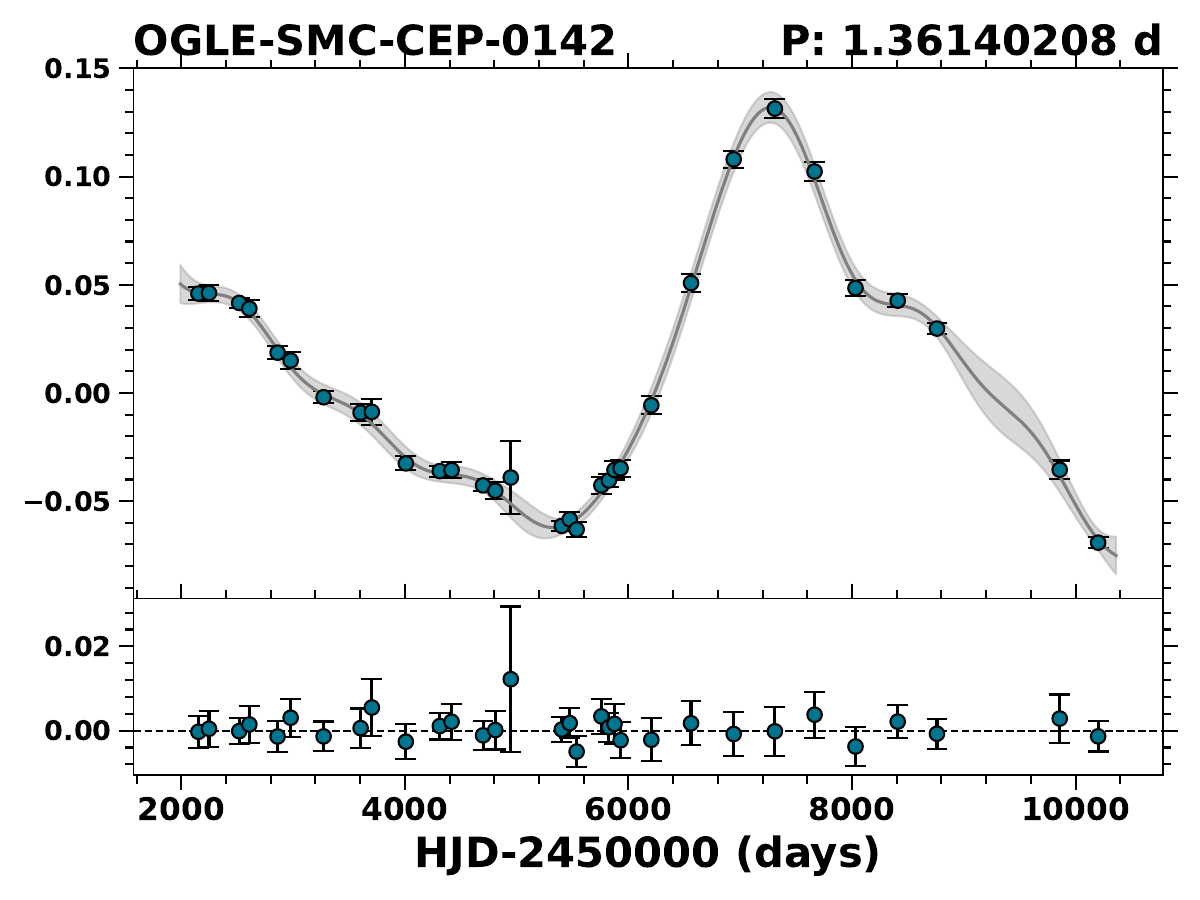}}
{\includegraphics[height=3.5cm,width=0.24\linewidth]{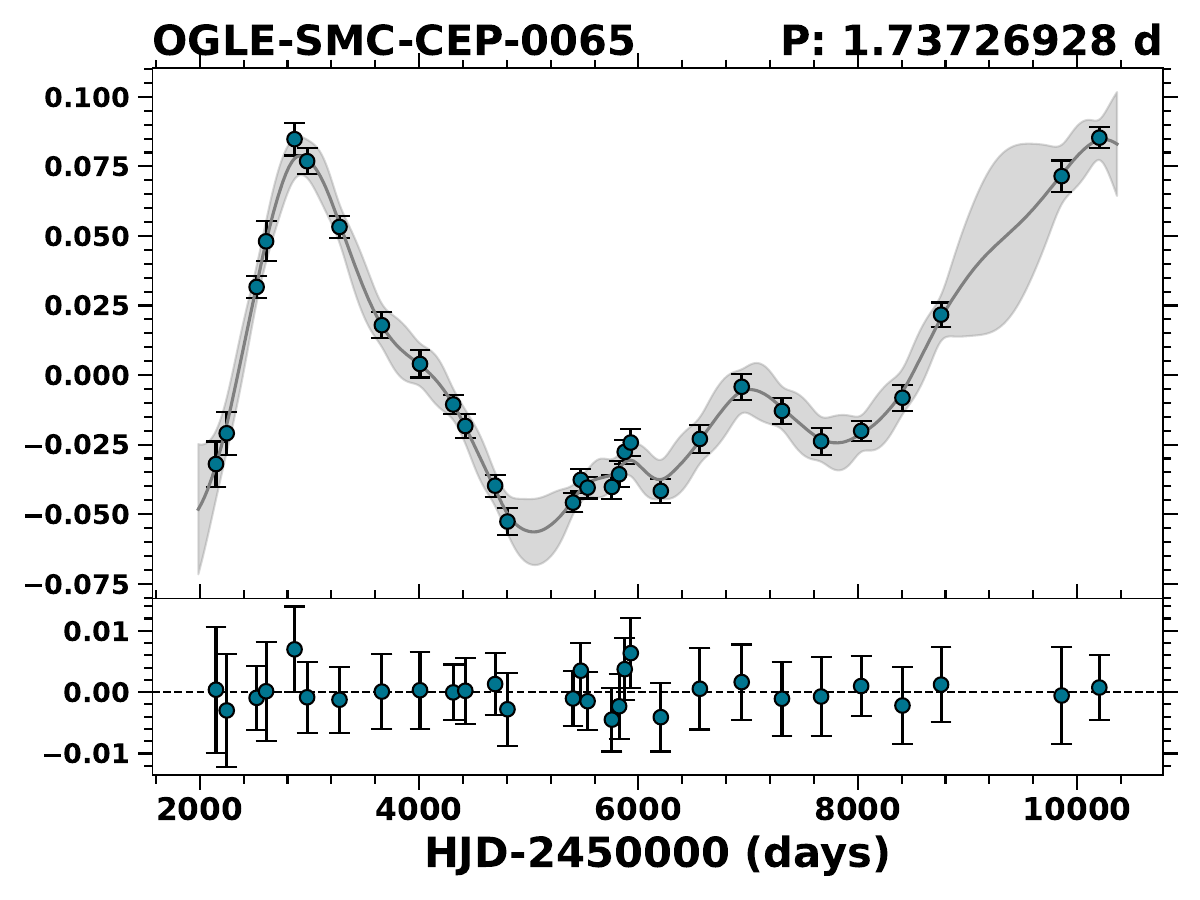}}
\caption{Examples of irregular shape $O-C$ diagrams (class 3) over-plotted with their GP fit solution (in gray) showing SMC 1O-mode candidates. Above each panel, the OGLE-ID and pulsation period are shown.}
\label{fig:ocplot_irregular_examples_SMC1O}
\end{center}
\end{figure*}


\begin{table}[ht]
\caption{Sample of class 1 Cepheid candidates.}
\label{tab:Class_1_list_mainpaper}
\centering\footnotesize
\begin{tabular}{lr}
\hline
\hline
\textbf{OGLE ID} & \textbf{P} \\
                &  \textbf{(d)}\\
\hline
\hline
\textbf{LMC F mode} & \\
\hline
OGLE-LMC-CEP-0180 & 3.1181490   \\
OGLE-LMC-CEP-3214 & 1.1347907   \\
OGLE-LMC-CEP-0827 & 1.1530167   \\
OGLE-LMC-CEP-1930 & 1.1591452   \\
\ldots & \ldots \\
\hline
\end{tabular}
\tablefoot{The two columns list the star's ID and its pulsation period in days. For the full list of candidates refer to section ~\ref{sec: Data availability}.}
\end{table}

\begin{table}[ht]
\caption{Sample of class 2 Cepheid candidates.}
\label{tab:Class_2_list_mainpaper}
\centering\footnotesize
\begin{tabular}{lrrr}
\hline
\hline
\textbf{OGLE ID} & \textbf{$P$} &  \textbf{d$P$/d$t$} &  \textbf{err d$P$/d$t$} \\
 & \textbf{(d)} &  \textbf{d/Myr)} &  \textbf{(d/Myr)} \\
\hline
\hline
\textbf{LMC F mode}        &             &           &            \\
\hline
OGLE-LMC-CEP-1617 & 0.99714957 & -6.621  & 0.569   \\
OGLE-LMC-CEP-3282 & 1.19690726 & 2.407   & 0.135   \\
OGLE-LMC-CEP-2358 & 1.74607748 & -3.917  & 0.389   \\
OGLE-LMC-CEP-0168 & 1.75437431 & 8.789   & 0.36    \\
\ldots & \ldots  & \ldots  & \ldots   \\
\hline
\end{tabular}
\tablefoot{The four columns give star's ID, pulsation period, period change rate and its error. For the full list of candidates refer to section ~\ref{sec: Data availability}.}
\end{table}

\begin{table*}[ht]
\caption{Sample of class 3 Cepheid candidates.}
\label{tab:Class_3_list_mainpaper}
\centering\footnotesize
\begin{tabular}{lrcccccc}
\hline
\hline
\textbf{OGLE ID} & \textbf{$P$} & \textbf{$\epsilon$} & \textbf{\textbf{Dominant}} & \textbf{Dominant} & \textbf{Stetson L index}  & \textbf{$\sigma$}  & \textbf{$\Delta P$} \\
 &  &  & \textbf{\textbf{Periodicity}} & \textbf{Amplitude} &   &   &  \\
\hline
\hline
\textbf{LMC F mode}  &  &  &  &  &  &  &  \\
\hline
OGLE-LMC-CEP-3107 & 1.0348131 & 0.00015 & 7659 & 0.0191 & 0.17589 & 0.0000070 & 0.000023 \\
OGLE-LMC-CEP-1594 & 1.2749185 & -     & 840  & 0.0005 & 0.03037 & 0.0000401 & 0.000132 \\
OGLE-LMC-CEP-1703 & 1.2901287 & 0.00053 & 7602 & 0.2580 & 0.80832 & 0.0002110 & 0.000831 \\
OGLE-LMC-CEP-1508 & 1.3088341 & 0.00082 & 5469 & 0.0269 & 2.46936 & 0.0000432 & 0.000133 \\
\ldots & \ldots  & \ldots  & \ldots & \ldots  & \ldots  & \ldots & \ldots  \\
\hline
\end{tabular}
\tablefoot{The columns give star's ID, pulsation period, $\epsilon$ parameter from E--P test; dominant variability period and amplitude from Wavelet method; Stetson L index; $\sigma$ and $\Delta P$ from instantaneous period method. For the full list of candidates refer to section ~\ref{sec: Data availability}.}
\end{table*}

\section{characterisation}
\label{sec: characterisation}

The analysis presented here is different and more explorative in nature from our previous work, \citetalias{Rathour2024A&A...686A.268R}, where non-evolutionary period change was due to binarity, and a Keplerian model could be used to characterise $O-C$ diagrams. In the absence of any physical mechanism or model regarding these irregular period changes, we resorted to multiple methodologies for characterising our targets, as detailed in the following sections.

\subsection{Eddington–Plakidis test}
\label{subsec: Eddington–Plakidis (E–P) test}

Fluctuations in the $O-C$ points can sometimes be misinterpreted as genuine non-linear period changes. To distinguish between true period changes and random phase fluctuations, we employ the well-established Eddington–Plakidis (E–P) test \citep{Eddington1929MNRAS..90...65E}. It has been used in various period change studies \citep[e.g.][]{Turner2003A&A...407..325T,Turner2006PASP..118..410T,Derekas2012MNRAS.425.1312D,Csornyei2022MNRAS.511.2125C} to test the random, cycle-to-cycle changes in the period. This method assumes that some of the variations in the $O-C$ diagram, characterised by fluctuation parameter, $\epsilon$, are caused by random cycle-to-cycle fluctuations in the period. To estimate $\epsilon$, we compute the absolute differences (delays) between $O-C$ residuals separated by $x$ cycles, represented as $u(x) = |a(r+x) - a(r)|$, where $a(r)$ is the residual $O-C$ value at the $r$th index. The equation below provides the linear relation between the period fluctuation parameter, $\epsilon$, and the mean of all accumulated delays, $\langle u(x)\rangle$:
\begin{equation}
\label{eq:EP}
\langle u(x)\rangle^2 = 2\alpha^2 + x\epsilon^2,
\end{equation}
where $\alpha$ characterises the random error of the measurement, and the slope of the above equation gives a measure of the square of random fluctuation in the period, $\epsilon$. If up to several tens of cycles the relation is linear, then the Cepheid pulsations indeed contain random period fluctuations. Importantly, before applying the test, it is necessary to remove quadratic trends to filter out possible effects due to secular evolution.

The challenging aspect of this calculation is identifying the appropriate cycle separation where the Eddington–Plakidis equation \eqref{eq:EP} can be fit. To address this, we construct a method combining two complementary approaches.
First, we apply an iterative linear fit to the data across a range of cycle separations, calculating the $R^2$  value at each step. The $R^2$ value, known as the coefficient of determination, quantifies how well the model explains the variance in the data. It is defined as:
\begin{equation}
\label{eq:Rsquare}
R^2 = 1 - \frac{ \Sigma_{\rm res} }{\Sigma_{\rm tot}},
\end{equation}
where $\Sigma_{\rm res}$ represents the sum of squared residuals (the squared differences between observed and fitted values), and $\Sigma_{\rm tot}$ represents the total sum of squares (the squared differences between observed values and their mean). This provides a broad view of where the data exhibits shifts in trends, with high $R^2$  values indicating better fits (the closer $R^2$  is to 1, the better the fit). These high $R^2$ regions guide the initial placement of segments for further analysis.

\begin{figure}
\begin{center}
{\includegraphics[height=6cm,width=\linewidth]{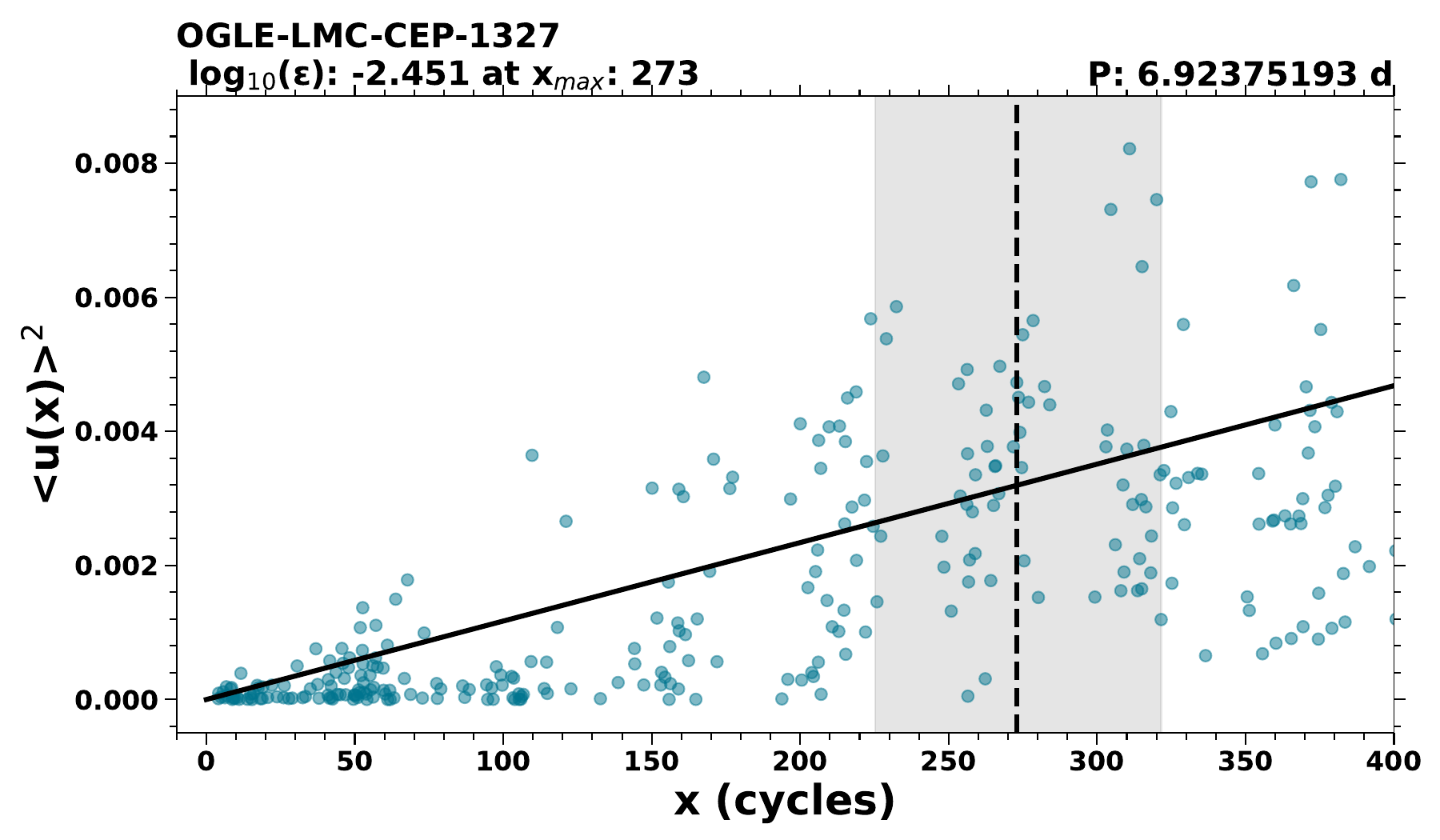}}
\caption{Eddington–Plakidis test for OGLE-SMC-CEP-1530. The plot shows the distribution of the mean accumulated delays as a function of cycle separation. The black vertical line indicates the break point, with the uncertainty shaded in gray. The linear fit to the distribution terminates at this break point. The cycle separation at the break point, along with the fluctuation parameter calculated from the E--P test, is displayed on the top-left corner.}
\label{fig:EP_plot}
\end{center}
\end{figure}

Next, we refine this analysis using the \texttt{piecewise regression} package \citep{Pilgrim2021}, which fits segmented linear models and identifies breakpoints where the slope changes. The method iteratively adjusts these breakpoints to minimize residuals and estimates shifts in slopes. To confirm significant breakpoints, the Davies test \citep{Davies10.1093/biomet/74.1.33} is applied, rejecting the null hypothesis of no breakpoints when the 
$p$-value is less than 0.05. In Fig.~\ref{fig:EP_plot} we show an example of the E--P test with the breakpoint indicated (vertical black line) with its margin of error (faded gray region). Once the breakpoint region is determined, it is straightforward to extract the parameters based on Eq.~\eqref{eq:EP}.

\subsection{Time-frequency analysis with wavelets}
\label{subsec: Time-Frequency analysis}
Another method we employed to characterise irregular $O-C$ diagrams is wavelet analysis. Wavelet analysis allows to detect and trace the evolution of periodic signals in time. We use it to determine the amplitude and time scale of the changes present in $O-C$ diagrams.

Before applying wavelet analysis, we first fit the $O-C$ curve of each Cepheid with a Gaussian Process Regression (GPR) model, implemented in the \texttt{scikit-learn} library \citep{Pedregosa2011scikit-learn}. GPR is a non-parametric, probabilistic approach that models the data with a mean function and a covariance function (kernel). Given the irregularity of our $O-C$ diagrams, for which we do not have any predefined function to model, we resorted to GPR to create a smooth profile for the wavelet analysis. For this, we construct a composite kernel consisting of three components. These are the Radial Basis Function (RBF) for capturing smooth variations, exponential sines squared for modelling periodic variations in the data, and finally a white noise component for handling noise in the data. These kernels in their mathematical form are described in Appendix~\ref{appendix: Gaussian Process Regression}.
\begin{figure}
\begin{center}
{\includegraphics[height=16cm,width=0.9\linewidth]{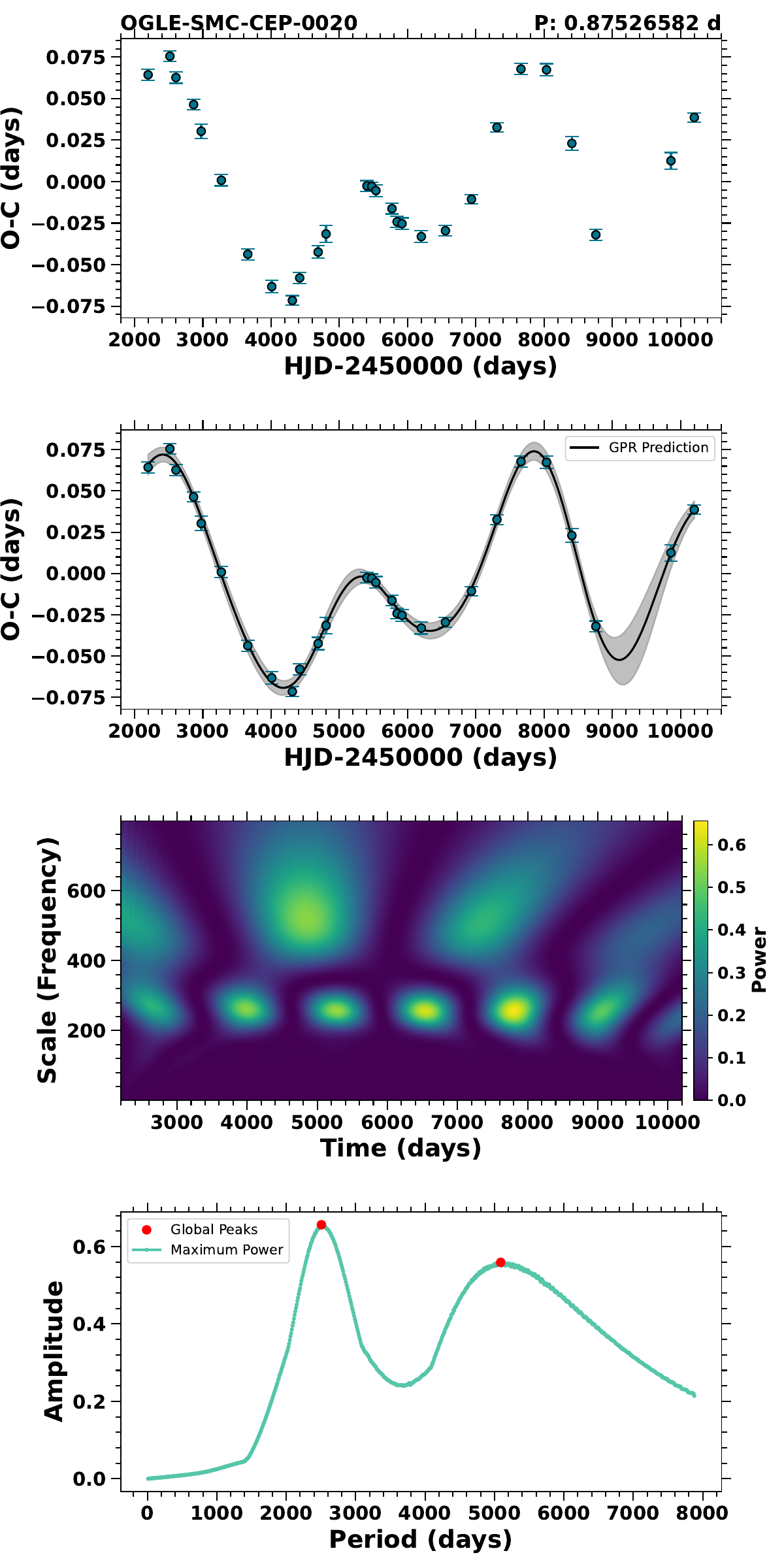}}
\caption{Wavelet analysis for OGLE-SMC-CEP-0020. The first panel shows the $O-C$ diagram, while the second panel presents the $O-C$ diagram with Gaussian Process Regression (GPR) applied (black), along with its uncertainty (gray). The third panel displays the wavelet transform of the smoothed GPR, with power represented by the colour bar. The fourth panel shows the maximum power (cyan) distribution, with global peaks highlighted in red. The highest peak determines the dominant variability timescale and amplitude.}
\label{fig:Time-Frequency_plot}
\end{center}
\end{figure}

Once the GPR is applied, the resulting smoothed $O-C$ fit becomes an input for the Continuous Wavelet Transform \citep[(CWT);][]{Farge1992AnRFM..24..395F,Torrence1998BAMS...79...61T} method, implemented using the \texttt{pywt} package \citep{Lee2019}, to analyze the time-frequency characteristics. The CWT is a powerful technique for analyzing signals that vary in both time and frequency, providing a detailed view of how different frequency components evolve over time. The CWT of a time series $x(t)$ is defined as:
\begin{eqnarray}
W(a, b) = \int_{-\infty}^{\infty} x(t) \psi^* \left( \frac{t - b}{a} \right) \, dt,
\end{eqnarray}
where  $\psi(t)$ represents the wavelet , $\psi^*$ is its complex conjugate, $a$ is the scale parameter that controls the frequency, and $b$ is the translation parameter that controls the time localization. We utilized the Morlet wavelet \citep{Morlet1982Geop...47..203M,Torrence1998BAMS...79...61T}, which is a complex sinusoid modulated by a Gaussian window:
\begin{eqnarray}
\psi(t) = \pi^{-1/4} \exp(i \omega_0 t)\exp(-t^2/2),
\end{eqnarray}

where $t$ is time and $\omega_0$ is the non-dimensional frequency. The Morlet wavelet offers a balance between time and frequency localization.

We compute the wavelet power spectrum, $|W(a, b)|^2$, and eventually identify dominant periods and their amplitudes. For a comprehensive view of how power is distributed across periods, we calculate the maximum power spectrum, which highlights the strongest signal component at each period and is thus more effective for identifying dominant periodicities.

The dominant periodicities and their amplitudes are located by peak-finding functions implemented after computing the cubic spline interpolated average and maximum power spectra profile. The above steps are represented by an example of OGLE-SMC-CEP-0020 in Fig.~\ref{fig:Time-Frequency_plot}.

\begin{figure}
\begin{center}
{\includegraphics[height=6cm,width=0.9\linewidth]{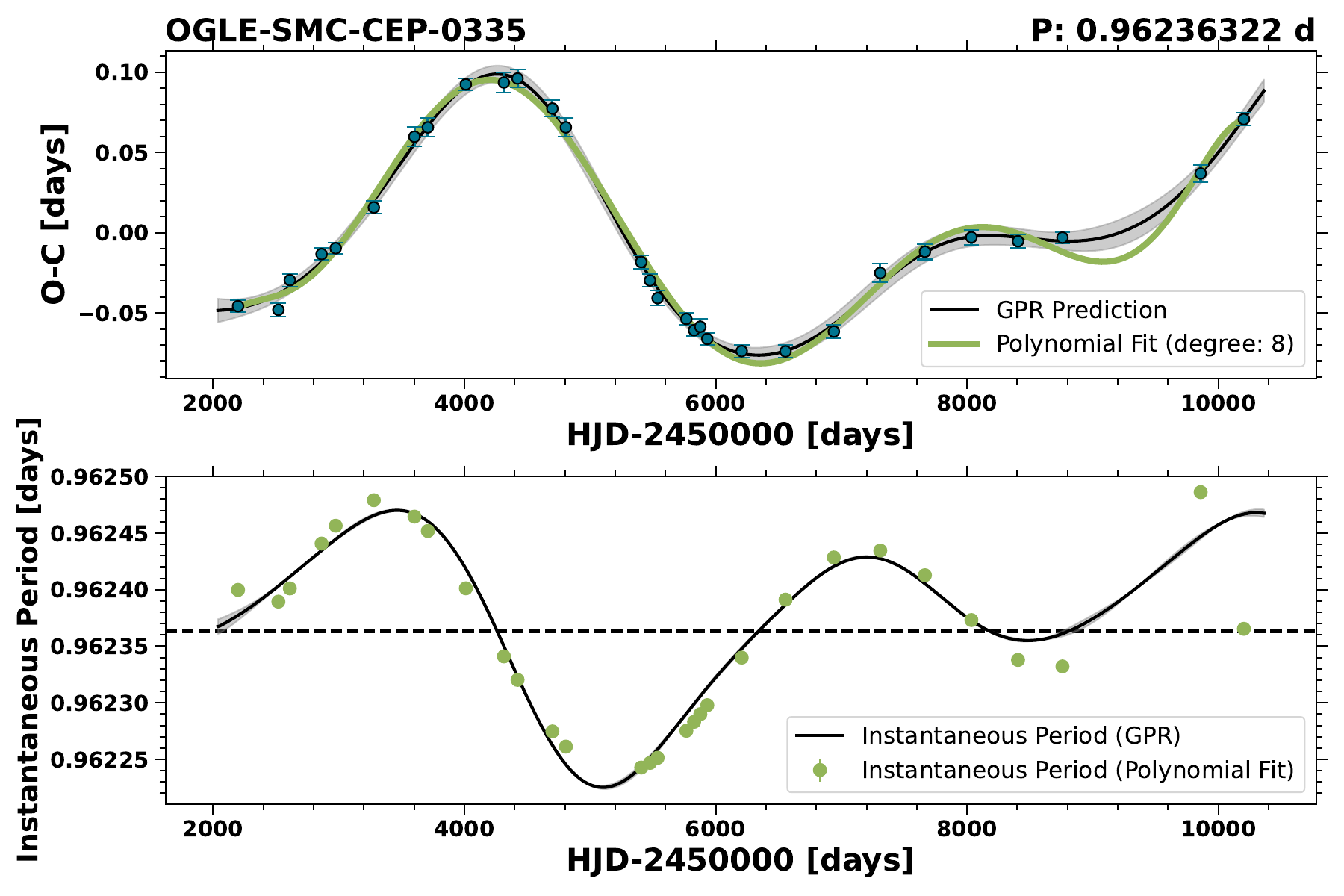}}
\caption{Calculation of the instantaneous period for OGLE-SMC-CEP-0335. The upper panel displays the $O-C$ diagram with the GPR prediction (black) and its uncertainty (gray), along with a polynomial fit (green). The lower panel presents the instantaneous periods derived from the fits in the upper panel. The horizontal dashed black line indicates the mean pulsation period of the Cepheid.}
\label{fig:Instantaneous_period_plot}
\end{center}
\end{figure}

\subsection{Instantaneous period method}
\label{subsec: Instantaneous period method}
In this method, we calculate instantaneous pulsation periods similarly to what was performed by \cite{Szeidl2011MNRAS.411.1744S} on Messier 5 RR Lyrae stars. We use their traditional method of obtaining temporal periods by fitting the $O-C$ with a polynomial equation:
\begin{eqnarray}
O - C = \sum_{i=1}^{k} c_i t^{i-1}\,.
\end{eqnarray}
The order of the polynomial is decided by iterative fitting of polynomials of orders 1--8 and calculating chi-square statistic, AIC and BIC. Each criterion yields a polynomial order and the smallest is adopted in the following. The instantaneous period, $P(t)$, is derived using the polynomial's derivative:
\begin{eqnarray}
P(t) = P_a \frac{d(O - C)}{dt} + P_a = P_a \sum_{i=2}^{k} (i - 1) c_i t^{i - 2} + P_a.
\end{eqnarray}
where $P_a$ is the period used to construct the $O-C$ diagram. In addition, we compute the instantaneous periods by calculating the local derivative of the smoothed GPR obtained in Sect.~\ref{subsec: Time-Frequency analysis}. This method is preferred due to its flexibility in handling the irregular nature of $O-C$ diagrams. 

The resulting instantaneous period statistics offer a detailed view of the $O-C$ variations. We record two quantities: $\Delta P/P$, where $\Delta P$ is the difference between the maximum and minimum instantaneous period and standard deviation for instantaneous period values calculated at $O-C$ points, $\sigma$. Fig.~\ref{fig:Instantaneous_period_plot} shows an example of how the method is applied to obtain the temporal variation of the pulsation period in OGLE-SMC-CEP-0335.

\subsection{Stetson variability index}
\label{subsec: Variability index test}
To characterise the variability in the $O-C$ curves, we also computed the Stetson $L$ index \citep{Stetson1996PASP..108..851S}. The variability index was originally developed to detect variable stars, such as Cepheids, by quantifying the coherence and amplitude of their variability in time-series data. The index evaluates correlations between deviations in consecutive observations, helping to identify patterns of consistent variability. Recently this variability index was used to identify Long Period Variables (LPVs) \citep[e.g.][]{Suresh2024PASP..136h4203S}. The Stetson $L$ index is defined as:
\begin{equation}
L = \frac{JK}{0.789}
\end{equation}
where
\begin{equation}
J = \sum_{n = 1}^{N-1}{\text{sign}(\delta_{n}\delta_{n+1})\sqrt{|\delta_{n}\delta_{n+1}|}}\,,
\end{equation} 
and
\begin{equation}
K = \frac{1/N \sum_{i = 1}^{N}{|\delta_i|}}{\sqrt{{1/N} \sum_{i = 1}^{N}{{\delta_i}^2}}}\,,
\end{equation} 
are the Stetson $J$ and $K$ indices \citep{Stetson1996PASP..108..851S}, with $\delta_i$ being the difference between the magnitude of the $i^{th}$ point and the mean magnitude of the lightcurve, scaled by the error of the $i^{th}$ point. The $J$ index captures the correlation between adjacent deviations, while the $K$ index captures the overall level of variability. Combining these into the $L$ index provides a comprehensive measure of both the strength of the variability and its coherence across adjacent data points. The amplitude of the $O-C$ is correlated with the $L$ index value and hence gives a measure of the $O-C$ variation for the variables with irregular period changes.

\section{Results}
\label{sec: Results}
In the following, we present our results on class 1 (Sect.~\ref{subsec: Candidates with negligible period change}), class 2 (Sect.~\ref{subsec: Candidates with secular PC-like feature}) and class 3 (Sect.~\ref{subsec: Irregular PC candidates}) candidates. Then we discuss in detail the irregular period change candidates, compiling the results from their individual characterisation.

\subsection{Class 1: Candidates with negligible period change}
\label{subsec: Candidates with negligible period change}

The first class of candidates were Cepheids with $O-C$ diagrams best described by a linear model (Sect.~\ref{sec: Classification Methodology}). These targets depicted a flat distribution of $O-C$ points across the time bins (both seasonal and finer resolution). This implies that the analysed time base is too short to reveal a secular period change and the pulsation period appears constant. Examples of $O-C$ diagrams for these candidates are shown in Fig.~\ref{fig:ocplot_linear_examples}. Altogether 2660 stars were classified as class 1 (56.3$\pm$0.7\% of the analysed sample). Considering all analysed F-mode stars in the LMC, 69.3$\pm$1.3\% were classified as class 1. The corresponding number for 1O stars is 19.8$\pm$1.6\%. In the SMC, the fractions of class 1 stars are 80.1$\pm$0.9\% for the F mode and 15.2$\pm$1.1\% for the 1O mode (Tab.~\ref{tab:data sample}). It clearly highlights the lower incidence rate of class 1 candidates among 1O-mode stars compared to F-mode stars in both the LMC and SMC. This indicates the difference in the detectability of period changes on the time scales covered by our data, with 1O-mode stars being more likely to exhibit detectable period variations. A list of class 1 candidates is compiled (see section~\ref{sec: Data availability}), the sample of which is shown in  Tab.~\ref{tab:Class_1_list_mainpaper}.

\subsection{Class 2: Candidates with secular PC-like feature}
\label{subsec: Candidates with secular PC-like feature}

In our analysis, we also classified 484 ($10.2\pm0.5$\%) candidates with parabolic $O-C$ shape (class 2). Examples of $O-C$ diagrams for these candidates are presented in Fig.~\ref{fig:ocplot_parabola_examples}. Of all analysed F-mode stars in the LMC 8.6$\pm$0.8\% were classified as class 2. For 1O stars, the incidence rate is 15.9$\pm$1.4\%. Corresponding numbers in the SMC are 7.5$\pm$0.6\% for F mode and 14.2$\pm$1.1\% for 1O Cepheids. From an evolutionary period change perspective, an upward parabolic $O-C$ indicates a period increase, when a Cepheid is evolving towards the red edge of the instability strip during either the first or third crossing. The observed period change rate helps distinguish between these two crossings. Conversely, a downward parabolic $O-C$ indicates a period decrease, showing that the Cepheids is on the second cross, as it evolves towards the blue edge of the instability strip. 

\begin{figure}
\begin{center}
{\includegraphics[height=6cm,width=0.9\linewidth]{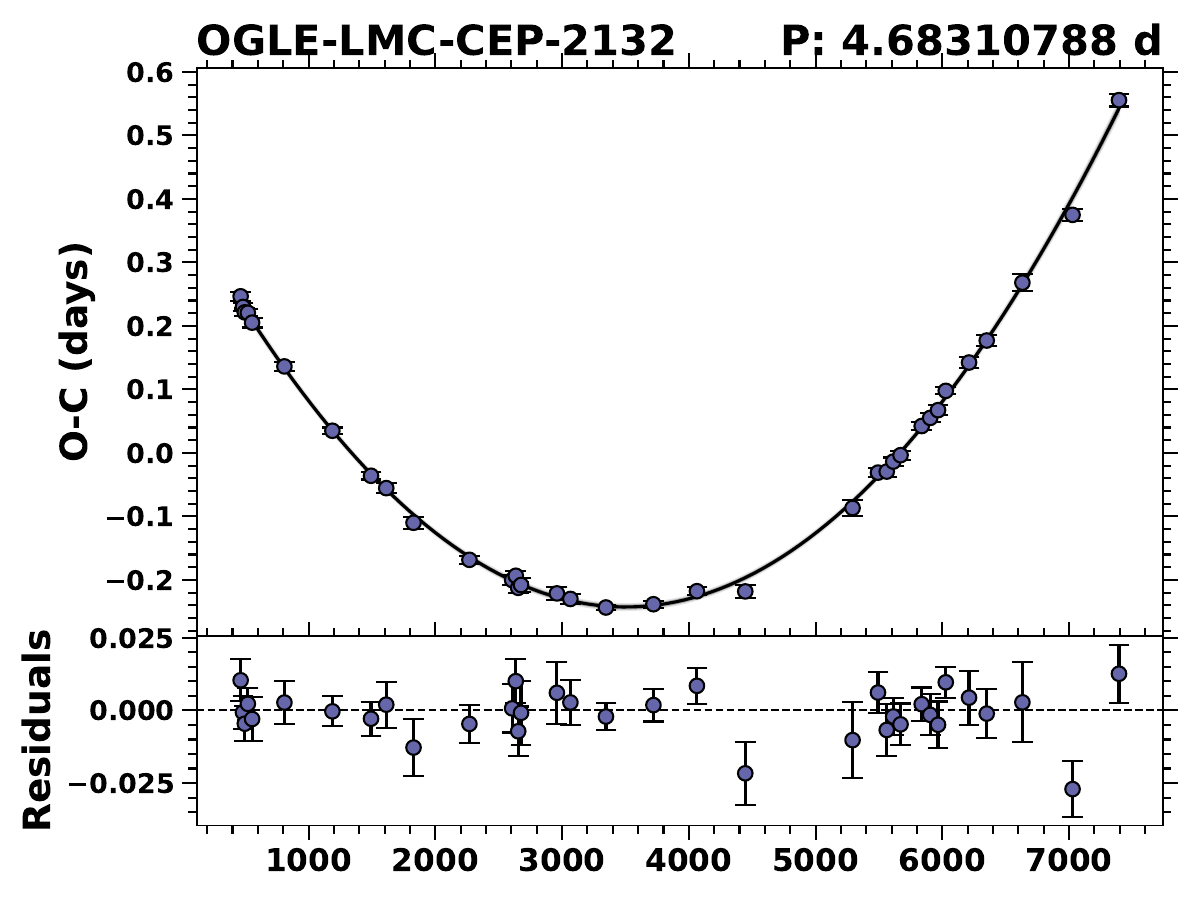}}
{\includegraphics[height=6cm,width=0.9\linewidth]{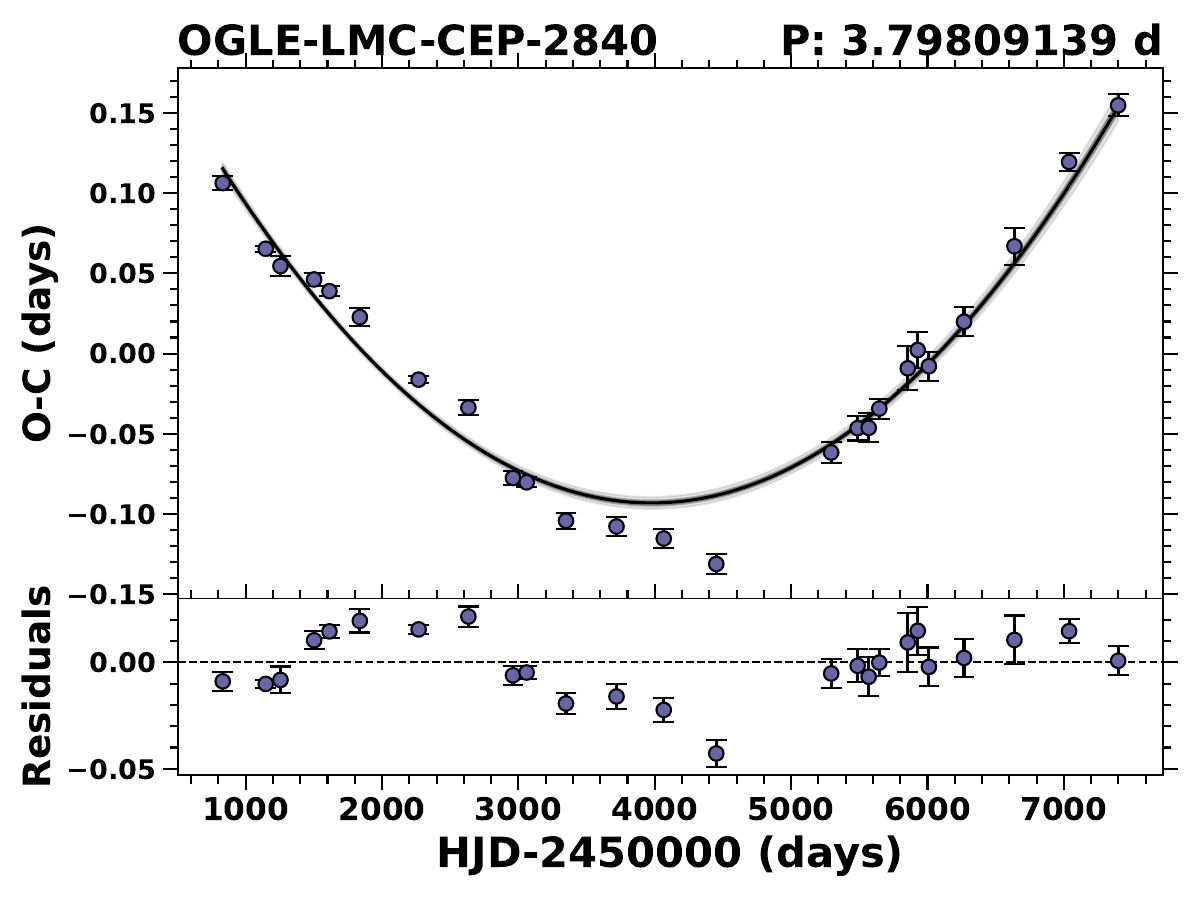}}
\caption{$O-C$ diagrams constructed for first crossing candidates reported by \cite{Rodriguez-Segovia2022MNRAS.509.2885R}.}
\label{fig:Segovia_examples}
\end{center}
\end{figure}

For decreasing period stars, we cannot conclude that the observed period change rate is due to evolution. The slower evolution on nuclear time scales during the second crossing means that approximately 20 years of photometry are insufficient to resolve such gradual changes well. In the non-secular regime, our class 2 parabolic shape candidates could instead reflect non-evolutionary period changes, such as those due to the LTTE in binary systems. These candidates were not covered in \citetalias{Rathour2024A&A...686A.268R} because, even if they are LTTE cases, the current data do not cover even one complete cycle to confirm an even longer periodicity. Alternatively, these can be non-evolutionary period changes, which is the primary focus of this work, but they might show a smooth parabolic $O-C$ shape because the fluctuations are over a much longer time scale. If that is the case, they should gradually become more irregular in upcoming observing seasons.
\begin{figure}
\begin{center}
{\includegraphics[height=6cm,width=0.95\linewidth]{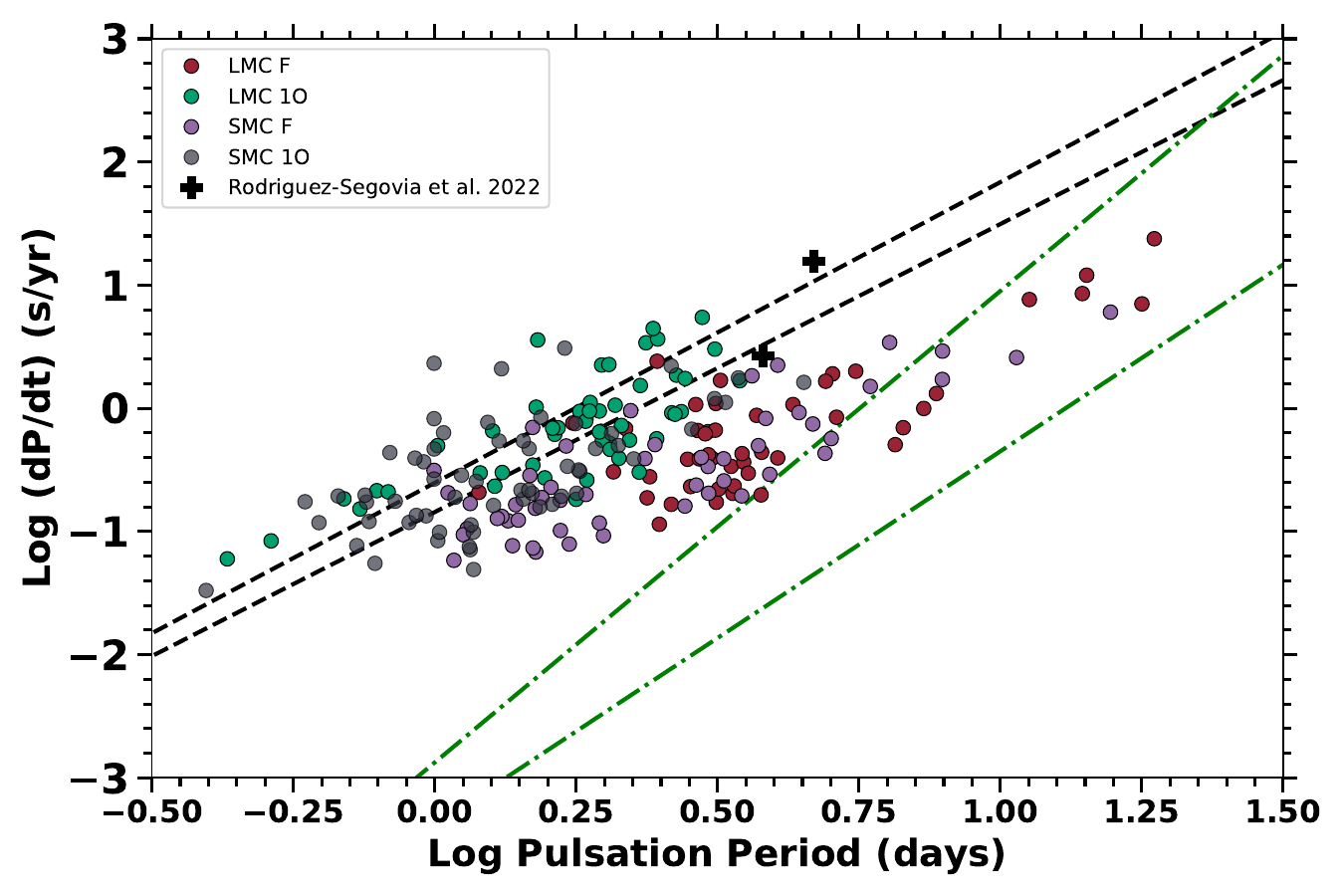}}
\caption{Distribution of log period change rate for first crossing candidates in class 2 sample as a function of log pulsation period. The scatter points represent different samples: LMC F (red), LMC 1O (green), SMC F (purple), and SMC 1O (gray). The theoretical regions for the first and third crossings, as described by \citet{Turner2006PASP..118..410T}, are indicated by black dashed and green dot-dashed lines, respectively. Two black plus symbols represent the first crossing candidates identified by \citet{Rodriguez-Segovia2022MNRAS.509.2885R}.}
\label{fig:PCR_class2}
\end{center}
\end{figure}

For the increasing period stars, while the interpretation of non-evolutionary (either LTTE or irregular) period change does hold, some of them can be plausible candidates for secular evolution during the first crossing. \cite{Rodriguez-Segovia2022MNRAS.509.2885R} reported two such fundamental mode candidates from the OGLE data study, OGLE-LMC-CEP-2840 and OGLE-LMC-CEP-2132 with period change rate (d$P$/d$t$) of 30.9 and 179.8\,d/Myr respectively. The $O-C$ diagram of these two Cepheids are shown in Fig. \ref{fig:Segovia_examples}. We note that the $O-C$ diagram of OGLE-LMC-CEP-2840 deviates from parabolic shape and the residuals show periodic nature. This target was subjected to LTTE analysis and results in an orbital period of $4758\pm249$ days (the details presented in appendix \ref{appendix: Binary Analysis of OGLE-LMC-CEP-2840}). After fitting a parabolic + LTTE model, we conclude that the $O-C$ variation is indicative of binarity signature in this first crossing candidate.

Finally, we corroborate our calculated period change rates for strictly parabolic $O-C$ Cepheids (see Fig.~\ref{fig:PCR_class2}) with the theoretically predicted rates for different evolutionary crossings \citep{Turner2006PASP..118..410T}.  While several tens of stars do overlap with the theoretical first crossing and appear as good candidates, it is not possible to unambiguously infer about the purely evolutionary nature of the observed period change and to assign a crossing number based solely on this comparison. A much longer time base is required for definitive confirmation, as non-evolutionary period changes may still be impersonating in these candidates. Such first crossings are particularly important for spectroscopic studies to understand the chemical abundances before the first dredge-up phase. Detailed analyses show that first-crossing Cepheids frequently display lithium enhancement \citep{Luck2001A&A...373..589L,Kovtyukh2019MNRAS.488.3211K,Catanzaro2020A&A...639L...4C,Ripepi2021A&A...647A.111R}

We may speculate that indeed, majority of class 2 sample represent non-evolutionary changes. This is indicated by a systematically larger incidence rates for 1O stars, as compared to F-mode stars, in agreement with incidence rates for irregular period change stars -- class 3 -- discussed in the next section.

The full list of class 2 candidates along with their computed period change rates is compiled (see section~\ref{sec: Data availability}), the sample of which is presented in Tab.~\ref{tab:Class_2_list_mainpaper}.

\subsection{Class 3: Irregular PC candidates}
\label{subsec: Irregular PC candidates}

\begin{figure}
\begin{center}
{\includegraphics[height=16cm,width=0.9\linewidth]{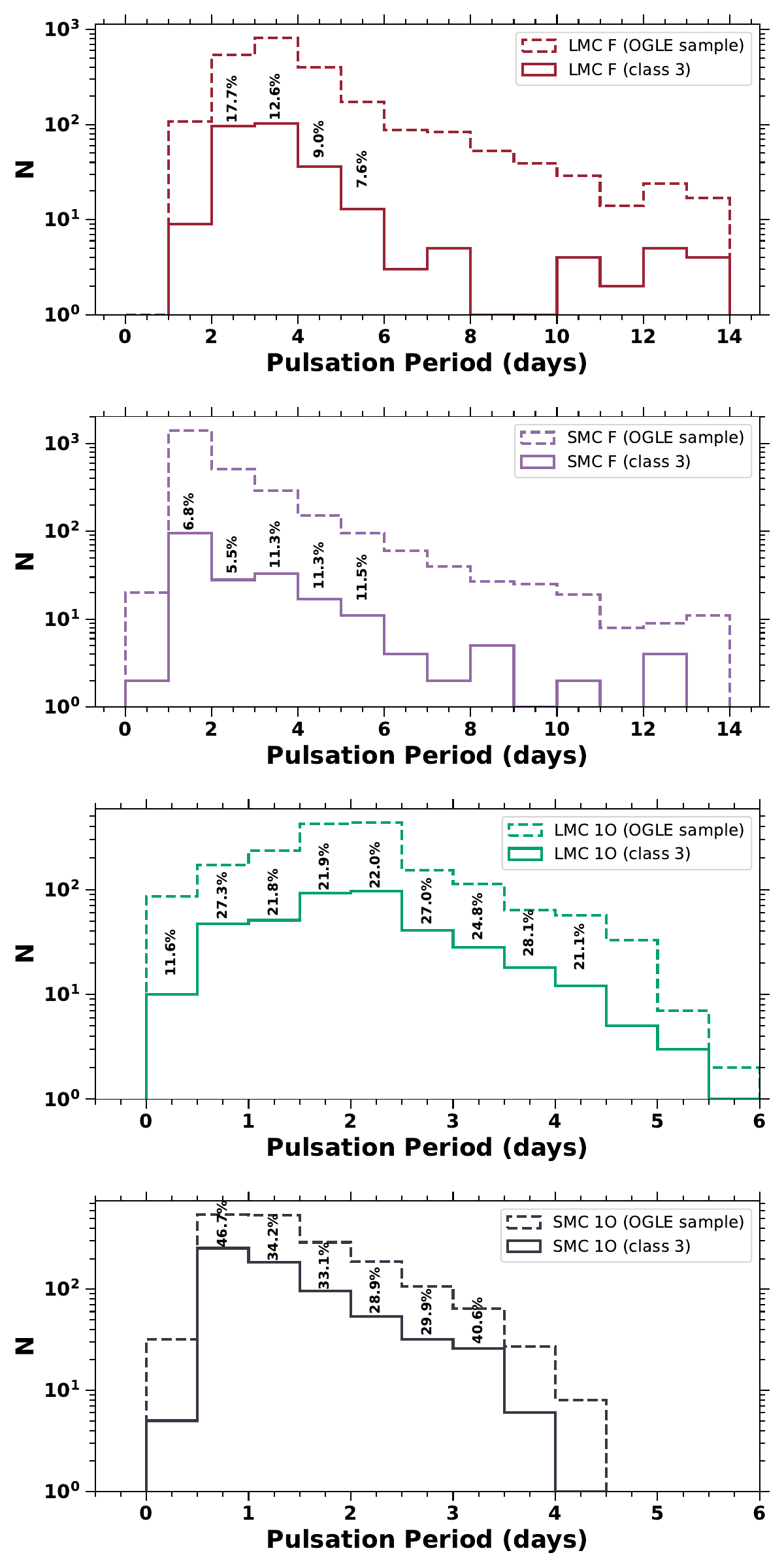}}
\caption{Pulsation period distribution for the parent OGLE sample (dashed lines) and irregular period change candidates (class~3, solid lines). From top to bottom, the panels show the distributions for LMC F, SMC F, LMC 1O, and SMC 1O mode Cepheids, respectively. The incidence rate for irregular period change candidates is displayed for bins containing at least 10 Cepheids and with an incidence rate exceeding 5\%.}
\label{fig:irregular_period_histogram}
\end{center}
\end{figure}

Our analysis resulted in 1585 ($33.5\pm0.7$\% of the analysed sample) irregular period change stars across both galaxies. The class 3 sample comprises 290 LMC F-mode (22.1$\pm$1.2\%), 405 LMC 1O-mode (64.3$\pm$1.9\%), 231 SMC F-mode (12.3$\pm$0.8\%), and 659 SMC 1O-mode (70.6$\pm$1.4\%) Cepheids. Examples of $O-C$ diagrams for these candidates are presented in Fig.~\ref{fig:ocplot_irregular_examples_LMCF},\ref{fig:ocplot_irregular_examples_LMC1O},\ref{fig:ocplot_irregular_examples_SMCF} and \ref{fig:ocplot_irregular_examples_SMC1O}.  Overall, the incidence rates are higher for the first overtone sample than for the fundamental mode sample in both galaxies, consistently with \cite{Poleski2008AcA....58..313P}, who reported that overtone Cepheids are more likely to undergo period changes. Additionally, the first overtone sample shows a higher incidence rate of irregular period changes in the SMC compared to the LMC, indicating that a lower metallicity environment might favor their occurrence in overtone Cepheids. However, the opposite trend is observed for the fundamental mode Cepheids, where their incidence rate is higher in the LMC.

We note however that the above incidence rates should not be considered representative for the whole population of Cepheids in the Magellanic Clouds. This is because we have applied several sample cuts that may be biased to a particular pulsation mode. For example, in steps 3--5 (see Tab.~\ref{tab:data sample}) we have rejected stars with additional low amplitude variability, which happen to be only first overtone stars, meanwhile modulated Cepheids are rejected from the fundamental mode sample. Investigation of period changes in the sample of rejected stars is beyond the scope of this paper (see also conclusions in Sect.~\ref{sec: Conclusions}). However, in a follow-up work we will investigate this sample of additional low amplitude variability and its connection with non-evolutionary period changes.

The distribution of the class 3 sample with pulsation period, with incidence rates reported within individual bins, is displayed in Fig.~\ref{fig:irregular_period_histogram}. For the LMC F-mode sample (top panel in Fig.~\ref{fig:irregular_period_histogram}) the distribution seems to follow the period distribution for all F-mode Cepheids. We observe a broad peak at periods of 2--4\,d and the highest incidence rate of $\sim18$\,\% in between 2 and 3\,d. There is a gap between 8--10\,d which is most likely due to sample cuts. We rejected Cepheids with periodic modulation of pulsation, which are numerous at around 10\,d. The same applies to the F-mode SMC sample. For the SMC F-mode sample (second panel in Fig.~\ref{fig:irregular_period_histogram}), the distribution peaks at pulsation periods 1--2\,d similar to the full OGLE sample, however, the incidence rate is the highest, $\sim11$\,\%, at higher pulsation periods of 3--6\,d. For the first overtone samples in both LMC and SMC (third and fourth panels in Fig.~\ref{fig:irregular_period_histogram}) we again observe that the distributions follow the parent distribution of the OGLE catalog. For the LMC the incidence rate is high $~\sim 20-30$\,\% in between 0.5 and 4.5\,d, with slightly higher incidence rates for longer (2.5--4\,d) periods. For the SMC, the incidence rates are highest (close to, or above 30\,\%) in between 0.5 and 3.5\,d, with the highest incidence rates in between 0.5 and 1\,d and then in between 3 and 3.5\,d.

\subsection{Characterisation: Eddington–Plakidis}
\label{subsec: Characterisation: Eddington–Plakidis (E–P)}

\begin{figure}
\begin{center}
{\includegraphics[height=9cm,width=1.\linewidth]{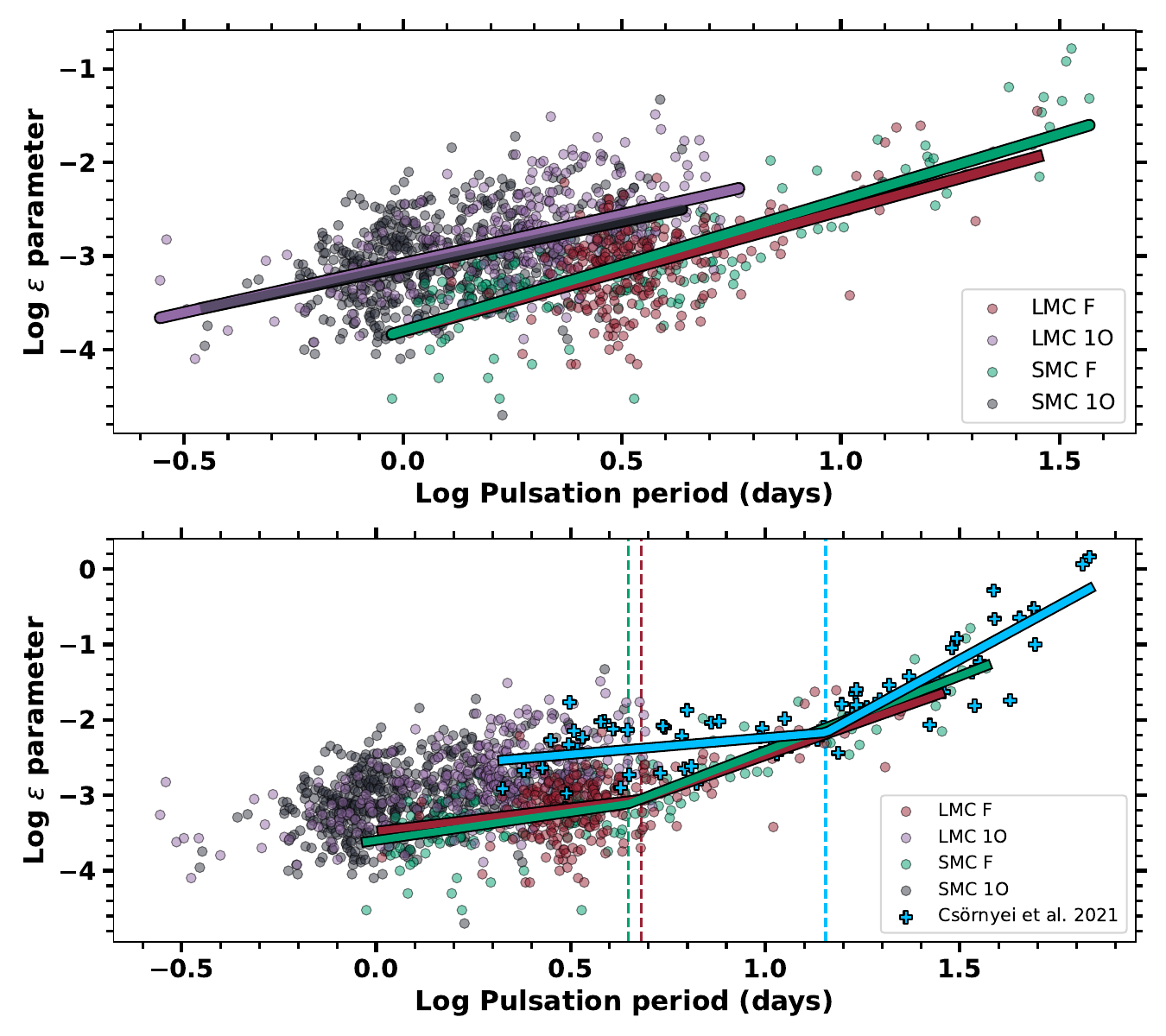}}
\caption{Distribution of logarithm of fluctuation parameter ($\epsilon$) as a function of $\log$ pulsation period shown in the upper panel. Scatter points represent LMC F (red), LMC 1O (green), SMC F (purple), and SMC 1O (gray) samples, with linear fits to each sample displayed in the corresponding colour. The lower panel includes the same sample, along with Galactic F-mode Cepheids from \citet{Csornyei2022MNRAS.511.2125C} (blue plus symbols). Broken linear regressions are shown for the Galactic, LMC, and SMC F-mode samples, respectively.}
\label{fig:EP_results}
\end{center}
\end{figure}

While investigating period changes of F-mode Galactic Cepheid, \cite{Csornyei2022MNRAS.511.2125C} hinted at a clear relation between the pulsation period of the Cepheids and the period fluctuation parameter, $\epsilon$. To investigate this, we applied the E--P test on all class 3 candidates, as described in Sect.~\ref{subsec: Eddington–Plakidis (E–P) test}. In Fig.~\ref{fig:EP_results} (upper panel) we plot period fluctuation parameter ($\epsilon$) as a function of the pulsation period. We observe that
$\epsilon$ increases as a function of the pulsation period, which we quantified with linear regressions displayed with solid lines. We note that on average, at a given pulsation period, $\epsilon$ is larger for the first overtone sample. This indicates that period fluctuations are stronger for the first overtone mode than for the fundamental mode. We note that this observation holds, even if we `fundamentalize' \citep[see e.g.][]{Moskalik2005AcA....55..247M, Pilecki2021ApJ...910..118P,Pilecki2024ApJ...970L..14P} the first overtone period, which corresponds to shifting the 1O relations by about 0.13 in $\log P$. For a given mode, fits are qualitatively similar for the LMC and the SMC.

In the bottom panel of Fig.~\ref{fig:EP_results}, for comparison, we also present Galactic classical
Cepheids sample from \cite{Csornyei2022MNRAS.511.2125C} which proves to be consistent with our results. The trends indicate that the random period fluctuation parameter positively correlates with the pulsation period for all Cepheid sub-samples across the metallicity environments and pulsation modes. This suggests that longer-period Cepheids exhibit more significant random cycle-to-cycle fluctuations in their periods.

In their Galactic Cepheid sample, \cite{Csornyei2022MNRAS.511.2125C} represented the relation between the period fluctuation parameter ($\epsilon$) and the pulsation period with a broken linear fit (with break at a pulsation period of $\sim$14\,d). To investigate this for our sample we use the \texttt{piecewise regression} tool \citep{Pilgrim2021}.  
As a test, we first apply the method to their Galactic data and indeed we find a break at  $\sim$14.4\,d with a $p$-value implying high statistical significance. On applying the methodology on our F-mode sample from both LMC and SMC, we found breaks at 4.8\,d (LMC) and 4.5\,d (SMC), both statistically significant, as supported by the corresponding $p$-values. In Fig.~\ref{fig:EP_results} (lower panel), we show these piecewise regressions to the discussed samples. Interestingly, we did not find a statistically significant piece-wise linear relation for our 1O-mode samples; the targets are distributed in a narrow period range with comparatively high scatter. 

\cite{Csornyei2022MNRAS.511.2125C} also considered other functional forms to represent the dependence of the period fluctuation parameter ($\epsilon$) on pulsation period. For the F-mode sample we may conclude that the relation is flatter for short period Cepheids and becomes steeper for long period Cepheids; the change appears at shorter pulsation periods in the Magellanic Clouds as compared to Milky Way.

\cite{Csornyei2022MNRAS.511.2125C} discussed that the amplitude of the fluctuation may have a minimum in the period range overlapping with the bump Cepheid regime, indicative of suppression of the mechanism causing the non-linear period change. We cannot investigate this with the MC sample, as in general our pulsation periods are shorter, and our sample is biased in this particular period range. When doing sample cuts we rejected Cepheids which show periodic modulation of pulsation. The modulation in F-mode Cepheids shows the highest incidence rate, that is $\sim$40\% in the pulsation period range of 12--16\,d for the SMC and $\sim$5\% in the range of 8--14\,d for the LMC \citep{Smolec2017MNRAS.468.4299S}. Consequently, we lack Cepheids in the period range of interest.

\subsection{Characterisation: Wavelet analysis}
\label{subsec: Characterisation: Wavelet analysis}
\begin{figure}
\begin{center}
{\includegraphics[height=10cm,width=.9\linewidth]{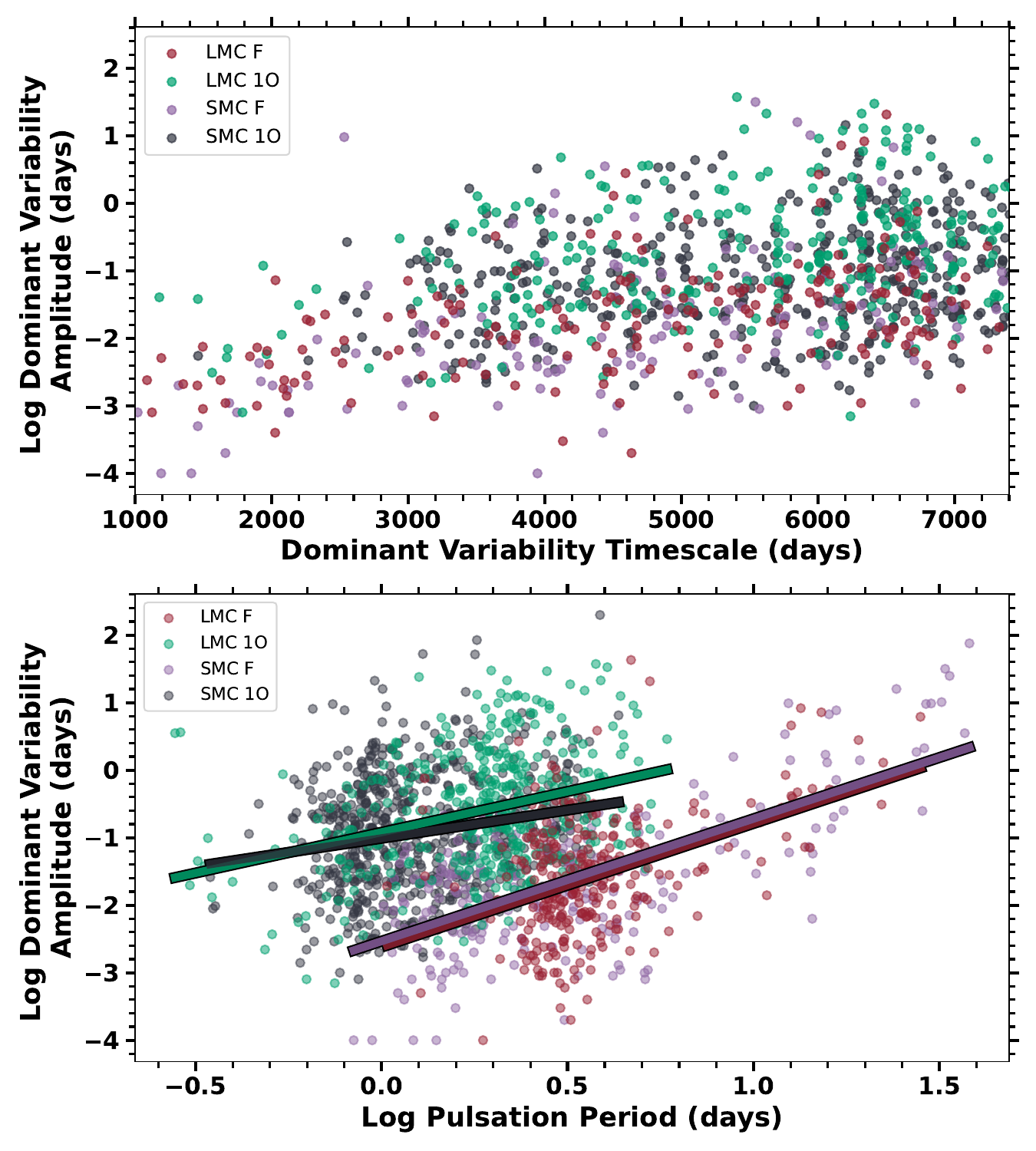}}
\caption{Distribution of the $\log$ dominant variability amplitude plotted against the dominant variability period of the $O-C$ variation, derived from wavelet analysis shown in the upper panel. The colour scheme is as follows: LMC F (red), LMC 1O (green), SMC F (purple), and SMC 1O (gray). The lower panel presents the $\log$ dominant variability amplitude as a function of the $\log$ pulsation period, with linear fits. The scatter points and fits follow the same colour scheme as in the upper panel.}
\label{fig:param_TF}
\end{center}
\end{figure}

We conducted wavelet analysis on class 3 candidates (as described in Sect.~\ref{subsec: Time-Frequency analysis}, with an example shown in Fig.~\ref{fig:Time-Frequency_plot}) to quantify the time scales and amplitude of the associated variations in the $O-C$ diagrams. In Fig.~\ref{fig:param_TF} (upper panel) we plot the variability amplitude as a function of its period. The periods range from 1000 to 7500\,d. 1O Cepheids show a larger scatter in the derived amplitudes as compared to the fundamental mode stars. The overall progression shows a marginal positive correlation between the variability amplitude and its period, ie. amplitude increases with period. In the lower panel of Fig.~\ref{fig:param_TF} we compare the variability amplitude with the pulsation period and note a clear positive correlation quantified with the linear fits. The correlation coefficients are 0.46 and 0.67 for the LMC and the SMC F-mode sample, respectively, reflecting moderate positive correlations. For the 1O sample, the correlation is weaker, as evidenced by the smaller correlation coefficients, 0.33 and 0.20 for the LMC and SMC, respectively. All correlations are significant, as evidenced by their $p$-values ($\ll$ 0.001). We observe lower variability amplitude of the F-mode candidates at a given period than of the 1O candidates. This behavior is similar to what we see with the period fluctuation parameter that is $\epsilon$ (see upper panel of Fig.~\ref{fig:EP_results})

\begin{figure}
\begin{center}
{\includegraphics[height=10cm,width=.9\linewidth]{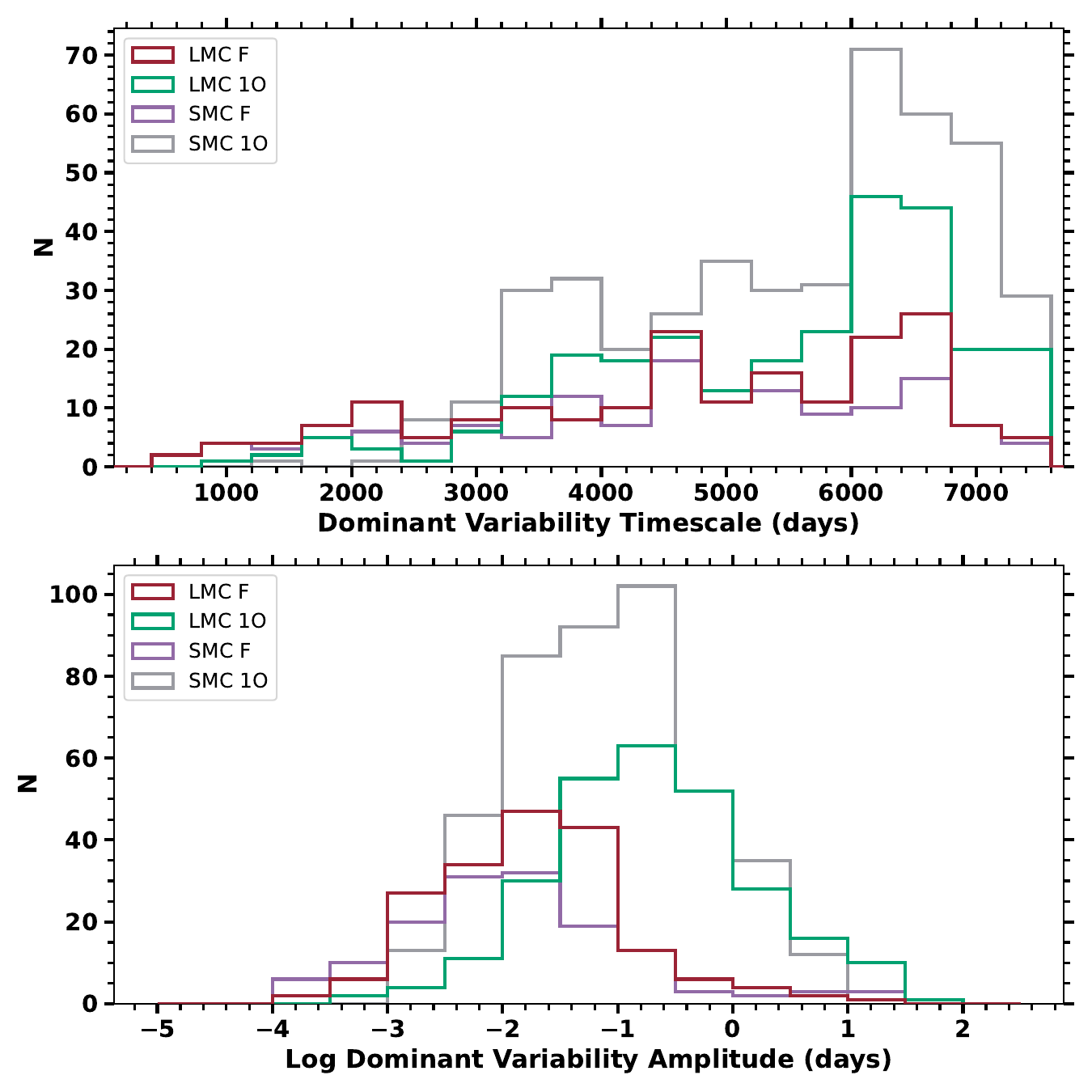}}
\caption{Distribution of the dominant variability period (upper panel) and $\log$ amplitude (lower panel) of the $O-C$ variation derived from wavelet analysis. The colour scheme is as follows: LMC F (red), LMC 1O (green), SMC F (purple), and SMC 1O (gray).}
\label{fig:TF_histogram}
\end{center}
\end{figure}

In Fig.~\ref{fig:TF_histogram} we show the distribution of the dominant variability period (upper panel) and amplitude (lower panel). We stress that in no case analysed here we observe a strict periodicity. The variability periods given below should be considered time scales for irregular variability detected in class 3 candidates.

The upper panel of Fig.~\ref{fig:TF_histogram} highlights that the overtone mode Cepheids from both the LMC and SMC exhibit a broad range of variability periods, with some peaks, that are not distinct however. For both the LMC and SMC 1O samples, dominant variability is most often recorded in the range of $\sim$6000-7000\,d. In the SMC, the secondary peaks are located at around 3500 and 5000\,d, but these peaks, in particular the latter, are not distinct. In contrast, fundamental mode Cepheids exhibit a more uniform distribution of variability periods in both galaxies, without prominent peaks. Typical variability periods are in between 4000 and 7000\,d.

The amplitude distribution (bottom panel of Fig.~\ref{fig:TF_histogram}) shows clear differences between the pulsation modes. The 1O stars exhibit a broader distribution of amplitudes, suggesting a wider range of variability in their $O-C$ curves. Distributions for both LMC and SMC 1O stars are qualitatively similar. In contrast, the fundamental mode Cepheids exhibit a much narrower amplitude distribution, centred at systematically lower amplitudes as compared to 1O stars. This again indicates that F-mode stars experience fewer and less significant $O-C$ irregularities compared to their overtone counterparts.

\subsection{Characterisation: Instantaneous period analysis}
\label{subsec: Results from instantaneous period analysis}

\begin{figure}
\begin{center}
{\includegraphics[height=10cm,width=.9\linewidth]{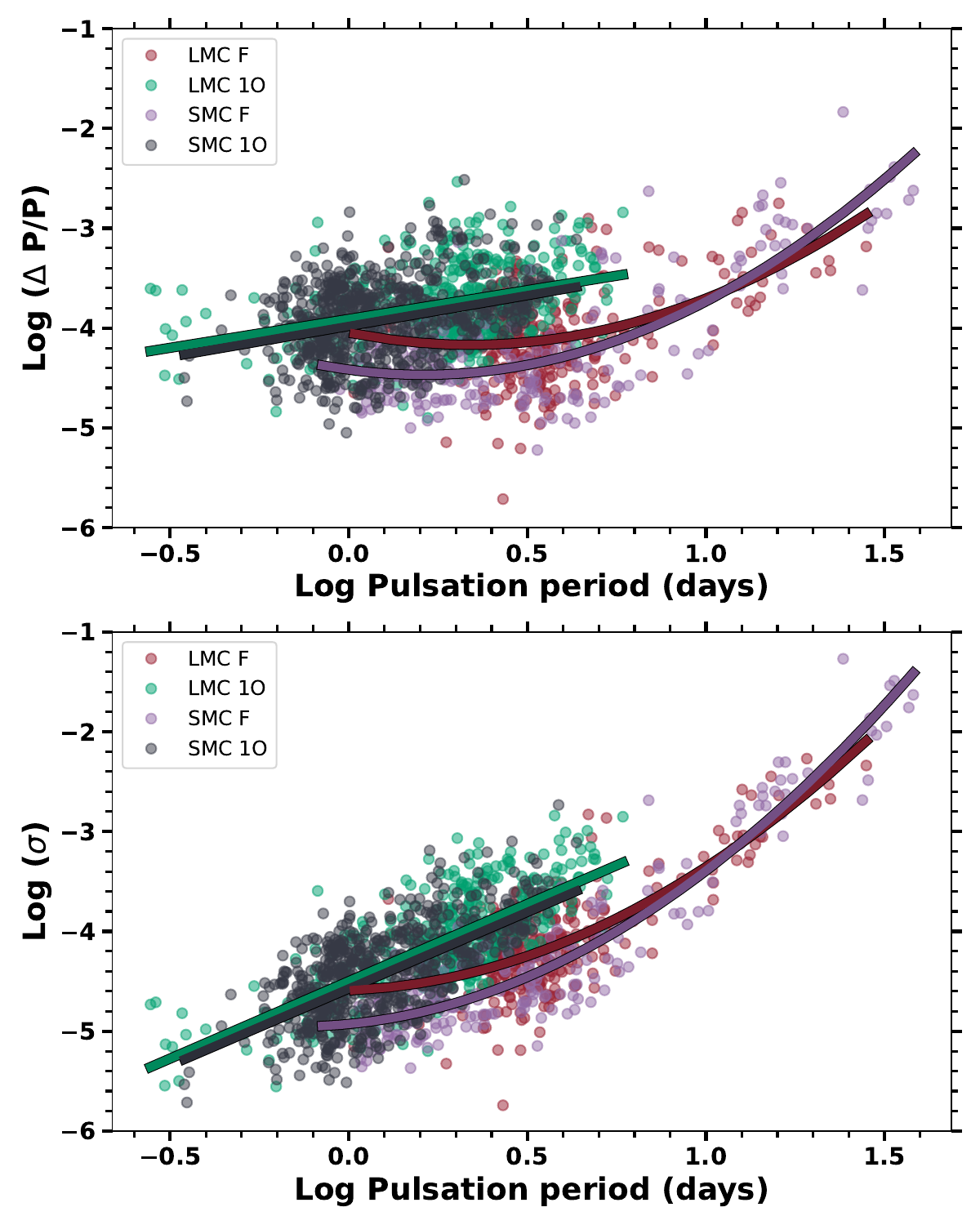}}
\caption{Logarithmic of $\Delta P/P$ (upper panel) and the standard deviation, $\log\sigma$ (lower panel), as functions of the log pulsation period. The scatter points represent LMC F (red), LMC 1O (green), SMC F (purple), and SMC 1O (gray) samples.}
\label{fig:instper_results}
\end{center}
\end{figure}

We applied the instantaneous period method (see Sect.~\ref{subsec: Instantaneous period method}) to all class 3 candidates. In Fig.~\ref{fig:instper_results} (top panel) we show the $\log \Delta P/P$ parameter plotted against the $\log $ pulsation period. In the bottom panel we also plot the standard deviation ($\log \sigma$) of the measured instantaneous periods as a function of the $\log$ pulsation period. We observe that both quantities increase with increasing pulsation period.
To represent these relations, we tested three models for each sub-population of Cepheids: linear, piecewise linear and quadratic. Based on AIC/BIC, we conclude that the linear model best describes the first overtone samples (in both SMC and LMC). However, for the fundamental mode Cepheids from both MC we arrived at the conclusion that a quadratic model best describes the data (see Fig.~\ref{fig:instper_results}). This applies to both $\Delta P/P$ and $\sigma$. The piecewise linear model is also  statistically significant. For this model, breaks are placed at 3.8 and 3.6\,d for the LMC and SMC and the $\Delta P/P$, and at 4.9 and 4.2\,d for the LMC and the SMC and $\sigma$, respectively. We note that these values, in particular for $\sigma$, are similar to the ones we recorded for $\epsilon$ in Sect~\ref{subsec: Characterisation: Eddington–Plakidis (E–P)}.

\subsection{Characterisation: Stetson $L$ index analysis}
\label{subsec: Results from Stetson L index analysis}
\begin{figure}
\begin{center}
{\includegraphics[height=10cm,width=.9\linewidth]{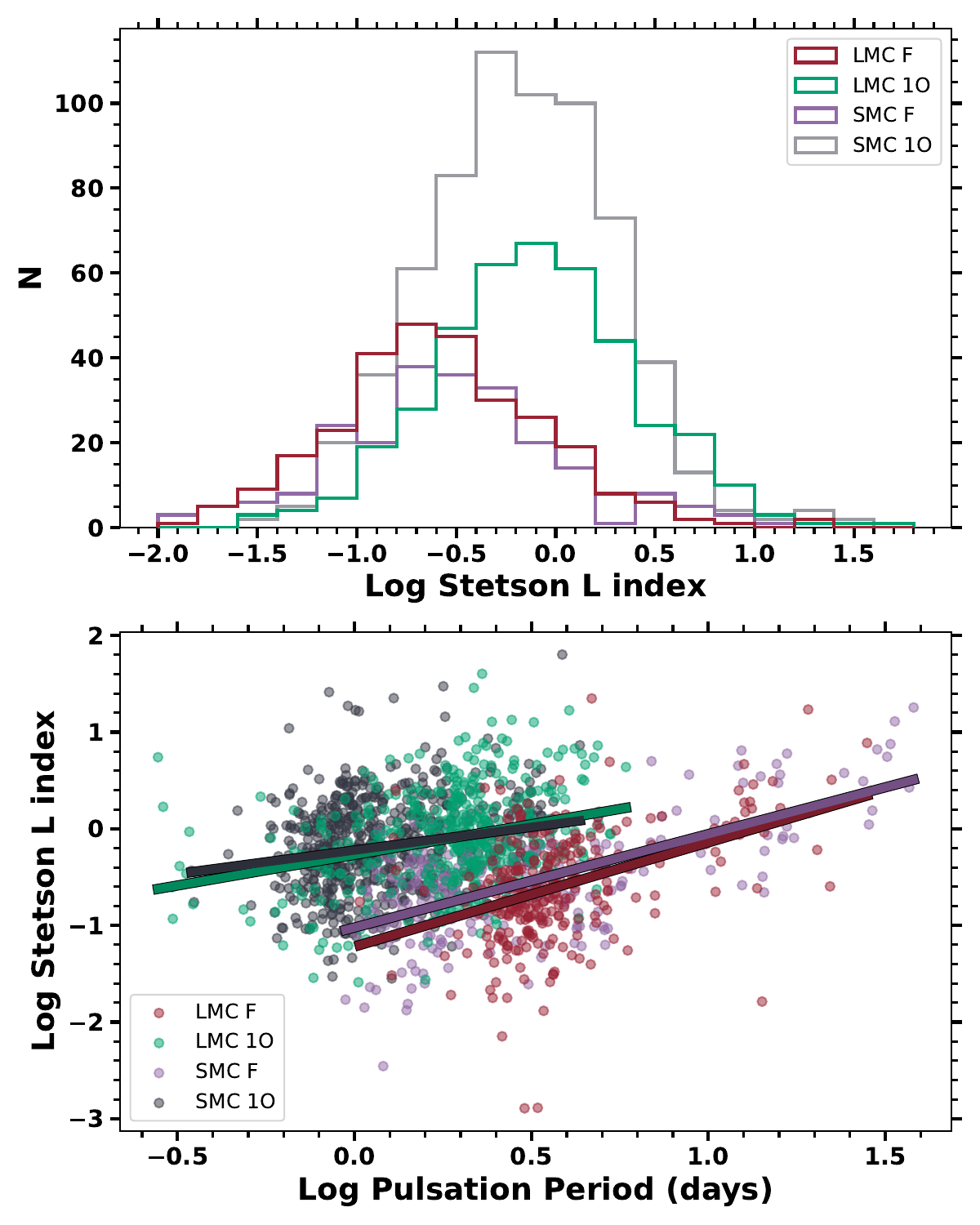}}

\caption{Distribution of the $\log$ Stetson $L$ index is shown in upper panel. The colour schemes are LMC F (red), LMC 1O (green), SMC F (purple) and SMC 1O (gray). The lower panel shows the $\log$ Stetson $L$ index as a function of the $\log$ pulsation period. The scatter points have the same colour scheme as in the top panel.}
\label{fig:histogram_stetson}
\end{center}
\end{figure}

Another measure of degree of variability in the $O-C$ diagrams is the Stetson $L$ index (see Sect.~\ref{subsec: Variability index test}). In Fig.~\ref{fig:histogram_stetson} (upper panel), we show the distribution of the Stetson $L$ index values. We see that distributions of first ovetone samples are broader and shifted towards higher values. The mean values of the $\log L$ index for the F mode samples are $-0.60$ and $-0.54$ for the LMC and the SMC, respectively, and are lower than the mean values for the 1O samples, $-0.11$ and $-0.18$, for the LMC and the SMC, respectively. This indicates a relatively stronger variability among first overtone mode samples. For reference, the mean value of $\log L$ for all class 1 candidates that do not show variability in their $O-C$ diagrams, is $-1.92$.

We also plot the Stetson $L$ index against the pulsation period in the lower panel of Fig.~\ref{fig:histogram_stetson}. There is a slight increase in the Stetson $L$ index value with increasing pulsation period across all sub-populations. The first overtone Cepheids show low correlation coefficient in the LMC (0.29) and the SMC (0.20). On the other hand, the fundamental mode candidates show moderate correlation coefficients, both in the LMC (0.42) and the SMC (0.65). We note that all correlations are statistically significant with $p$-value $\ll$0.001.

\section{Discussion}
\label{sec: Discussion}

\subsection{characterisation cross-validation}
\label{subsec: characterisation technique cross-validation}

We have used multiple methods to characterise the $O-C$ diagrams. Even though in their own different methodology they represent different quantities, at the core the idea is always to quantify either the amplitude and/or the time scales of the irregular $O-C$ shapes. 

\begin{figure}
\begin{center}
{\includegraphics[height=16cm,width=1.\linewidth]{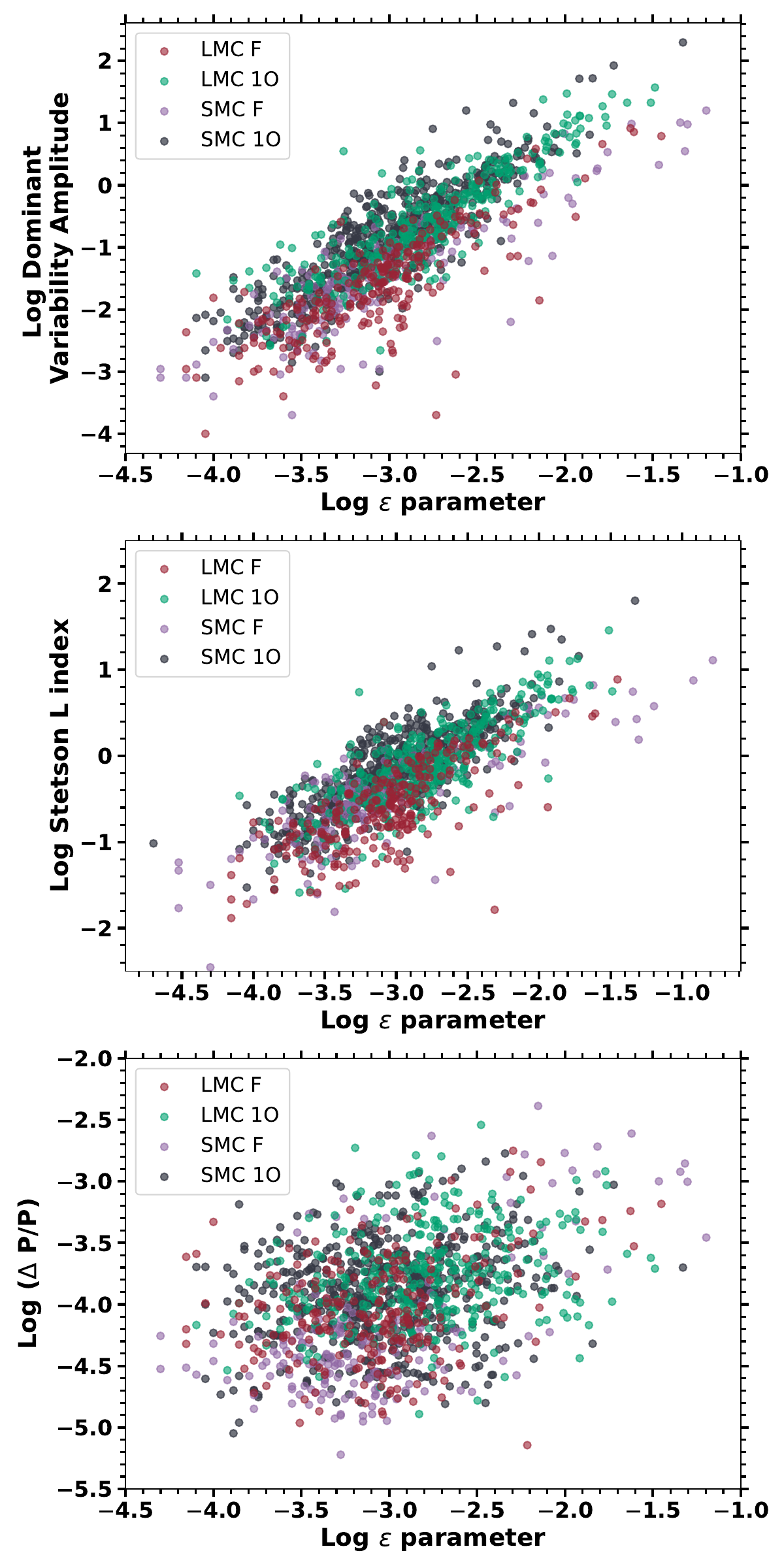}}

\caption{Comparison of $\log$ $\epsilon$ parameter with $\log$ dominant variability period (top), $\log$ Stetson L index (middle), and $\log \Delta P$ (bottom) for LMC F (red), LMC 1O (green), SMC F (purple), and SMC 1O (gray) samples.}
\label{fig:param_EP_TF_SL_IP}
\end{center}
\end{figure}

 To quantify the amplitude of irregularities we have computed the $\epsilon$ parameter (from the Eddington-Plakidis test) which quantifies the size of the random fluctuations in the pulsation period, dominant variability amplitude (from wavelet analysis) which measures the most significant fluctuation across all identified time scales, the Stetson $L$ index that estimate coherent variability trends, and the $\Delta P$ parameter which assesses the extent of pulsation period fluctuations over time (from the instantaneous period method). In Fig.~\ref{fig:param_EP_TF_SL_IP}, the latter three quantities are plotted against $\epsilon$. Overall, we observe a positive correlation between the parameters. For the three plotted relations, the correlation coefficients are $0.88$, $0.83$, and $0.24$, for dominant variability amplitude, Stetson $L$ and $\Delta P/P$, respectively, all correlations statistically significant ($p$-value $\ll$ 0.001). The techniques we have used may vary in sensitivity and may capture different aspects of variability, but overall provide a consistent picture of $O-C$ changes in class 3 candidates, as we discuss in more detail in the next subsection.

\subsection{Non-evolutionary PC across pulsation modes and environments}
\label{subsec: Nonevolutionary PC across pulsation modes and environments}
We compiled a sample of 1585 Cepheids showing non-evolutionary period changes in different metallicity environments (LMC and SMC) as well as pulsating in different modes (fundamental and first overtone). Overall, they constitutes 33.5\% of the analysed sample. Undoubtedly, irregular period changes are a common property of classical Cepheids.

Considering the environment, $35.8\pm1.1$\% of the analysed Cepheids in the LMC show irregular period changes. The incidence rate is only slightly lower in the SMC at $31.9\pm0.9$\%. Considering the pulsation mode, not differentiating between the environment, irregular period changes were detected in $16.5\pm0.7$\% of the analysed F-mode stars and in $68.1\pm1.2$\% 1O stars -- more than half of the analysed sample. Without any doubt, we can conclude that irregular period changes significantly more often affect Cepheids pulsating in the first overtone. The susceptibility of the mode to irregular period changes does depend on the environment. Within the LMC, $22.1\pm1.2$\% and $64.3\pm1.9$\% of the analysed F and 1O mode stars, respectively show irregular period changes. In the SMC, the corresponding numbers are $12.3\pm0.8$\% and $70.6\pm1.4$\%. Therefore, F-mode Cepheids are more susceptible to irregular period changes in the more metal rich environment (LMC) as compared to the lower metallicity environment. The opposite is true for the 1O mode, although the difference is not as pronounced as for the fundamental mode.

We note that these incidence rates cannot be considered representative for the whole population of classical Cepheids in the Magellanic Cloud, as several sample cuts were applied during the analysis. Earlier work by \cite{Poleski2008AcA....58..313P} indicated that period changes in LMC Cepheids are more frequent in first overtone mode (41\%) as compared to fundamental mode (18\%) Cepheids. Together, this indeed points to 1O pulsators being less stable than fundamental mode Cepheids, and more likely to show period change behavior (including non-evolutionary period changes).

All used characterisation methods indicate that the degree of variability in the $O-C$ curves increases with the pulsation period (see Figs.~\ref{fig:EP_results}, \ref{fig:param_TF}, \ref{fig:instper_results} and \ref{fig:histogram_stetson}). In this regard, the differences between the two investigated environments (LMC and SMC), are subtle, if any. Considering the pulsation mode, the differences are significant, and cannot be explained by the difference in original and fundamentalised 1O periods. First, with all methods we observe that at a given pulsation period, the amplitude of variability, as quantified with various considered parameters, is larger for the 1O stars, than for the F mode stars. Second, with the increase of pulsation period, the increase in the amplitude of the $O-C$ variability is linear for the 1O stars (in a logarithmic parameter space). In contrast, for the F-mode stars, the relation is more complex, flatter at shorter pulsation periods and then more steep at longer pulsation periods. This is particularly evident for the parameters $\epsilon$ (Fig.~\ref{fig:EP_results}), $\Delta P/P$ and $\sigma$ (Fig.~\ref{fig:instper_results}), for which we quantified the relation with pulsation period with quadratic and piece-wise linear fits. Both representations are statistically significant. In case of piece-wise linear fits for $\epsilon$/$\sigma$, the break points are located at periods of 4.8/4.9\,d in the more metal rich LMC and at slightly shorter periods of 4.5/4.2\,d in the SMC. These values are significantly shorter than the 14\,d reported by \cite{Csornyei2022MNRAS.511.2125C} for Galactic F mode Cepheids, which on average, are the most metal rich environment. Interestingly (see Fig.~\ref{fig:EP_results}, bottom panel), while for longer period F-mode Cepheids $\epsilon$ values are comparable within all three environments, for periods shorter than 14\,d the fluctuation parameter is clearly larger in the Galactic Cepheids. 

While the functional form may be debated, a steeper increase of $O-C$ variability amplitude at longer pulsation periods for F-mode stars is unquestionable.

\subsection{Implications for underlying mechanism and models}
\label{subsec: Disscussion for modelling??}

It is now well established that pulsations in classical Cepheids are not clockwork. Irregularities are apparent at various time scales. At the shortest time scale, of order of single pulsation period, subtle cycle-to-cycle changes in the lightcurves and pulsation period were detected with space telescopes, see e.g. \cite{Derekas2012MNRAS.425.1312D,Derekas2017MNRAS.464.1553D,Evans2015MNRAS.446.4008E}. In the present work our focus was on a much longer time scales of a few hundreds to thousands of days. Whether the same mechanism may be at action is unknown. Unfortunately, systematic analysis of the period variations across many time scales is not yet possible with the observations at hand.

As summarised in the introduction, a few mechanisms were proposed to explain irregularities in the $O-C$ diagrams on time scales investigated in this study. Unfortunately, these are qualitative rather than quantitative models. The lack of mathematical formalism does not allow making predictions on the expected amplitude of changes, distribution of variation time scales or on the dependence on pulsation mode or metallicity. We are not at a position to validate or invalidate any of the proposed ideas. 

Our study provides however crucial constraints that a successful model should meet, which is essential for future work in this field. The strongest constraints to be met are the prevalence of irregular period variations and the strong dependence on the pulsation mode: the phenomenon is significantly more frequent and stronger in 1O Cepheids. Another constraint is the dependence on pulsation period. The longer the pulsation period, the stronger the irregularities. The increase of the amplitude of irregularities is much stronger at longer pulsation periods in F-mode Cepheids. Our study provides some measures of the degree of the irregularity that can be confronted with models.

Probably the key to the puzzle is the dependence on pulsation mode, which hints to where the mechanism must be operational. Surely, it is the envelope; deep interior of the star is excluded, as mode amplitudes are negligible there. Then, the obvious difference between the two modes is the presence of a pulsation node for the 1O. At the node, the amplitude of pulsation is low and hence we may speculate that the mode is more prone to perturbations. The pulsation node is located below the envelope convection zone \citep{Smolec2008AcA....58..233S,Paxton2019ApJS..243...10P}, which does not implicate the velocity field is only due to pulsation there. Convective motions are still possible due to overshooting and the velocity field is expected to be turbulent. These background flows couple to and affect the pulsations. This indicates that 3D hydrodynamical modelling of the turbulent convective motions along with pulsations is necessary to address the problem of pulsation period changes on a more quantitative level. This is however, a challenging task, that not only needs immense computational resources, but also developments on the theory side. The models that are currently being developed \citep[e.g.][]{Geroux2015ApJ...800...35G, Mundprecht2015MNRAS.449.2539M, Muthsam2016CoKon.105..117M} are still rather exploratory and are not robust enough to solve this problem.

We also note that 1O Cepheids are more prone to excitation of additional low-amplitude variabilities, which are essentially not detected in F-mode Cepheids \citep[see e.g.][]{Smolec2023MNRAS.519.4010S,Suveges2018b,Suveges2018a}. This also points at the 1O mode being less stable against perturbations than the F mode.

\section{Summary and conclusions}
\label{sec: Conclusions}

We have analysed OGLE data for classical Cepheids in the Magellanic Clouds with the goal of investigating pulsation period changes. Analysed sample counts 4729 Cepheids, of which 1943 are in the LMC (1313 F-mode and 630 1O mode) and 2786 in the SMC (1852 F-mode, 934 1O mode). We summarise our findings below:

\begin{itemize}

\item In 2660 stars, which constitutes $56.3\pm0.7$\% of the sample, no period changes are detected. Their $O-C$ diagrams are {\it flat} (class 1 stars). Interestingly, most of these stars, $\sim$90\%, are fundamental mode Cepheids, which is a first indication that this pulsation mode is more stable against irregular period changes than the first overtone.

\item A total of 484 Cepheids ($10.2\pm0.5$\% of the analysed sample) with parabolic $O-C$ diagrams (class 2) were detected, with no evidence for additional variation, indicative of linear period change. Data we analyse are too short to unambiguously attribute these changes to secular evolution. These changes may still be of non-evolutionary origin.

\item A part of the above class 2 sample, with positive linear period change (or upward parabolic $O-C$ shape) may correspond to Cepheids during the first crossing of the instability strip, which is much faster than the crossings during the blue loop phase.

\item Our systematic search for irregular period change candidates resulted in a sample of 1585 Cepheids ($33.5\pm0.7$\% of the analysed sample). Irregular period changes are a common property of classical Cepheids. 

\item Considering the environment, the incidence rates for irregular period changes are similar: $35.8\pm1.1$\% in the LMC and slightly lower in the SMC, $31.9\pm0.9$\%.

\item Considering the pulsation mode, irregular period changes were detected in $16.5\pm0.7$\% of the analysed F-mode stars and in $68.1\pm1.2$\% of the 1O stars -- more than half of the analysed sample. Unambiguously, we can conclude that irregular period changes affect 1O Cepheids significantly more often. We note, however, that the susceptibility of the mode to irregular period changes does depend on environment. Within the LMC, $22.1\pm1.2$\% and $64.3\pm1.9$\% of the analysed F and 1O mode stars, respectively, show irregular period changes. In the SMC, the corresponding numbers are $12.3\pm0.8$\% and $70.6\pm1.4$\%. Therefore, F-mode Cepheids are more susceptible to irregular period changes in the more metal rich environment (LMC) as compared to low metal environment. The opposite is true for the 1O mode, although the difference is not as pronounced as for the fundamental mode.

\item For Cepheids showing irregular period changes, the Eddington-Plakidis test was used to quantify the amplitude of random period fluctuations with fluctuation parameter, $\epsilon$. In a logarithmic parameter space, $\epsilon$ increases with the pulsation period. The increase is linear for the 1O stars, while for the F-mode stars a more complex relation, represented with either quadratic, or piece-wise linear function, is observed. For F-mode stars, on a qualitative level, the same picture emerges form analysis of Galactic Cepheids \citep{Csornyei2022MNRAS.511.2125C}; however, the increase of fluctuation parameter becomes more steep at longer pulsation periods. For shorter pulsation periods the amplitude of fluctuations is larger in Galactic Cepheids.

\item At a given pulsation period,  the amplitude of fluctuations, as quantified with the fluctuation parameter, is significantly larger for 1O stars. 

\item The other measures of the amplitude of variability in the $O-C$ diagrams corroborate these results. In particular,  time-frequency characterisation of $O-C$ diagrams using wavelet analysis yields amplitude that is increasing with pulsation period and, at a given pulsation period,  is systematically higher in first overtone mode stars. The time scales of variability range from hundreds to a few thousands of days.

\end{itemize}

Through a systematic investigation of irregular period change in classical Cepheids, we provide empirical constraints on the underlying mechanisms. The key constraints that models should meet are the common occurrence of irregular period changes in classical Cepheids and strong dependence on the pulsation mode: the phenomenon is significantly more frequent in first overtone stars. The models should also quantitatively predict the amplitudes and time scales of the variation to allow comparison with observational constraints.

A difficult problem, that still needs to be addressed is the decoupling of the secular and irregular variations. A good understanding of how these two may interact is necessary to correctly compare observed period change rates with predictions of stellar evolution modelling. We plan to address this problem in a subsequent study. Another research avenue that we will follow is to investigate irregular period changes in Cepheids with additional low-amplitude variability, such as non-radial modes and periodic modulations of the pulsation, which are common, in particular among the first overtone mode stars.

\section{Data availability}
\label{sec: Data availability} 
Full data to Tables \ref{tab:Class_1_list_mainpaper}, \ref{tab:Class_2_list_mainpaper} and \ref{tab:Class_3_list_mainpaper} in electronic form are made available at Zenodo link: \href{url}{https://doi.org/10.5281/zenodo.14637987}.

\begin{acknowledgements}
We thank all the OGLE observers for their contribution to the collection of the photometric data over the decades. RSR thank Géza Csörnyei for discussion on Eddington-Plakidis test. RSR, RS and OZ are supported by the National Science Center, Poland, Sonata BIS project 2018/30/E/ST9/00598. GH, VH and PK acknowledge grant support from the European Research Council (ERC) under the European Union's Horizon 2020 research and innovation program (grant agreement No. 695099). IS acknowledge support from the National Science Centre, Poland, grant no. 2022/45/B/ST9/00243. For the purpose of Open Access, the author has applied a CC-BY public copyright license to any Author Accepted Manuscript (AAM) version arising from this submission. This research made use of the SIMBAD and VIZIER databases at CDS, Strasbourg (France) and the electronic bibliography maintained by the NASA/ADS system.
\end{acknowledgements}


\bibliographystyle{aa} 
\bibliography{aanda} 

\begin{appendix}

\section{Analysis workflow}
\label{appendix: Analysis workflow} 
\begin{figure*}
\begin{center}
{\includegraphics[height=5.5cm,width=0.65\linewidth]{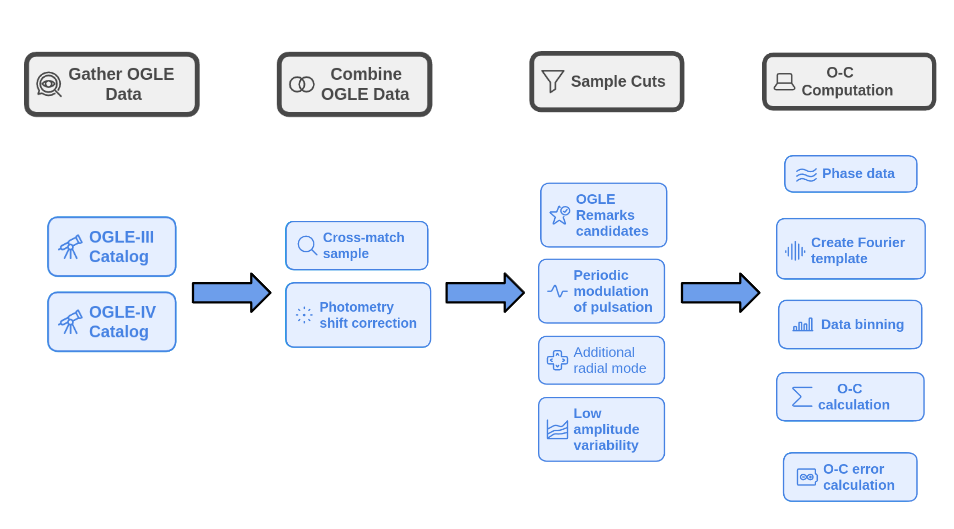}}
\caption{Schematic workflow of the $O-C$ analysis pipeline.}
\label{fig:pipeline1}
\end{center}
\end{figure*}

\begin{figure*}
\begin{center}
{\includegraphics[height=6cm,width=0.8\linewidth]{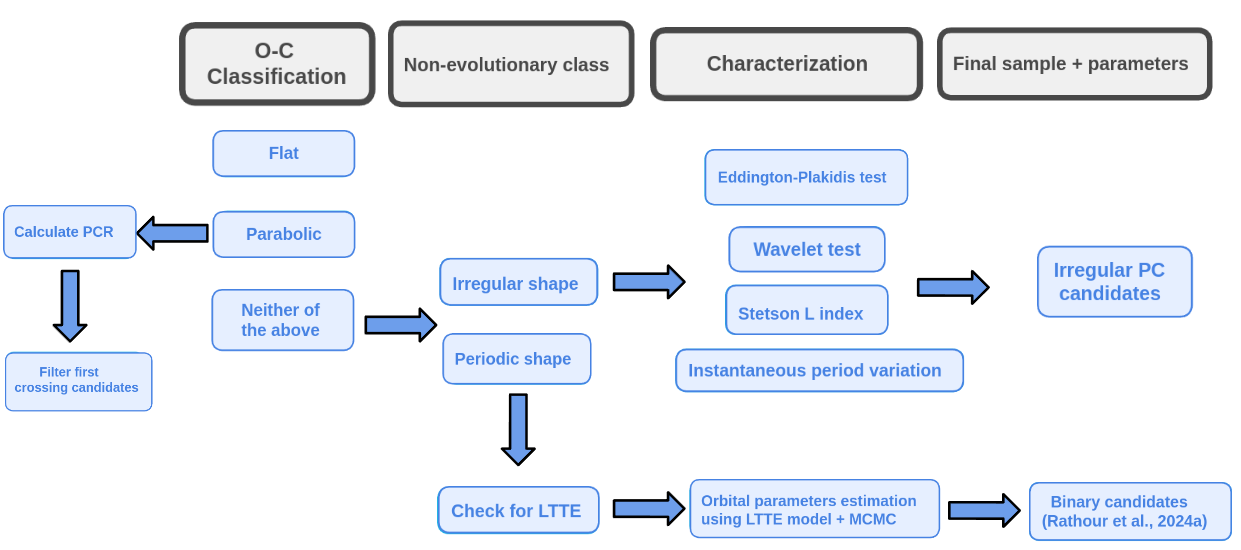}}
\caption{Schematic workflow of the $O-C$ characterisation pipeline.}
\label{fig:pipeline2}
\end{center}
\end{figure*}

Here we outline the analysis workflows mainly consisting of two parts: (i) $O-C$ analysis; and (ii) $O-C$ characterisation. The schematic of the workflows is presented in Fig.~\ref{fig:pipeline1} and ~\ref{fig:pipeline2}.

\section{Bayesian formalism}
\label{appendix: Bayesian formalism} 

In Bayesian formalism, the probability of parameters ($\Theta$) given the data (D) and model (M) is defined as:
\begin{eqnarray}
\label{eq:ultranest_eq1}
Pr(\Theta|D,M) = \frac{Pr(D|\Theta,M) \hspace{0.1cm} \text{x}\hspace{0.1cm} Pr(\Theta|M)}{Pr(D|M)}
\end{eqnarray}
or
\begin{eqnarray}
\label{eq:ultranest_eq2}
    \text{Posterior} = \frac{\text{Likelihood \hspace{0.1cm}x\hspace{0.1cm} Prior}} {\text{Evidence}}
\end{eqnarray}

In a test case run, in the linear model, \texttt{UltraNest} conducted 14,849 likelihood evaluations, achieving an efficiency rate of approximately 74.83\%. In contrast, the parabolic model required 33,900 evaluations, with an efficiency rate of 53.33\%, reflecting the increased complexity of the parameter space. The effective sample size (ESS) was 1605.4 for the linear model and 1902.3 for the parabolic model, both significantly exceeding the minimum requirement of 400, indicating that the posterior distributions were well-sampled. The Kullback-Leibler test \citep{Kullback1951} confirmed good convergence, with values of 0.45 $\pm$ 0.10 for the linear model and 0.46 $\pm$ 0.06 for the parabolic model, demonstrating that the posterior uncertainty strategy was satisfied.

\section{Gaussian Process Regression}
\label{appendix: Gaussian Process Regression} 

The kernel used for our Gaussian Process Regression using \texttt{scikit-learn} library \citep{Pedregosa2011scikit-learn} is composed of the following kernels:
\begin{eqnarray}
k1_{\text{RBF}}(x, x') = \sigma^2 \exp\left( -\frac{(x - x')^2}{2 \ell^2} \right)
\end{eqnarray}

\begin{eqnarray}
k2_{\text{ExpSineSquared}}(x, x') = \sigma^2 \exp\left( -\frac{2 \sin^2\left( \frac{\pi |x - x'|}{p} \right)}{\ell^2} \right)
\end{eqnarray}

\begin{eqnarray}
k3_{\text{White Noise}}(x, x') = \sigma^2 \delta(x, x')
\end{eqnarray}
where $\sigma$ is a scaling factor; $\ell$ is the length scale parameter; $p$ is the periodicity parameter. For the white noise kernel, $\sigma$ is the noise level and $\delta(x, x')$ is the Kronecker delta function, which is 1 if $x=x'$ and 0 otherwise. The values of the kernel parameters are kept fixed throughout the analysis (RBF: length scale=10.0, length scale bounds=(1e-5, 1e5); ExpSineSquared: length scale=10.0, length scale bounds=(1e-5, 1e5), periodicity=1.0, periodicity bounds=(1e3, 1e6); WhiteKernel: noise level=0.01, noise level bounds=(1e-10, 1e2)). The parameter \texttt{n restarts optimizer} is set to 10, which indicates the number of times the optimiser will restart with different initialisations to improve finding optimal kernel hyperparameters in GPR. The overall kernel is a combination written as:
\begin{eqnarray}
k(x, x') = k_{\text{RBF}}(x, x') + k_{\text{ExpSineSquared}}(x, x') + k_{\text{White Noise}}(x, x')
\end{eqnarray}

\section{Binary analysis: OGLE-LMC-CEP-2840}
\label{appendix: Binary Analysis of OGLE-LMC-CEP-2840} 
\begin{figure}[!h]
\begin{center}
{\includegraphics[height=4.5cm,width=0.95\linewidth]{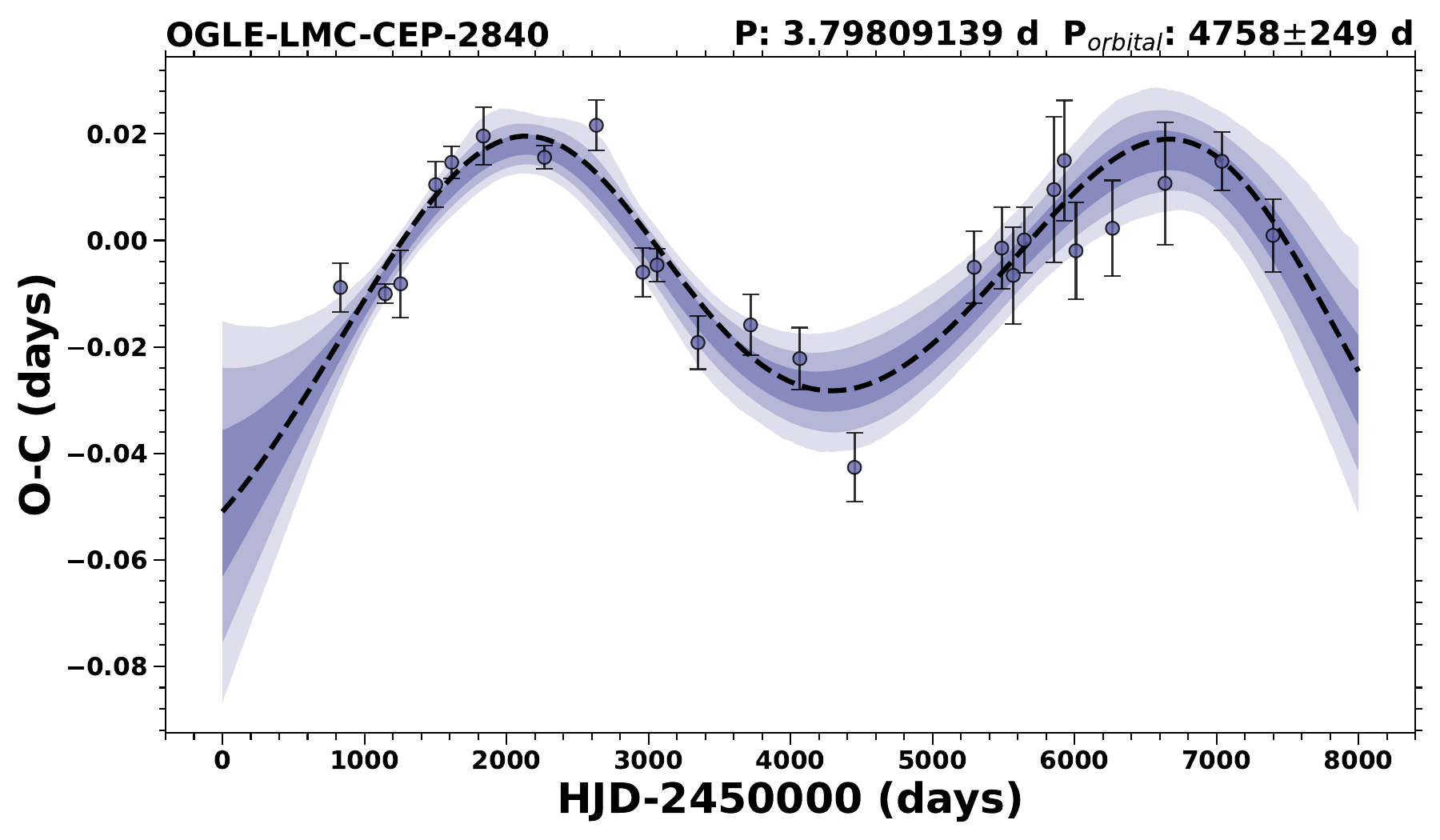}}
\caption{Binary model fit (following the method described in \citepalias{Rathour2024A&A...686A.268R}) of the residual O-C diagram of OGLE-LMC-CEP-2840 (also see Fig. \ref{fig:Segovia_examples}). The black dashed line represents the orbital solution obtained from the median parameter values of the posterior distribution. The blue-shaded regions contain the ranges (credible intervals) of individual MCMC solutions according to one, two, and three standard deviation at a given point in time, in order of decreasing transparency. Above the panel the OGLE-ID, adopted pulsation period to construct the $O-C$ diagram and the orbital period are shown.}
\label{fig:Binary_LTTE_plot}
\end{center}
\end{figure}

The $O-C$ diagram of OGLE-LMC-CEP-2840 indicated a departure from the parabolic shape, therefore, the residuals were investigated further. These show a periodic signature which is reminiscent of the signal expected of binarity. These residuals were fit with a parabola + LTTE model, following \citepalias{Rathour2024A&A...686A.268R}, and this model is shown in Fig.~\ref{fig:Binary_LTTE_plot} (upper panel).

Our model suggests an orbital period of $4758.3\pm249$ days with a low eccentricity of $0.11\pm0.08$. The derived mass function $f(m)$ is $0.446\pm0.100$ M$_{\odot}$. The Cepheid OGLE-LMC-CEP-2840 is a first crossing candidate as reported by \cite{Rodriguez-Segovia2022MNRAS.509.2885R} and we conclude that it is also a strong binary Cepheid candidate.

\begin{figure}[!h]
\begin{center}
{\includegraphics[height=8.5cm,width=\linewidth]{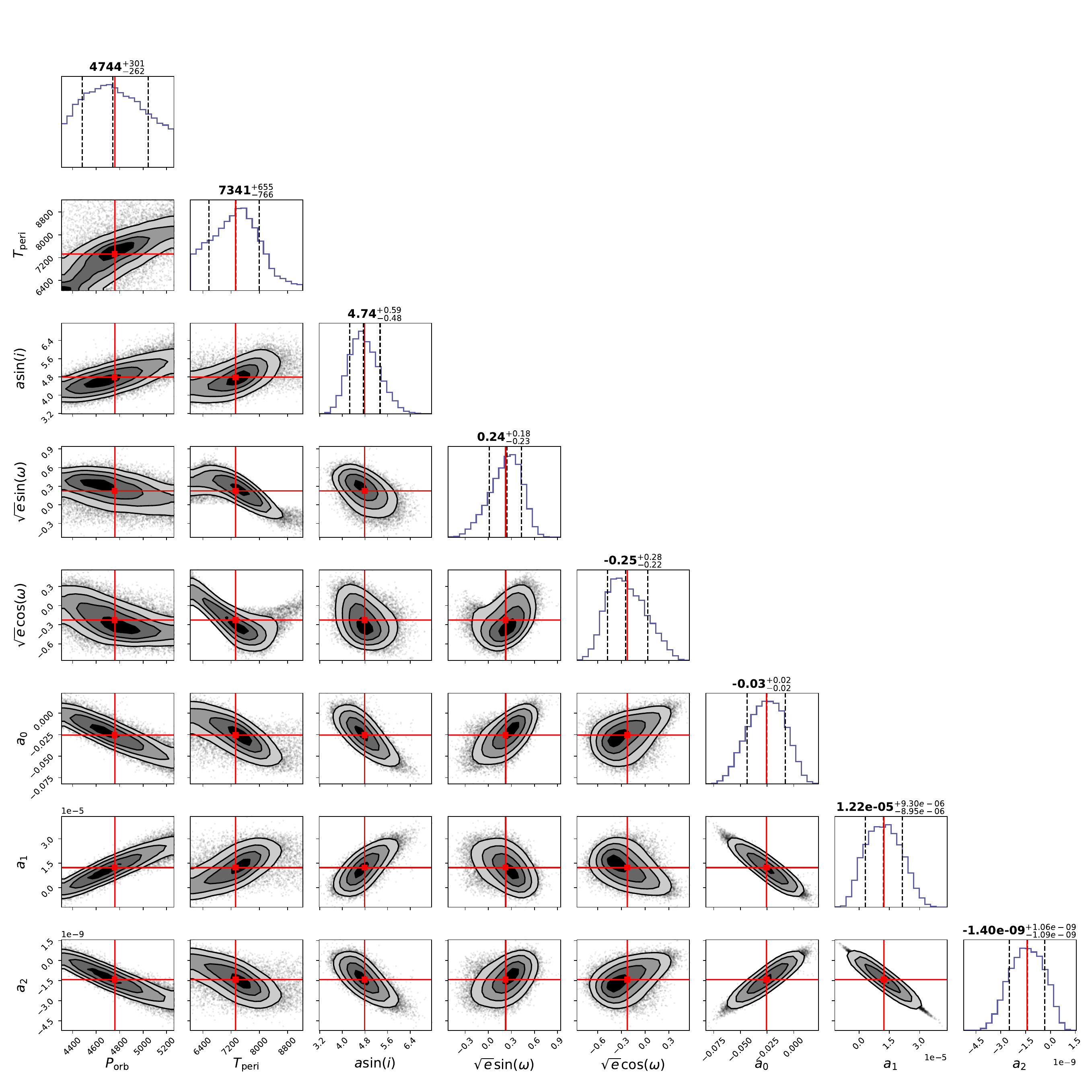}}
\caption{Posterior distribution of orbital parameters of OGLE-LMC-CEP-2840.}
\label{fig:Binary_MCMC_plot}
\end{center}
\end{figure}

\end{appendix}
\end{document}